# Physical and Mechanical Properties of Cu-Fe System Functionally Graded and Multimaterial Structures after the DED

Konstantin I. Makarenko [a, *], Oleg N. Dubinin [a, b], Stepan D. Konev [a], and Igor V. Shishkovsky [a]

[a] Skolkovo Institute of Science and Technology, Bolshoy Boulevard 30, bld. 1, Moscow, 121205, Russian Federation

[b] World-Class Research Center «Advanced Digital Technologies», Saint Petersburg State Marine Technical University, Lostmanskaya street 3, Saint Petersburg, 190121, Russian Federation

*Corresponding author. E-mail: konstantin.makarenko@skoltech.ru

**Abstract:** This paper is devoted to experimental characterisation of linear thermal expansion coefficient (LTEC) and mechanical characteristics of the laser deposited Cu-Fe system multilayer functionally graded (FG) structures and binary Cu-Fe alloys, fabricated from the tin, aluminium, and chromium bronze with 89–99 wt.% of copper and stainless steel (SS) AISI 316L with 1:1 and 3:1 bronze-to-steel ratio. The best tensile mechanical strength of as-built parts is demonstrated by the aluminium bronze – stainless steel 1:1 alloy and reaches 876.4 MPa along with low elasticity modulus (11.2 GPa) and 1.684±0.3 $K^{-1}$ LTEC. Contrarily, the worst values of the mechanical characteristics are exhibited by parts created from the chromium bronze and SS, which failed at 294.0–463.3 MPa ultimate stress, showed the highest elasticity modulus (up to 42.4 GPa) and comparatively high average LTEC (up to 1.878±0.3 $K^{-1}$). The aluminium bronze – stainless steel binary and FG alloys are discussed in the light of prospective application as the part of gradient materials, created by additive manufacturing (AM) technologies via the gradient path method and the alternating layers technique, with expected possibility of application in aerospace, nuclear, and electronic industry due to advantageous combination of the antifrictionality, heat conductivity, and oxidation resistance of the bronze, and the high mechanical strength, corrosion and creep resistance of the stainless steel.

**1. Introduction**

The natural environment all around is saturated of numerous examples of materials and structures with unique gradient of arrangement, dimension, orientation, distribution and other properties, which continuously or intermittently variate within the boundaries of an indiscerptible object. These examples include integration of multiple gradients of the teethes of people and animals, hierarchical gradients of their bones [1], skin [2], radial gradients of the stems of trees [3] and plants, hydration gradient of the horse hoof, mineralization gradient of the crayfish mandible, structural gradients of the pangolin scale, elk antler, wheat awn, and so on [4-13]. Due to growing interest and the expanding request in the different types of such gradient materials from a side of the real industry, and a lot of practical applications that endure a shortage in parts with the gradient properties, a new group of the engineering materials has evolved: the functionally graded materials (FGMs) [2-3,14-16], also known as compositionally graded materials (CGMs) [17], and, at whiles – multi-material structures [18-19]. According to the chemical composition, they may be divided into the six broad groups [20] that have the highest opportunities and importance for the practical implementation: Ni-based FGMs; Ti-based FGMs; Al-based FGMs; Fe-based FGMs; Cr-based FGMs; and Cu-based FGMs. The current research concerns all of them in a certain degree, but is mainly devoted to investigation of the Cu-Fe FGMs that correspond to the last group. These materials due to their antifrictionality [21], high heat conductivity, low thermal expansion, good oxidation resistance of bronze [20,22], and high mechanical strength, corrosion and creep resistance of steel [23] are expected to be good candidates in such areas of industry as aerospace, steam turbine and power nuclear plants and electronic components production [22-24]. Possibility of additive manufacturing of steel-copper and steel-bronze FGMs with improved characteristics in comparison with base metals, or unique particularities in their thermal and mechanical properties, is described in a great amount of recent works:

a) C. Tan et al. [25] reported about fabrication of maraging steel-copper FGM with high interfacial strength via laser power bed fusion (LPBF); moreover, it showed hardness in the transition region much higher than center of steel area (3.54 GPa vs. 2.6-2.9 GPa); this increase of hardness in the border area between steel and bronze was also demonstrated in the study [26];

b) Z.H. Liu et al. [27] developed a multimaterial structure from UNS C184000 copper alloy and SS 316L via selective laser melting (SLM) with 310±18 MPa tensile strength and obtained a good metallurgical bonding at the interface of steel/copper laminates;

c) SLM was also applied for fabrication of the SS 316L and C52400 copper alloy-based FGM in the study [28]; it was reported about excellent metallurgical bonding between alloys due to their mixing at the interfaces. The several defects in the FGM were observed: cracks, oxides and pores near the interface. Cracking extending to the SS 316L side was associated with the crystal lattice parameter mismatch and stress mismatch because of difference in physical characteristics (melting points, heat conductivity and thermal expansion) of the SS and C52400.

d) X. Zhang et al [29] described the laser deposition of copper on the SS 304L that was conducted directly and through the interlayers of a nickel alloy Deloro 22 (D22). The poor tensile strength was shown in the first case (about 124.4 MPa), the transition zone was free of cracks, but had porosity. The bonding strength was higher in the second case, it reached 226.2 MPa for Cu-D22 interface, and 648.2 MPa for D22-SS transition. A 100% increase of thermal conductivity of Cu-D22-SS system was obtained in comparison with pure SS 304L.

e) The maraging steel (MS) 300 – T2 copper functional bimetal was fabricated via hybrid SLM combined with a subtractive process (computer numerical control (CNC)-machining) [30]. It demonstrated 557 MPa flexural strength, what is higher than that of the pure T2 copper. The interface area was found to be free of defects and well metallurgical bonded. Generally, in case of MS-based FGMs, the additional treatment after the DED is often recommended, because the decrease of porosity usually provides the increase of residual stresses and lowing of such important parameter as the fatigue strength [31].

The most technologically simple direct joining (direct deposition [32]) method [14] of the FGMs fabrication can be successfully applied almost only in the case of the similar materials such as different kinds of stainless steel, or for deposition of the reinforced hardening coatings on the metal substrate, if the chemical composition and thermomechanical characteristics of the metal matrix of the coating and the material of the substrate are significantly close to each other. A usability of the direct joining method in the mentioned cases was demonstrated in the following studies:

a) B. Heer and A. Bandyopadhyay reported in the work [17] about fabrication of the binary SS 340-SS 316 functionally graded material deposited on the SS 316 surface; the resulted structure demonstrated functionally graded both magnetic (demonstrated on the SS340 side) and non-magnetic (demonstrated on the SS316 side) properties;

b) C.H. Zhang et al. [33] created a multilayer gradient stainless steel-based structure deposited on the 316 stainless steel substrate via direct joining method. The structure consisted of the five 5 mm-width areas with different Cr to Ni ratio: 1.4; 1.7; 2.3; 3.7; 9.7, which increased towards the build direction. The part had no obvious defects, proved that a good metallurgical bonding can be obtained between the as-deposited part and the stainless steel substrate, and demonstrated the graded change of the microhardness with the variation of the chemical composition (the ultimate value was found to be equal to 6.90 GPa);

    c) The direct joining was applied for the metal matrix composite (MMC) fabrication for niobium carbide – stainless steel 304 system [34]: T. Gualtieri and A. Bandyopadhyay demonstrated a direct deposition of NbC reinforced SS304 via LENS$^{TM}$ resulted in the strengthening of the matrix metal, increase of its hardness, and a 75% decrease of the wear rate of the coating compared with the pure stainless steel;

    d) The metal-ceramic composite of CaP and Ti was fabricated by A. Bandyopadhyay et al. via LENS$^{TM}$ [35]: The CaP-Ti coating increased the wear resistance and strength of the commercially pure (CP) titanium and Ti64:

    – up to 92% when the 10% wt. CaP was added to the CP Ti metal matrix;

    – up to 70%, when the 5% wt. CaP reinforced the Ti64 alloy.

The predominating issue with fabrication of FGMs based on two or more initial materials with rather different thermomechanical characteristics (melting point; laser radiation absorption coefficient; heat conductivity; thermal diffusivity; thermal expansion coefficient; elasticity modulus), such as stainless steel and copper; stainless steel and bronze; stainless steel and aluminium; nickel superalloys and bronze; etc. is a physical inability to create them via the direct joining method because of embrittlement and intense liquation and solidification cracking in the transitional area between initial materials due to several key factors:

    – limited miscibility of the base materials and their crystal lattice parameters mismatch;

    – overheating of the material that has significantly lower melting point than the another one;

    – forming of stress concentrators and residual thermal stresses caused by different factors, one of which is a stepwise change of thermal expansion coefficient and elasticity modulus by crossing of the interface;

    – difference in thermal history of separate layers that were fabricated from different materials using varied treatment conditions;

    – forming of the brittle phases, including intermetallics, near the interface between base components of the gradient alloy.

The above listed phenomena were shown for the system of Ni-Ti in the study [32] (direct deposition of GRCop-84 and Inconel 718 showed poor diffusion and low surface quality because of metal lumps and balling); for the system of Cu-Fe – in the studies [24,26] (direct fabrication of Cu-Fe functionally graded structure demonstrated local increase of microhardness up to 266 HV and elasticity modulus up to 43 GPa that resulted in intense cracking in the border area); cracking was observed in the transitional area between 100% CoCrMo and 100% Ti-6Al-4V [36]; for Fe-Ti system – in the following works:

a) cracking in the transitional area between directly joined Ti64 and SS 410 was demonstrated in the work [37] – authors suggested it appeared because of residual stresses caused by difference in thermal characteristics of the base materials;

b) M. Ghosh et al. [38] reported about the presence of the brittle intermetallics (σ, $Fe_2Ti$, FeTi, $Fe_2Ti_4O$) in diffusion bonds fabricated from Ti-5.5Al-2.4V and SS 304, stated that these intermetallics are responsible for lowering of the mechanical strength, and observed the failure in the region between FeTi and β-Ti phases during the tensile tests;

c) Diffusion bonding between a titanium alloy (TA17) and steel (0Cr18Ni9Ti) was also discussed in the study [39]: the σ, $Fe_2Ti$, Fe-Ti intermetallic phases, and β-Ti were observed in the reaction zone, and the decrease of the mechanical strength (the maximal value was equal to 307 MPa) and plasticity were associated with the brittle intermetallic Fe-Ti phase.

Several techniques attributed to different manufacturing technologies (such as SLM; spark plasma sintering – SPS; thermal spraying; electron beam melting – EBM; selective laser sintering – SLS; direct energy deposition – DED) – were suggested to resolve the issue of direct joining method in case of base materials with rather different thermomechanical characteristics. These techniques include:

– intermediate section method [14] (compositional bond layer build strategy [32]) that proved itself:

a) in the study of H. Sahasrabudhe et al. [37] who demonstrated advantages of Ni-Cr applied as a 1.5 mm length intermediate section (bond layer) between SS and Ti64, observed elimination of cracking and showed that the only brittle intermetallic phase in the transitional region of 3-component alloy detected by the X-ray diffraction (XRD) analysis was $Cr_2Ti$;

b) being conducted by S. Kundu et al [40]: commercially pure titanium and SS 304 were joint through the pure nickel 300 μm layer under 3 MPa excessive pressure in the 800…950 ℃ temperature range via solid state diffusion bonding; the resulted parts demonstrated ~311 MPa maximum tensile strength and ~236 MPa shear strength along with absence of the brittle intermetallic Fe-Ti phases;

c) in the similar work [41] also devoted to diffusion bonding of FGM: pure titanium and AISI-1008 low carbon steel were joint through 100 μm intermediate section of the copper-based alloy (Cu–12Mn–2Ni, wt.%). Hard and brittle Ti-C and Fe-Ti phases didn't precipitate the interface area of the resulted alloy, what was proved by the results of XRD analysis;

d) in the study [42] of W. Li et al, where three intermediate sections of V, Cr, and Fe were applied simultaneously to fabricate a functionally graded structure between Ti6Al4V and SS 316 using laser metal deposition (LMD). The result showed no cracking, no brittle intermetallic phases within the interface areas between nearby layers of different materials, absence of σ phase at the border between chromium and iron. The results of scanning electronic microscopy (SEM) showed presence of elongated lathy microstructure with tiny epitaxial grains caused by the high rate of cooling;

e) YanBo Bi et al. [43] demonstrated the possibility TC4 titanium alloy and SS 304 joining through the 300 μm width intermediate sections of V and V+Cu; this technique excluded the formation of brittle Ti-based intermetallics and allowed to fabricate a FGM with maximal tensile strength of 92 MPa in case of V intermediate section, and 181 MPa in case of V+Cu.

– alternating layers (multilayer transition) technique [24,26], characterized in the following studies:

a) C. Tan et al. [44] fabricated an FG structure from two high-strength steels: maraging steel C 300 and SS 420. The resulted alloy showed ultrahigh tensile strength equal to 1.32 GPa with 7.5% elongation. Such growth of mechanical strength in comparison with initial materials was associated with hetero-deformation induced strengthening;

b) D. Melzer et al. [45] produced the alternating layers gradient structure from SS 316L and Inconel 718 via the DED, and reported about yield and ultimate tensile strength of the interface areas close to these values of stainless steel, without dependence from the transition type (gradient smooth transition or sharp border between nearby layers);

c) absence of cracking in the transitional zone between aluminium bronze and stainless steel, deposited using this technique, was discussed in [26]. The Cu-Fe FGM, fabricated on the same operation conditions via direct joining, showed intense cracking in its border area due to precipitation of the hardening phases such as AlNi and $Cr_{0.7}Fe_{0.3}$;

d) the alternating layers printing sequence was applied to study a cracking process at the interfaces between DED-fabricated SS 316L and nickel alloy Inconel 625 [46-47]. N. Chen et al. [47] demonstrated that two types of the resulted interfaces (Inconel under steel and vice versa) have unique morphologies (first type was characterized by epitaxial growth of Inconel grains on the grains of steel, the second type had lack of the epitaxial growth and showed intense solidification, liquidation, and further ductility dip cracking).

e) N. Koga et al [48] discussed accumulative roll bonding of Cu-Fe sheets (3, 7, 50, 100 and 1000 alternating layers). It was shown that soft Cu layers may play a damping role between hard Fe layers and suppress the cleavage cracking in the Fe layers. The maximal tensile strength was associated with 1000-layer sheet (above 500 MPa). The positive effect of number of layers on the tensile mechanical properties was associated with grain refinement strengthening. It was also observed that the layered structure exists when the number of sheets is less than 100, and the 1000-layer sheet shows the network-like morphology pattern instead as the dual phase Cu-Fe cast alloy [49].

– gradient path method [14] (compositional gradation build strategy [32]) that successfully proved itself in the great amount of experimental studies. In particular:

a) this method was conducted by J. Park et al. [50] during the fabrication of a functionally graded ZnS: Cu, Cl using spark plasma sintering (SPS);

b) X. Lin et al. [51] applied this technique to create a SS 316L/Rene88DT superalloy FGM with linear change of the Rene88DT from 0% to 100% within 100 layers (40 mm) using laser freeform fabrication system;

c) B.E. Carroll et al. [52] reported about fabrication of the gradient SS 304L-Inconel 625 (INC) material using gradient zone consisted of 6 areas: 96% SS + 4% INC; 83% SS + 17% INC; 67% SS + 33% INC; 55% SS + 45% INC; 43% SS + 57% INC; 27% SS + 73% INC, but this material demonstrated cracking observed in the high-magnification image in the zone near the second, enriched by Nb and Mo;

d) the results of similar technique were provided by L. Li et al [53] during laser melting deposition of functionally graded $TiC_p$/Ti6Al4V with gradual increase through variation of feed rate of TiC from 0 vol.% up to 50 vol.% within 37 mm transitional zone (the ultimate tensile strength was improved by 12.3% in comparison with Ti6Al4V);

e) gradual increase of the Co-Cr-Mo coating on the Ti-6Al-4V substrate was demonstrated by B. Vamsi Krishna et al. [36]; the resulted FGM showed absence of cracking along with nontoxicity and biocompatibility, what is a key advantage of such materials that are widely used in total hip replacement (THR) and total knee replacement (TKR) surgeries;

f) T. Gualtieri and A. Bandyopadhyay [54] discussed functionally graded SS 304-vanadium carbide (VC) coating with 5 to 100 wt% ranging of VC fabricated by LENS[TM];

increase of hardness and wear resistance was achieved due to composite coating of SS 304 and VC, and 100% additionally deposited VC increased these parameters in excess;

g) B. Onuike et al. [32] applied laser engineered net shaping (LENS$^{TM}$) for development of graded GRCop-84-Inconel 718 alloy with a compositional gradation through the 200 μm compositional section of 50 wt.% Inconel 718 and 50% wt.% GRCop-84 premixed powders, and reported about the increase of its thermal conductivity by ~300% more than Inconel 718 and 250%-improved thermal diffusivity.

The combined gradient path-intermediate section method was applied by L.D. Bobbio et al. [55]: joint of 100% Ti-6Al-4V and SS 304L via DED was conducted through five intermediate subsections with varied vanadium (V) percentage: 75 vol% Ti-6Al-4V + 25 vol% V; 50 vol% Ti-6Al-4V + 50 vol% V; 25 vol% Ti-6Al-4V + 75 vol% V; 25 vol% SS 304L + 75 vol% V; 50 vol% SS 304L + 50% V. The deposition was stopped at the last section because of intense cracking at the transition between 25 vol% Ti-6Al-4V + 75 vol% V and 25 vol% SS 304L + 75 vol% V, and also between 25 vol% SS304L + 75 vol% V and 50 vol% SS 304L + 50 vol% V, what was associated with precipitation of the hardening brittle Fe-Ti and σ-FeV phases.

Almost the same result was provided by A. Reichardt et al [56]: transition from the pure Ti-6Al-4V to SS 304L was conducted by addition of intermediate sections with the same vanadium percentage as in the previously discussed study [55]. The authors finished the deposition process at the point 25 vol% SS 304L + 75 vol% V, one stage earlier than it was shown in [55], due to cracking between 25 vol% Ti-6Al-4V + 75 vol% V and 25 vol% SS 304L + 75 vol% V zones associated with precipitation of the brittle Fe-V-Cr σ phase. The results of these two works [55-56] allow to make a suggestion that such technique is not well appropriate to fabrication of Ti-6Al-4V + SS 304 L FGM. The results of the study [42], discussed above, are more successful, and triple intermediate section technique, applied in this research, gave the better result without cracking and other visible defects.

As for the gradient path method, as for the combined gradient path-intermediate section technique, there is an opportunity to fabricate a Cu-Fe FGM through the intermediate binary alloy consisting of the two base components – bronze and steel – taken in the various ratios. Such binary alloys, created via the AM, are the actually new materials with promised physical and mechanical properties (flexural and tensile strength, ductility, rigidity, microhardness, thermal expansion, heat conductivity, thermal diffusivity, magnetic susceptibility and permeability, etc.), phase and structure composition (including crystal lattice parameters, grain shape and size, dimensions of pores and amount of porosity). The first priority task and the main purpose of the conducted research is evaluation of the several most critical items in the mentioned row of parameters, and determination of dependencies between them and the chemical composition of base materials, laser treatment conditions, and the deposition schemes, applied for fabrication of the binary alloy. The results of such analysis can be used for the further prediction of the physical and mechanical properties of the binary sections in the gradient alloys, selection of the most prospective combinations of steel and bronze that could be implemented in AM of FGMs. Moreover, not only gradient structures are the object of interest; bimetallic Cu-Fe alloys themselves also have a wide field of practical applications due to significant influence of copper on mechanical properties of steel. In particular, it is possible to increase creep strength of low carbon austenitic steel by inclusion of 4 wt% copper in its structure [57-58] along with reduction of ductility of the material; temperature strength of austenitic steels in the area of ~ 650-750 °C can be also improved by addition of Cu, what was demonstrated for SUS 304H and 3 wt% Cu in the huge amount of works [58-67].

What is more, it should be mentioned that there is a possibility to improve the DED technology for Cu-Fe FGMs and increase the laser radiation absorption coefficient in the infrared area of spectrum, which is significantly low in case of bronze. Wavelengths of 1.05–1.50 μm are the most common for the industrial laser sources – fiber lasers (such as doped by Er or Yb). Therefore, it is reasonable to use auxiliary dopants to the bronze sections of the gradient parts to increase the percentage of absorbed radiation and cause the improved heat consumption, thus increasing the production efficiency. These dopants also form the new Cu-Fe materials with uninvestigated exploitation parameters, so their optimal percentage, and physical properties of the resulted complex binary alloys created via the DED, should be studied specifically.

## 2. Materials and Methods

In this research the stainless steel (SS) 316L–5520 (fraction 50-150 μm) manufactured by Höganäs Belgium SA, and three different kinds of bronze: aluminium bronze («ПР-БрАЖ9,5-1», similar to UNS C61800 aluminium bronze, fraction 45-125 μm), tin bronze («ПР-БрО10», similar to copper-tin alloy CuSn10-B (CB480K), fraction 100-140 μm), and chromium bronze («ПР-БрХ», similar to UNS C18400 chromium copper, fraction 63-125 μm), manufactured by joint stock company Polema (Tula, the Russian Federation), were applied. The chemical compositions of bronzes are presented in **Table 1**. The chemical composition of SS 316 L is described in **Table 2**.

**Table 1.** Chemical compositions of applied bronzes

| Material | Cu, % | Al, % | Cr, % | Fe, % | Ni, % | $O_2$, % | P, % | Pb, % | S, % | Sb, % | Si, % | Sn, % | Zn, % |
|---|---|---|---|---|---|---|---|---|---|---|---|---|---|
| Aluminium bronze | Base | 9.5 | - | 1.0 | - | - | - | 0.02 | - | 0.05 | 0.1 | 0.05 | 0.05 |
| Tin bronze | Base | 0.05 | - | 0.1 | 0.1 | - | - | 0.05 | - | 0.05 | 0.05 | 9.96 | 0.11 |
| Chromium bronze | Base | - | 0.76 | 0.05 | 0.05 | 0.053 | 0.018 | - | 0.0043 | - | - | - | - |

**Table 2.** Chemical composition of SS 316L

| Fe, % | C, % | Cr, % | Mb, % | Mn, % | N, % | Ni, % | P, % | S, % | Si, % |
|---|---|---|---|---|---|---|---|---|---|
| Base | 0.03 | 16.00–18.00 | 2.00–3.00 | 2.00 | 0.10 | 10.00–14.00 | 0.045 | 0.030 | 0.75 |

The research was aimed to investigation of three different kinds of the transitional zone for each implemented type of bronze (Al-type, Cr-type, and Sn-type), targeted to application in the Cu-Fe FGMs. These kinds included 25 wt.% and 50 wt.% stainless steel, added to the bronze base, what is common concentrations for the gradient path method, and the multilayer gradient steel-bronze structure, that is used in case of the alternating layers technique. Therefore, totally 8 groups of specimens were suggested for the research (**Table 3**) (the multilayer combination of Sn bronze and stainless steel was excluded from consideration, because the pure Sn bronze layers showed poor laser manufacturability, low heat consumption, bad adhesion with steel, and didn't obtain an appropriate shape during the fabrication). Here and below, "alt." means "alternating layers", and refers to the functionally graded Cu-Fe alloy.

**Table 3.** Experimental specimens' groups

| Group 1 | Group 2 | Group 3 | Group 4 | Group 5 | Group 6 | Group 7 | Group 8 |
|---|---|---|---|---|---|---|---|
| Al bronze + SS (1:1) | Cr bronze + SS (1:1) | Sn bronze + SS (1:1) | Al bronze + SS (3:1) | Cr bronze + SS (3:1) | Sn bronze + SS (3:1) | Al bronze + SS (alt.) | Cr bronze + SS (alt.) |

Direct energy deposition (DED) was conducted using MX-1000 (InssTek, Daejeon, Republic of Korea) technological installation working in a direct tooling mode (DMT) [26]. A 1 kW ytterbium fiber laser was applied as a source of laser radiation. The operation conditions for all groups are listed in **Table 4** (the last two columns in the table show the processing parameters of the bronze layers of functionally graded specimens of the Groups 7 and 8; conditions of SS 316L layers of specimens of these Groups are particularly specified in **Table 5**) [20,24,26]. Geometrical parameters of the deposited powder beds are shown in **Figure 1.**

**Figure 1.** Scheme and geometrical parameters of the laser deposited powder beds (not to scale)

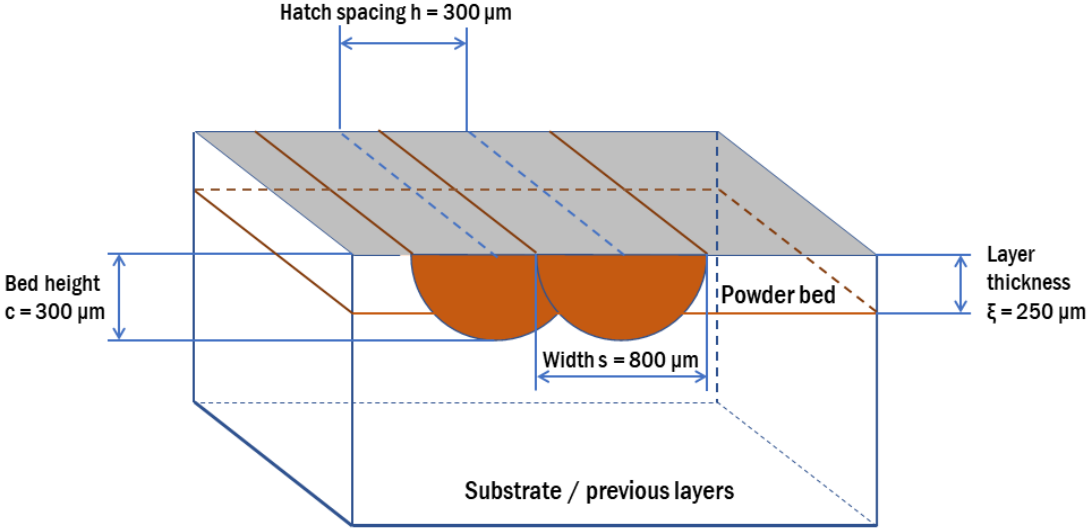

**Table 4.** Operation conditions of all specimen groups

| Parameter | Group 1 | Group 2 | Group 3 | Group 4 | Group 5 | Group 6 | Group 7 | Group 8 |
|---|---|---|---|---|---|---|---|---|
| Average laser power, W | 308 | 550 | 500 | 500 | 500 | 500 | 500 | 500 |
| Ultimate laser power, W | 450 | 800 | 750 | 750 | 750 | 750 | 750 | 750 |
| Scanning speed, m/min | 0.85 | 0.85 | 0.85 | 0.85 | 0.85 | 0.85 | 0.85 | 0.85 |
| Powder rate, g/min | 3.15 | 3.5 | 3.5 | 3.5 | 3.5[a] | 3.5 | 3.5 | 3.5 |
| Coaxial gas flow, L/min | 9.0 | 9.0 | 9.0 | 9.0 | 9.0 | 9.0 | 9.0 | 9.0 |
| Powder gas flow, L/min | 2.0 | 2.0 | 2.0 | 2.0 | 2.0 | 2.0 | 2.0 | 2.0 |
| Shield gas flow, L/min | 10.0 | 10.0 | 10.0 | 10.0 | 10.0 | 10.0 | 10.0 | 10.0 |
| Cooling time between layers, s | 8 | 8 | 5 | 18 | 18 | 5 | 5 | 5 |

[a] 4.0 g/min at first layers powder rate was applied to increase thickness of first layers, which were printed on the cold substrate.

**Table 5.** Operation conditions of SS 316L layers

| Average laser power, W | Ultimate laser power, W | Scanning speed, m/min | Powder rate, g/min | Coaxial gas flow, L/min | Powder gas flow, L/min | Shield gas flow, L/min |
|---|---|---|---|---|---|---|
| 308 | 450 | 0.85 | 3.5 | 9.0 | 2.0 | 10.0 |

The scheme of the laser heat spot movement within the even and odd layers is shown in the **Figure 2** [20,24,26], the similar deposition patterns are also discussed in the related works [68-69].

**Figure 2.** The scheme of the tracks within the single layer

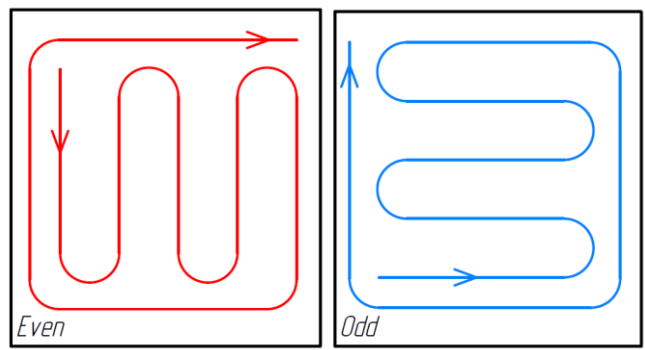

Specimens for tensile tests were prepared via electron discharge machining (EDM) using Mitsubishi-MV-1200R (Mitsubishi Electric Europe B.V., Ratingen, Germany). Tensile tests were performed in accordance with ASTM E8/E8M-16a [70] at a rate of 2.70 mm/min using INSTRON 5969 dual column testing system and an INSTRON Bluehill® Universal materials testing software (Norwood, Massachusetts, United States), later the results were analyzed via Vic-3D Digital Image Correlation (DIC) package (Correlated Solutions, Inc., Irmo, South Carolina, United States). Numerical simulation of the stress-strain state in homogeneous ideal specimen was provided via Dassault Systèmes® (Vélizy-Villacoublay, France) Abaqus/CAE of ABAQUS (Abaqus FEA) software suite. The macrofractography analysis was performed using optical microscope Altami MET 1C (OOO Altami, Saint Petersburg, the Russian Federation). Linear thermal expansion coefficients (LTECs) were estimated using the heating plate and a hand caliper with 0.01 mm uncertainty as a length measuring tool. Each specimen underwent 10 measurements of its length and width on prepared surfaces before and after heating from +21 ℃ (initial temperature was equal to the room temperature) to +220 ℃, respectively. The maximum heating temperature was chosen to be lower than the temperature of low-temperature annealing of bronze; otherwise it could incorporate changes in the results of later mechanical tests. Moreover, this temperature is lower than Curie temperature of SS 316L, which is equal to 279 ℃ [71]; heating over the Curie temperature significantly changes the trend of LTEC due to changes in the material magnetic properties [71].

## 3. Results and Discussion

### 3.1. As-built Deposited Parts

The 8 groups of the experimental specimens were prepared for the research using 8 different corresponding parts. These parts in as-biult state appeared as rectangular parallelepipeds 21x6x75 mm nominal size (**Figure 3**).

**Figure 3.** As-built parts (white squares are 10x10 mm size)

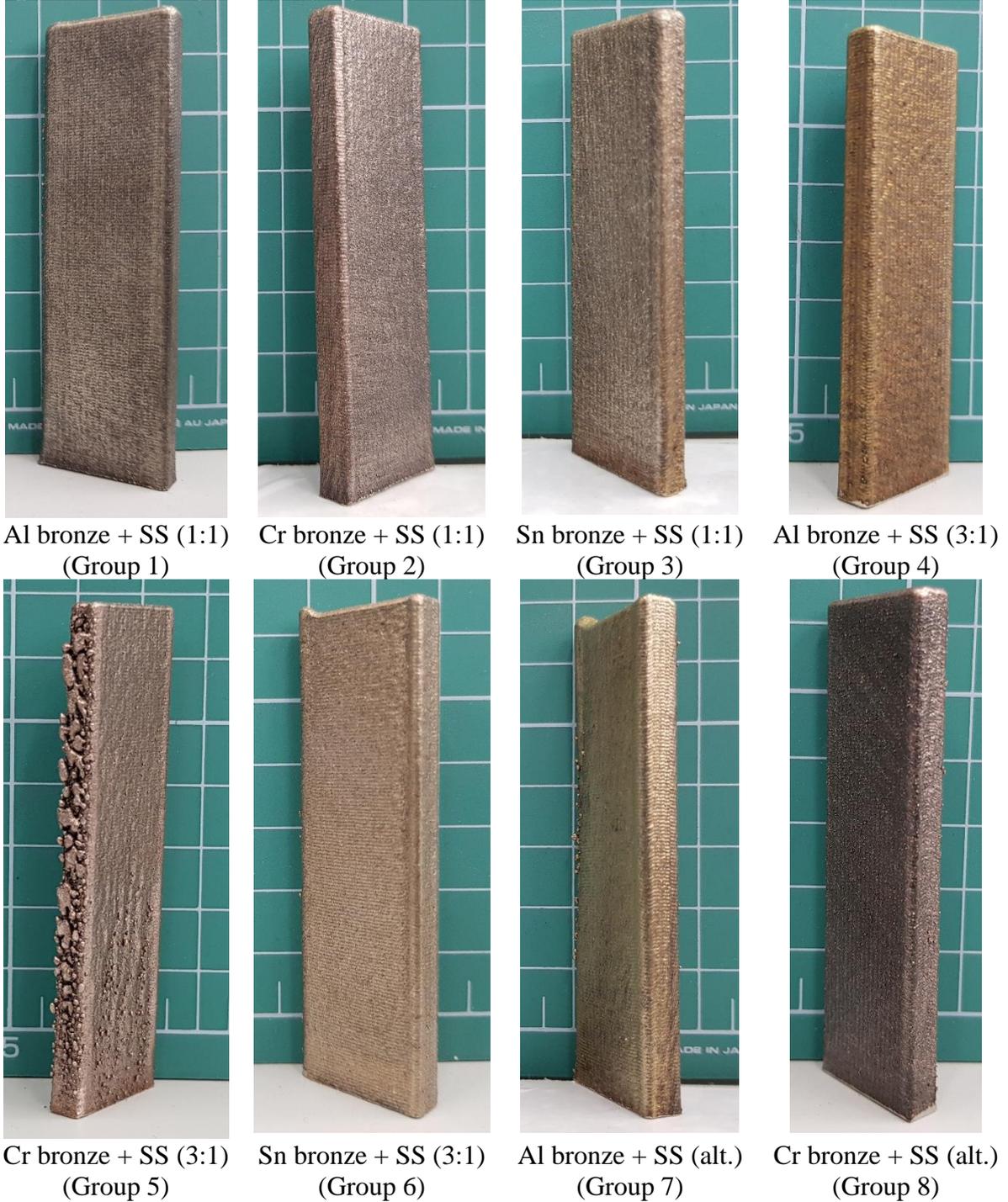

Al bronze + SS (1:1) (Group 1)  Cr bronze + SS (1:1) (Group 2)  Sn bronze + SS (1:1) (Group 3)  Al bronze + SS (3:1) (Group 4)

Cr bronze + SS (3:1) (Group 5)  Sn bronze + SS (3:1) (Group 6)  Al bronze + SS (alt.) (Group 7)  Cr bronze + SS (alt.) (Group 8)

It is seen from Figure 3 that parts of Groups 6 and 7 demonstrate the same defect: a small knobble at the top of the part. It could be associated with powder that falls down at the surface of the overheated part in the end of each layer while laser is turned off, but nevertheless, this defect was not observed in all other parts, and the reason of it is not evident. Presumably, the source of such defect could be an imperfection of substrate of these two parts. The same surface satellite defect as a balled-up protrusion was discussed by M. Liu et al. [72]; it was pointed that a such defect is typically observed at the overlapping points between the start and end points of metal deposition and it could be typically machined off. The non-optimal powder feed rate that delivers too much powder and provides a spatter was mentioned as an another possible source of appearance of such defect [72-73].

Another type of defect is seen in the part of Group 5: agglomerations of the sintered powder particles on the side surface of the part (similar to the balling, which is described by the Plateau-Rayleigh capillary instability [71,74-76] and occur when the melt track tends to break up into spherical balls and becomes discontinious [71]). This part was created from chromium bronze; this material have the highest percentage of copper (about 99%, see Table 1). It was the reason why this part was printed with issues at the beginning of each layer: the laser energy density was lower that its critical value $E_c$, and the surface tension was accompanied with decreased wettability [71]. Corresponding operation conditions for defected and non-defected parts can be found in Table 4. Mentioned defects didn't have influence on the conducted experiments, because the specimens for the mechanical tests were cut from the centre, and measurements of LTEC were conducted on the prepared surfaces, as it was mentioned before.

**3.2. Linear Thermal Expansion Coefficient**

In case of additive manufacturing, it is known that thermomechanical characteristics, which are very significant for Cu-based FGMs, strongly influence on exploitation parameters of the material and depend on, i.a., processing method and operation conditions [29,77]. Therefore, for the developed binary materials, these characteristics should be studied specifically. In this research, LTECs of laser deposited Cu-Fe alloys were experimentally determined and compared with characteristics of pure metals and theoretically estimated values. For a first approximation, it is possible to figure out dependence between concentration of bronze and steel in the binary alloy and its LTEC, using a simple rule-of-mixtures equation (1) [78-79]:

$$\alpha_{Br-SS} = \sum_i \alpha_i \cdot V_i = \frac{N_{Br}}{N} \cdot \alpha_{Br} + \frac{N_{SS}}{N} \cdot \alpha_{SS}, \qquad (1)$$

where $V_i$ denotes a volume fraction of each component of the binary alloy; $N_{Br}$, % is a percentage of the bronze, $N_{SS}$ – of stainless steel, $N = 100\%$, $\alpha_{Br}$ and $\alpha_{SS}$ are the LTECs of the pure bronze and pure stainless steel respectively. It should be noted that this simple formula doesn't take into account the appearance of new phases after 3D printing, which have LTECs different from initial materials, and don't consider the variation of porosity of the parts. This parameters, if necessary, can be taken into account during more precise calculations.

In the temperature range 0–100 °C, $\alpha_{SS} \cong 1.60 \cdot 10^{-5}$ K$^{-1}$ [80]; $\alpha_{Br}^{Al} \cong 1.68 \cdot 10^{-5}$ K$^{-1}$ [80] (aluminium bronze); $\alpha_{Br}^{Cr} \cong 1.65 \cdot 10^{-5}$ K$^{-1}$ [80] (chromium bronze); $\alpha_{Br}^{Sn} \cong 1.80 \cdot 10^{-5}$ K$^{-1}$ [81] (tin bronze). Here $\alpha_{SS}$, $\alpha_{Br}^{Sn}$, $\alpha_{Br}^{Al}$, and $\alpha_{Br}^{Cr}$ are relevant to these materials created via traditional methods, not additively manifactured (analytical estimation for the 3D-printed material using the example of stainless steel 316L will be discussed in this section later). Further calculations consider the assumption that LTEC in the whole range from 21 °C to 220 °C doesn't depend on the temperature (in general, this dependence exists and has a substantially complex form: for each binary intermetallic compound of the laser deposited alloy it can be expressed as a sum of $\alpha_{lat} + \alpha_{el} + \alpha_{mag}$ [82], where $\alpha_{lat} \sim T^3$ is a lattice contribution that can be described in terms of Debye theory; it becomes temperature independent approximately above the Debye temperature; $\alpha_{el} \sim T$ is an electronic contribution, and $\alpha_{mag}$ is a magnetic contribution that is temperature independent in the paramegnetic range and tends to zero above the Kurie temperature).

For 1:1 mixtures (Groups 1-3) $N_{Br} = N_{SS} = 50\%$; for 3:1 mixtures (Groups 4-6) $N_{Br} \cong 75\%$; $N_{SS} \cong 25\%$. In case of alternating layers gradient structures (Groups 7-8), their LTECs are also gradient. The dependence between α and the vertical coordinate *z*, μm, can be defined, in general, using another form of the rule of mixture [79,83]:

$$\alpha(z) = V(z) \cdot \alpha_1 + (1 - V(z)) \cdot \alpha_2, \tag{2}$$

where *V(z)* is volume fraction of constituents (stainless steel, bronze) that obeys a piecewise-defined function (3):

$$V(z) = \begin{cases} 1, \text{if } 500 \cdot n < z < 250 \cdot (2 \cdot n + 1), \\ 0, \text{if } 250 \cdot (2 \cdot n + 1) < z < 500 \cdot (n + 1), \\ \phantom{0,}n = 0, 1, 2, 3, \ldots 149. \end{cases} \tag{3}$$

Within the certain neighboorhoods $\varepsilon(z)$ of the irregular points $z = 250 \cdot n$, the behaviour of the function depends on the phase composition at the interface areas between steel and bronze (most of all, on the forming of intermetallic phases). *V(z)* equation takes a more complicated form if the shrinkage caused by pervasion of each sequential layer into the previous one and their partial intermixing is also considered. But the volume-mean average LTEC of FGMs of Groups 7 and 8 still can be estimated using the equation (1) in the assumption of $N_{Br} = N_{SS} = 50\%$ (same as for Groups 1-3).

Therefore, the results of the theoretical estimation of LTECs of all Groups look like the following (see **Table 6**):

**Table 6.** Estimated LTECs for materials of all groups based on LTECs of pure metals created via traditional methods, $K^{-1}$

| Group 1 | Group 2 | Group 3 | Group 4 | Group 5 | Group 6 | Group 7* | Group 8* |
|---|---|---|---|---|---|---|---|
| $1.640 \cdot 10^{-5}$ | $1.625 \cdot 10^{-5}$ | $1.700 \cdot 10^{-5}$ | $1.660 \cdot 10^{-5}$ | $1.638 \cdot 10^{-5}$ | $1.750 \cdot 10^{-5}$ | $1.640 \cdot 10^{-5}$ | $1.625 \cdot 10^{-5}$ |

*The volume-mean average.

In **Figure 4**, the experimentally obtained thermal coefficients of Groups' 1-8 parts are demonstrated. Four calculations of LTEC in total were made for each part (each calculation was based on length/width measurements made on separate couple of prepared surfaces); accuracy of single LTEC calculation was estimated to be not less than $\pm 3 \cdot 10^{-6}$ $K^{-1}$, so the uncertainty of the arithmetic mean values $\alpha_{Br-SS} = \pm 3 \cdot 10^{-6}/\sqrt{4} = \pm 1.5 \cdot 10^{-6}$, where «4» is a number of calculations as it was mentioned before. This uncertainty can be significantly optimized in the subsequent research using a more precise measurement method and equipment. Different colors in Figure 4 show the results of the different measurements: light-green, blue, yellow and green columns are the measurements №№1-4, and the dark blue color is attribiuted to the arithmetic mean.

**Figure 4.** LTECs of parts of Groups 1-8

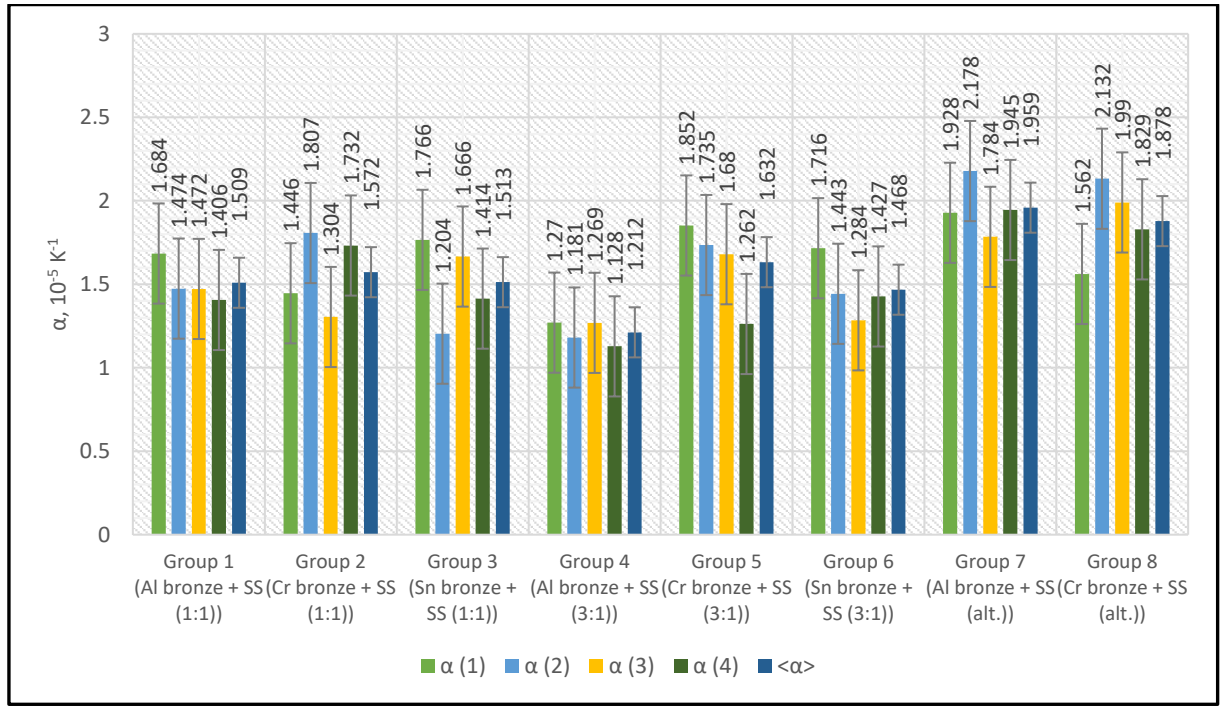

It is seen that the lowest values of LTECs are associated with parts of Groups 4 and 6. Both of them have the highest bronze percentage (75%). Within these two Groups, a binary non-gradient alloy based on aluminium bronze and stainless steel in ratio of 3:1 (Group 4) has the lowest LTEC: 1.212±0.15 K$^{-1}$. It indicates that this material will have the best resistance to linear thermal deformation under influence of high temperatures. Part of Group 5 (Cr bronze + stainless steel), which also has the same bronze percentage (75%), nevertheless, didn't show such a low LTEC. The hypothetic reason of it may be associated with forming of additional Fe-Cr phases in this alloy and the total absence of Fe-Zn, Fe-Sn, and Fe-Al.

All parts created via alternating layers technique (Groups 7-8) have the highest LTEC ((1.878–1.959) ±0.15 K$^{-1}$). It means that they will suffer the highest deformations under the high temperatures what is an undesired property in many practical applications. And, finally, parts with 1:1 steel-to-bronze ratio (Groups 1-3) have intermediate (and approximately similar) LTEC: from 1.509 to 1.572 (±0.15) K$^{-1}$.

It is known that there is a difference between LTECs on the basis of a manufacturing method (traditional technologies vs AM) provided by chemical, phase, and structural composition distinction. Comparing the resulted LTECs with characteristics of pure materials created via traditional methods (for SS 316L α ≅ (1.65–1.75)·10$^{-5}$ K$^{-1}$; for tin bronze: (1.85–1.92)·10$^{-5}$ K$^{-1}$; for aluminium bronze: (1.62–1.70)·10$^{-5}$ K$^{-1}$; for chromium bronze: (1.64–1.70)·10$^{-5}$ K$^{-1}$), and with theoretically estimated results shown in Table 6, we see that the real values are lower in case of Groups 1-6 (1:1 and 3:1 alloys), and higher for Groups 7-8 (FG multilayer materials). Also it is seen that LTECs of Groups 1-2 significantly differ from Groups 7-8, but according to the equation (1) they were expected to be equal to each other. This phenomenon is associated, first of all, with forming of new phases, i.a. intermetallics (see [26] and [84]) in the interface areas between different metals, which have thermal characteristics (i.a. LTEC) different from pure initial materials – steel and bronze.

A consideration of the dependence of LTEC on the temperature and parameters of the laser treatment is the rather more complicated task, that is particularly discussed below regarding to the pure laser deposited stainless steel. A three-way interaction regression model considering the parameters of laser treatment, developed using Minitab software and described by M. Yakout et al [71] in the case of SLM, with could be accurately applied for the DED process too. According to this model, the relative thermal expansion $\varepsilon_T$, [μm/m], for SS 316L:

$$\varepsilon_T = a \cdot T^2 + b \cdot T + c \tag{4}$$

$$a = 0.009714; \tag{5}$$

$$b = 16.954 - 0.00577 \cdot P - 14.98 \cdot h + 0.0599 \cdot P \cdot h; \tag{6}$$

$$c = 452 - 3.557 \cdot P - 1.350 \cdot v - 6925 \cdot h - 0.001838 \cdot P^2 - \\ -10\,399 \cdot h^2 + 0.005893 \cdot P \cdot v + 40.28 \cdot P \cdot h + 11.47 \cdot v \cdot h - 0.05238 \cdot P \cdot v \cdot h. \tag{7}$$

Here $P$, $v$, and $h$ are the processing parameters: $P$ is a laser power, W; $v$ is a scanning speed, mm/s; and $h$ is a hatch spacing, mm (see Figure 1). $P$, $v$, and $h$ are the independent variables, $P^2$, $v^2$, and $h^2$ are the two-way interactions, $T$ and $T^2$ are the temperature variables, and $P \cdot T$, $v \cdot T$, $h \cdot T$, $P \cdot v \cdot T$, $P \cdot h \cdot T$ are the interaction terms. According to [71], the described model was accepted as statistically significant and has a standard deviation $S=9.64627$, coefficient of determination $R^2=99.99\%$, and the predicted residual error sum of squares PRESS=11785.3. Calculations based on this model give the resulted values of $b$, $c$, and the $\alpha_T$ (LTEC of SS 316 L, deposited on the above mentioned processing parameters $P_{average} = 308$ W and $v = 0.85$ m/min $\cong 16.67$ mm/s, Table 5; $h = 300$ μm $= 0.3$ mm, Figure 1), as the following: $b = 16.217600$, $c \cong -125.021612$, and $\varepsilon_T = 0.009714 \cdot T^2 + 16.217600 \cdot T - 125.021612$. Therefore, it is possible to estimate the value of $\varepsilon_T$ of the SS at the concrete temperature point; particularly, $\varepsilon_T \cong 219.8$ μm/mm at the room temperature 21°C; $\varepsilon_T \cong 3913.0$ μm/mm at 220 °C what is close to results provided by SLM at the critical laser energy density in the study [71] ($\varepsilon_T \cong 3900$ μm/mm at 220 °C), and is higher than thermal expansion for SLM at the same laser power and scanning speed, but lower hatch spacing ($h = 120$ μm, $\varepsilon_T \cong 3100$–$3200$ μm/m [71]).

### 3.3. Tensile Tests

*3.3.1. Specimens*

For the purpose of the research, 17 specimens for tensile tests were produced using EDM technology. Number of specimens of each experimental Group (1-8) was not equal. This inequality was caused by defects of initial part, problems with realization of technology, and other technical difficulties. Therefore, Groups 1-4 and 7 finally consisted of two specimens each, Group 5 – of one, and Groups 6 and 8 – of three. The example of resulted tensile test specimen shape is demonstrated in **Figure 5**. Its thickness *s* (1.6 mm) and width *b* (6 mm) allow observation of Poisson's ratio and realization of Digital Image Correlation (DIC) optical method. Length of the specimen work zone (middle area) *L* (45.2 mm) was calculated using the equation (8) [70]:

$$L = l + 2 \cdot 1.6 \cdot b + 2, \tag{8}$$

where $l = b \cdot 4$ (in accordance with ASTM E8/E8M-16a [70]), two summands $1.6 \cdot b$ were added for averaging of stresses in agreement with St. Venant's principle [85], and 2 mm (1 and 1 from both sides) were sunk in the grips of the INSTRON 5969 [86].

**Figure 5.** Tensile tests specimen sketch and external view

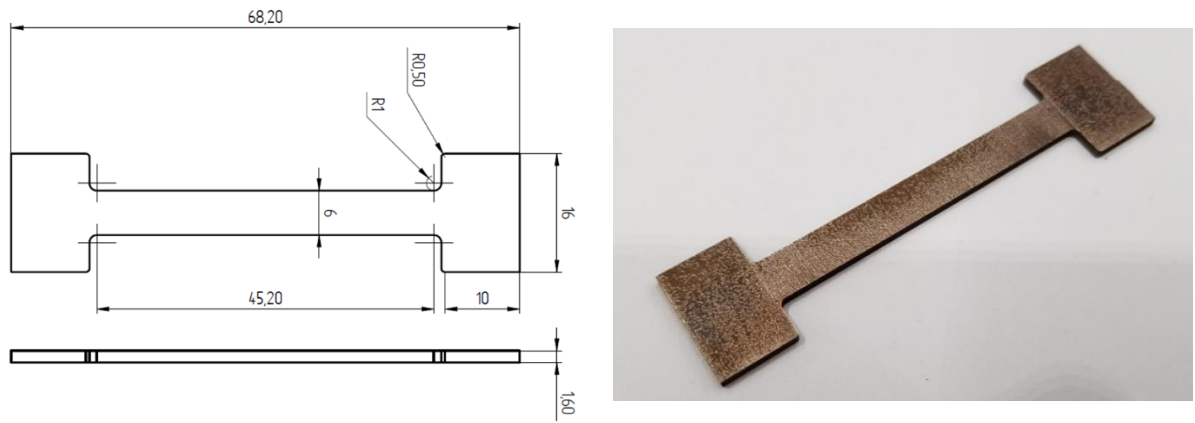

If such specimen is created from a homogeneous material and treated by distributed tensile load, its internal stress distribution looks like it is shown in **Figure 6**. This result was obtained via the numerical simulation and compared with data of the real tests showing the actual disruption areas and stress distribution. The Cauchy problem was solved based on the boundary conditions: the first tale of the specimen – pinned (U1=U2=U3=0), the uniformly distributed load with a ramp amplitude acted on the second tale. The standard element library of 3D stress family with linear geometric order was applied, C3D8R (8-node linear brick with reduced integration and a hourglass control) was chosen as an element type. Basic parameters of the mesh: approximate global size of the seeds – 0.0011 m, curvature control – applied, maximum deviation factor – applied, minimum size – controlled by fraction of 0.1 of the global size. Parameters of the final simulation step: type – static, general; maximum number of increments – 100, initial size – 1, minimum – $10^{-5}$, maximum – 1. Parameters of the equation solver: method – direct, matrix storage – solver default, solution technique – full Newton. Mass density of the material – constant, uniformly distributed. Calculation method: Abaqus CAE 2018. Calculation time: 55 s. Parameters of the PC: CPU – Intel® Core™ i7-7700HQ 2.81 GHz. RAM: 16,0 GB. OS: Windows x64.

Blue areas in the image (Figure 6) show regions with lowest expected stresses and strains, green areas are associated with higher values, yellow – with even higher, and red areas – with highest. It should be noted that such simulation can be also used for prediction of concrete numbers ultimate mechanical strength, values of internal stresses and strains, but it can be done only when the characteristics of material (elasticity modulus, Poissons' ratio, mass density) are known; in our experiments these parameters are a priori undefined.

**Figure 6.** Stress distribution in the specimen developed via numerical simulation

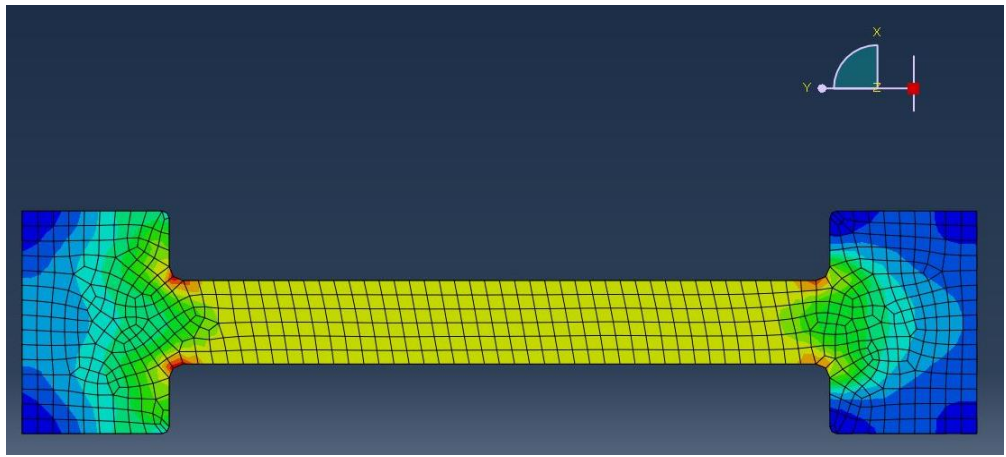

The conducted tensile tests, which are described below, were aimed to acquisition of: a) ultimate tensile stress, b) breaking stress; c) yield stress; d) energy of proportional deformation; e) residual plastic, elastic deformation, and their ratio; f) elasticity modulus, g) Poisson's ratio, h) Bulk modulus and Lamé parameters, of all materials under discussion. The results of all measurements and calculations are presented later on.

*3.3.2. Areas of Breaking*

**Figure 7** demonstrates frames taken by camera in the moment of breaking of all specimens of all groups. Using these frames, we can find common areas of breaking.

**Figure 7.** Frames of breaking of all groups

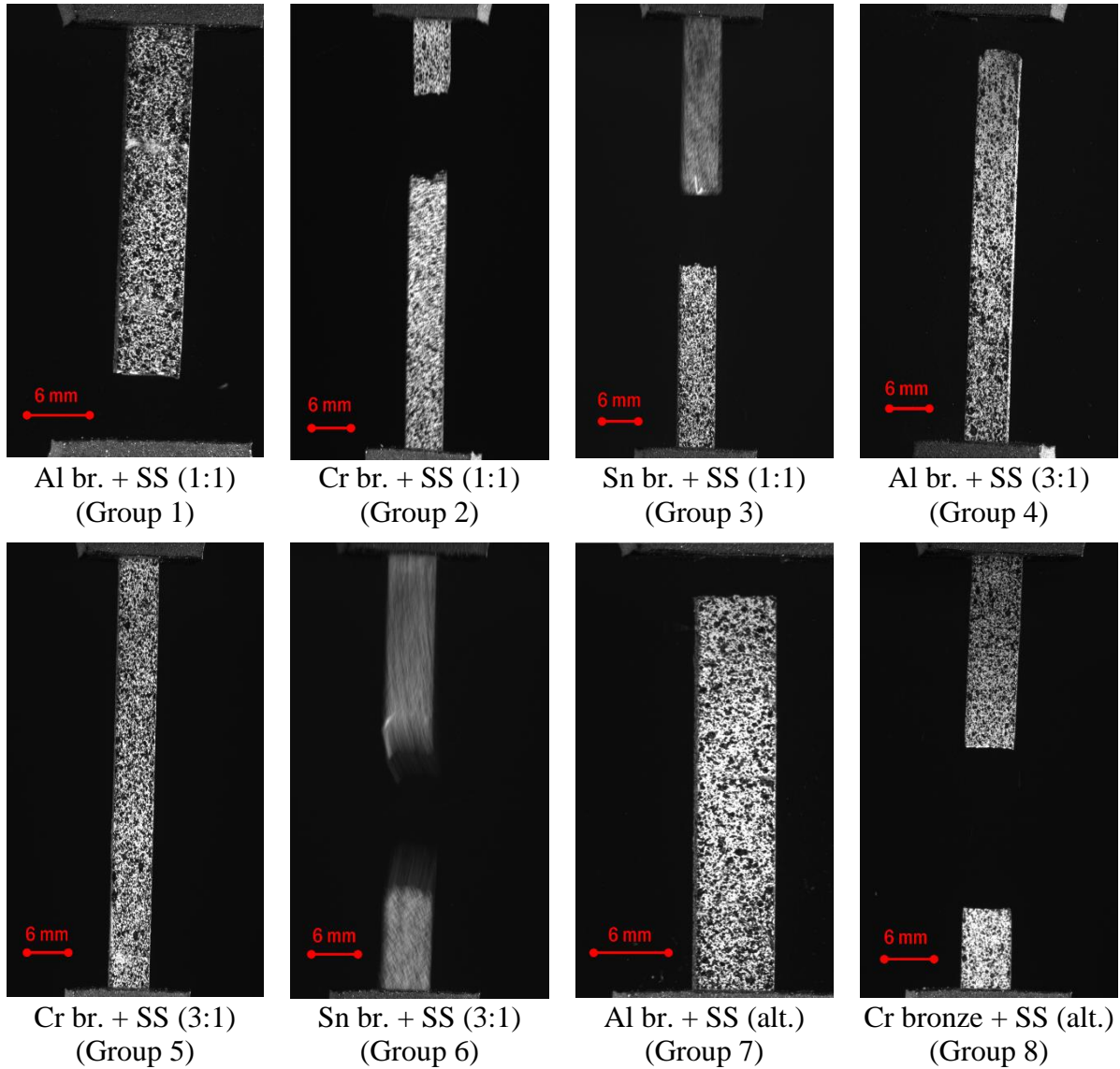

| Al br. + SS (1:1) (Group 1) | Cr br. + SS (1:1) (Group 2) | Sn br. + SS (1:1) (Group 3) | Al br. + SS (3:1) (Group 4) |
| Cr br. + SS (3:1) (Group 5) | Sn br. + SS (3:1) (Group 6) | Al br. + SS (alt.) (Group 7) | Cr bronze + SS (alt.) (Group 8) |

It is observed three main dependencies from all obtained breaking frames.

a) The major part of the 1:1 specimens (Groups 2 and 3), except aluminium bronze-based, failed in the middle area, what is a deviation from the simulation model proposed previously (Figure 6). This deviation could be refered to the micro-void defects or porosities, that are known to be one of the common reasons of the mechanical strength degradation in the AM-fabricated parts [68,87-90]. The two main types of porosity may be responsible for transition of the expected area of ultimate stresses from the tails of the specimens to their middle zones: the interlayer porosity (caused by insufficient powder delievery that leads to lack of fusion in the solidified metal), and the intralayer (vaporization) porosity that is attributed to a shielding, feeding, or coaxial gas and has a random localization within the volume of a deposited part [68].

b) The results closest to the numerical simulation were demonstrated by the specimens of Group 5, most of which failed inside the grips. It points at higher homogeneity of these alloys, absence of huge porosity and significant crystalline defects in their middle areas. This suggestion was proved by the results of the optical microscopy (**Figure 8**) that demonstrated the intermixed structure without visible cracking and small amount of porosity with the highest size of a single pore was about 50 μm as it is shown in the Figure 8. Islands of steel (light odd-shaped areas) are almost regularly distributed within the copper-chromium matrix. The microstructural pattern allows seeing the movement direction of the laser heat spot according to the shape of crystallization front (the paraboloid-type convexities are directed contrarily the vector of a scanning speed $\vec{v}$ [20] as it is seen form the Figure 8). The build direction of the part is normal to the plane of the picture. It is seen that stainless steel islands precipitated mainly between the nearby tracks, all of which have opposite direction of the scanning speed vector according to the deposition scheme of a single layer (see Figure 2 in the section 2 "Materials and Methods").

**Figure 8.** The optical microscopy image of Cr bronze + SS 3:1 alloy. The build direction is normal to the image plane

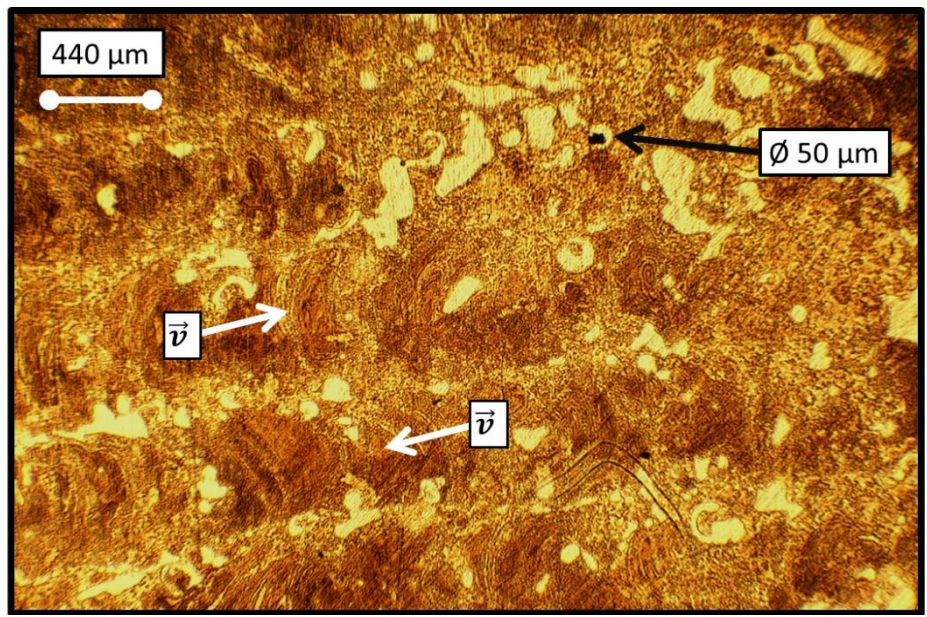

c) There were no aluminium bronze-based specimens, that were broken in their middle areas. This fact also (as it was stated for Group 5 alloy) points at the absence of significant defects such as interlayer and intralayer porosity that could cause appearance of the crack nucleuses inside the middle areas of these specimens. Comparison with results of the pure SS 316L specimens deposited using different building schemes [68] with nanoscale porosity show that all of them failed in their middle areas without clear dependence from the building direction, but differences in the specimen geometry (two times higher rounding radius near the tail of the specimen – 1 mm in our case, 2 mm in case of [68], and three times more narrow middle zone – 6 mm in the current research, 2 mm in [68]) points at the transition of the theoretically expected ultimate stress areas of the specimens discussed in [68] to the middle area, contrarily with the current case, where the predicted areas of ultimate stresses are located near the tails as it is shown in the Figure 6. Therefore, it is reasonable to assume that as aluminium bronze-based specimens of Groups 1, 4, and 7, and Cr-based specimens of Group 5, as defectless SS 316L laser deposited specimens, have the real failure areas the same as it can be predicted theoretically before testing.

*3.3.3. Distribution of Major Deformation*

Surface distribution of major deformation (oriented in vertical direction) of specimens of all groups immediately prior to breaking created using Vic-3D is demonstrated in **Figure 9**.

**Figure 9**. Distribution of major deformation in specimens of all groups before breaking

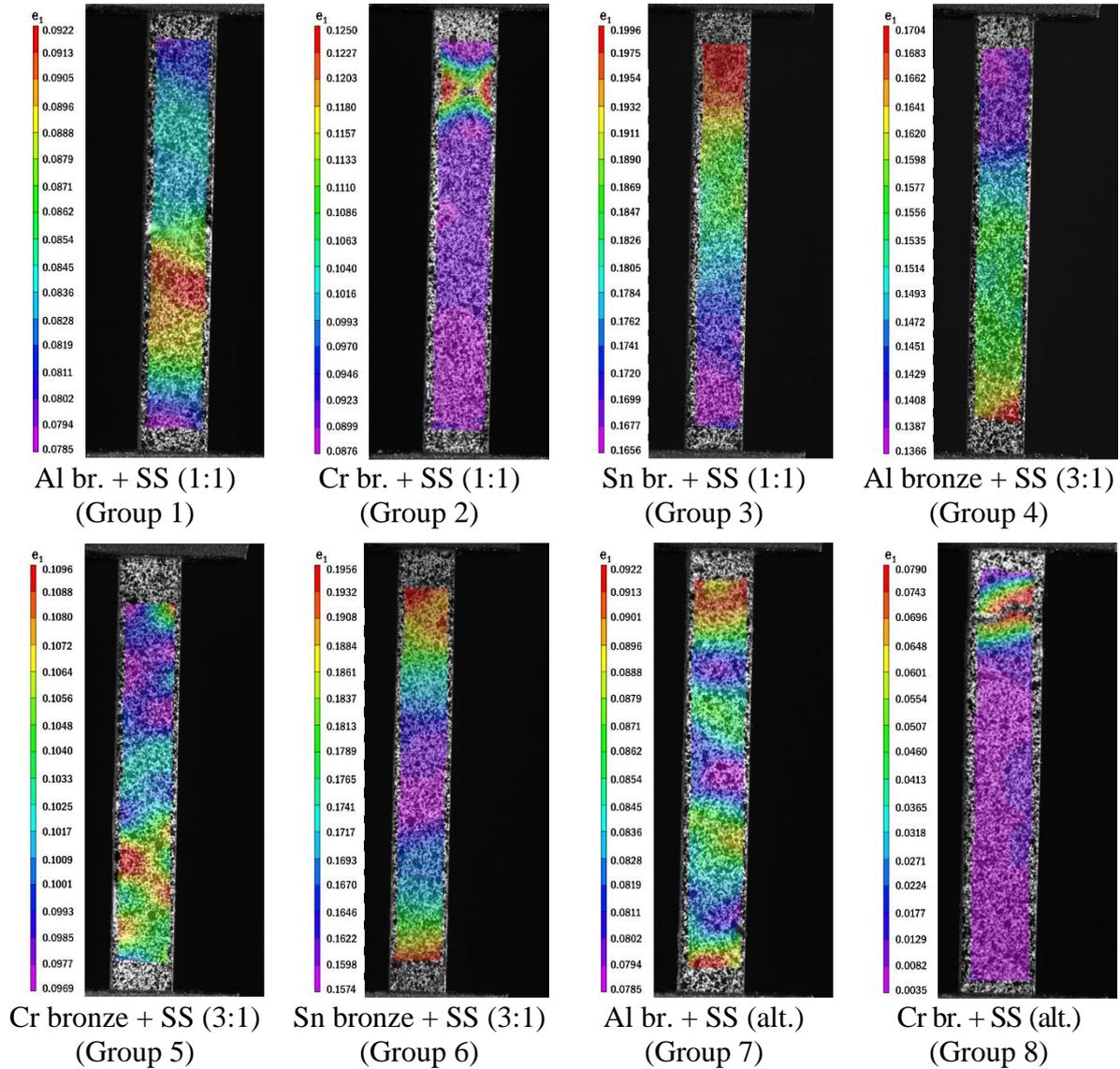

It is seen from Figure 9 that FG Group 7 of an alternating layers alloy has a double-sided (up and down) increase of a major deformation alongside the vertical axis with evident alternating of deformation value and two areas of stress and deformation concentration. Two areas of ultimate stress were also predicted by numerical simulation (Figure 6), the breaking frame of this group (see Figure 7) is also in accordance with this result. The DIC strain maps of the similar alternating layers structure fabricated from two kinds of the stainless steel (420 SS and C300 MS) [44] also exhibited the strain alternating between different layers of the specimen, but the period of this alternating was rather less. This difference could be associated with higher intermixing of the aluminium bronze and SS 316L in comparison with 420 SS and C300 MS, what could be also seen from the microstructure patterns provided by C. Tan et al. [44], and with the more significant degree of the epitaxial growth between the layers, inherent for the alternating-layered DED-processed multimaterial structures [47]. The possible increase of intermixing was also approved by the $e_{xx}$ strain distribution of the FG structure created from 50 alternating Cu-Fe laminated sheets via the accumlative roll bonding [48], where the evident alternating character of strain distribution is also erased, and the high intermixing, leading from the layered to the network-like structure, was observed with increase of the number of layers. Another group with alternating layers (Group 8) contrarily didn't exhibit the alternating strain pattern, but showed an intense ellipsoid-like concentrator in the top area, which probably corresponded to the defect in this specimen that was seen during the fracture analysis. One specimen of this group didn't break entirely, but suffered cracking in the area of maximal stress and strain concentration. The same deformation distribution, as in Group 8, was observed in Group 2, also created from stainless steel and chromium bronze (defect caused a hyperboloid-like shape of major deformation distribution near the stress concentrator).

Except Group 7, only Group 6 also has two areas of ultimate strains, but the homogeneous structure of the material doesn't provide any alternating of internal stress value as it is seen in the FG structure of Group 7. It is also seen that there are no specimens created from chromium bronze that have more than one areas of maximal deformation.

Additionally it should be noted that the stepless character of the major deformation variation is inherent to all Groups with 1:1 ratio of steel and bronze, to Group 4 (3:1 ratio of aluminium bronze and stainless steel), and to Group 8 (alternating layers created from chromium bronze and stainless steel). It points at the higher homogeneity of these aloys and absence of the significant defects in their structure, that could cause the multiareal stress concentration map.

*3.3.4. Engineering Stress-Extension Curves and Ultimate Tensile Stress*

Based on the results of mechanical testing, the two types of experimental dependencies were obtained: the engineering stress-extension (stress-absolute deformation) (**Figures 10-12**), and the engineering stress-strain (**Figures S1-S17** in the chapter «Supporting Information»). The engineering stress-extension curves are meaningful for the direct fixation of the residual plastic part of the absolute deformation $\Delta L_{plast}$, which persists in the specimen after its failure, the elastic part $\Delta L_{elastic}$ that disappears after the rupture, and their ratio $\Delta L_{plast} / \Delta L_{elastic}$ that indicates in which stage of loading the specimen was broken. The engineering stress-strain curves were used for acquisition of the elasticity modulus (angle between linear part of the curve and the positive abscissa axis), and the specific energy of the proportional deformation (the triangle area under the stress-strain curve in the region of the proportional deformation). The data attributed to the values of breaking, ultimate, and yield stress, can be obtained both from the engineering stress-extension curves, and from the engineering stress-strain curves. The engineering stress-extension curves presented in Figures 10-12 are divided into three broad classes according to the proportion of materials. Figure 10 demonstrates specimens of Groups 1-3 with 1:1 mixture of stainless steel and bronze; Figure 11 – mixture 3:1 (Groups 4-6), and Figure 12 – all specimens with alternating layers of steel and bronze (Groups 7-8). Numerical indices (*i*, *j*) near the $\Delta L_{plast}$ and $\Delta L_{elastic}$ show the number of Group (1–8) and specimen (1–3) respectively.

**Figure 10.** Engineering stress-extension curves of Groups 1-3

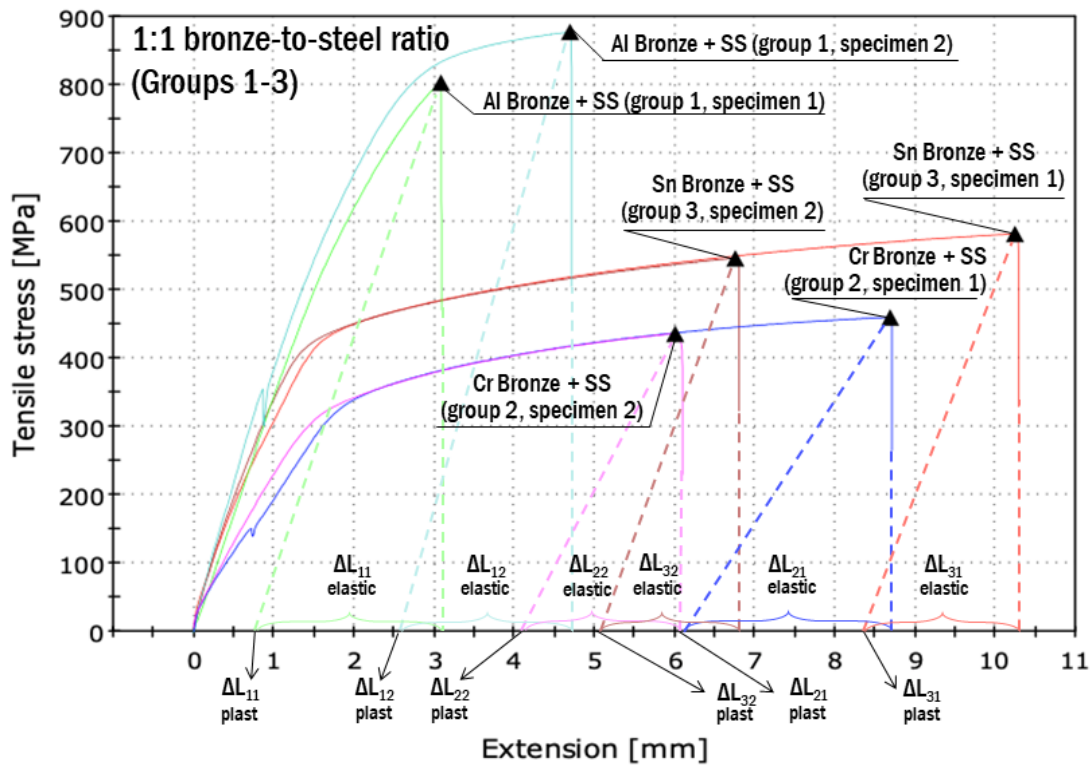

**Figure 11.** Engineering stress-extension curves of Groups 4-6

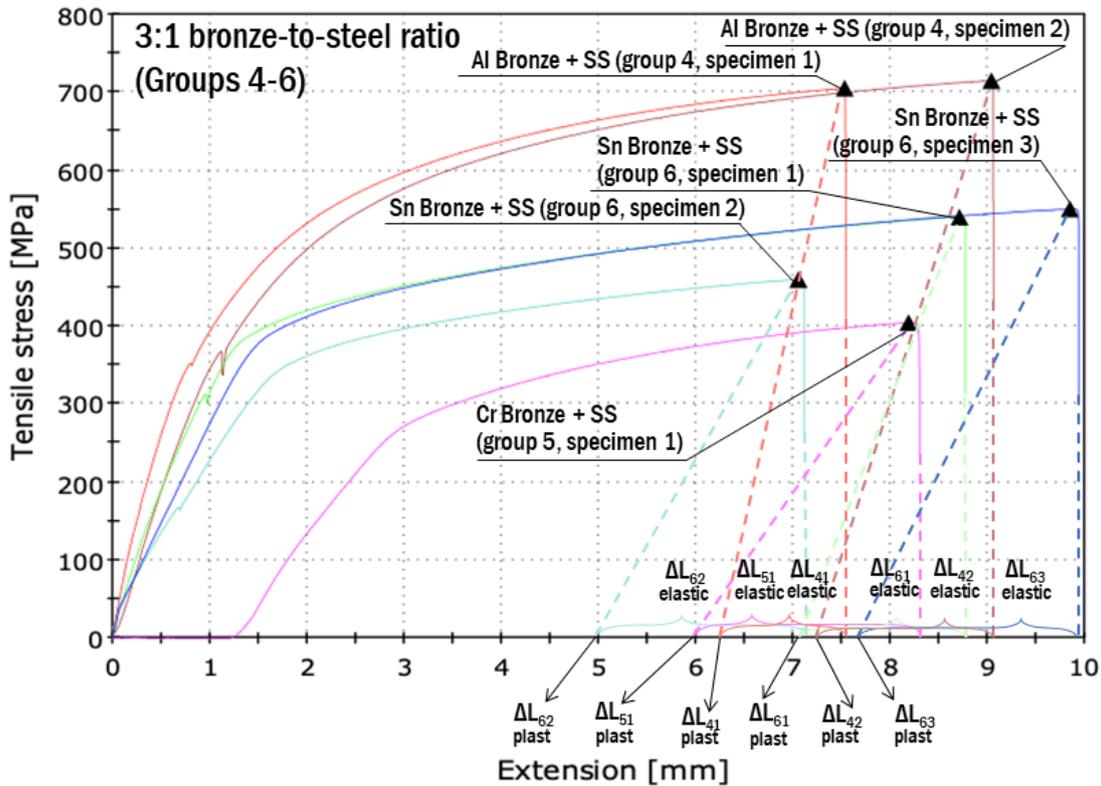

**Figure 12.** Engineering stress-extension curves of Groups 7-8

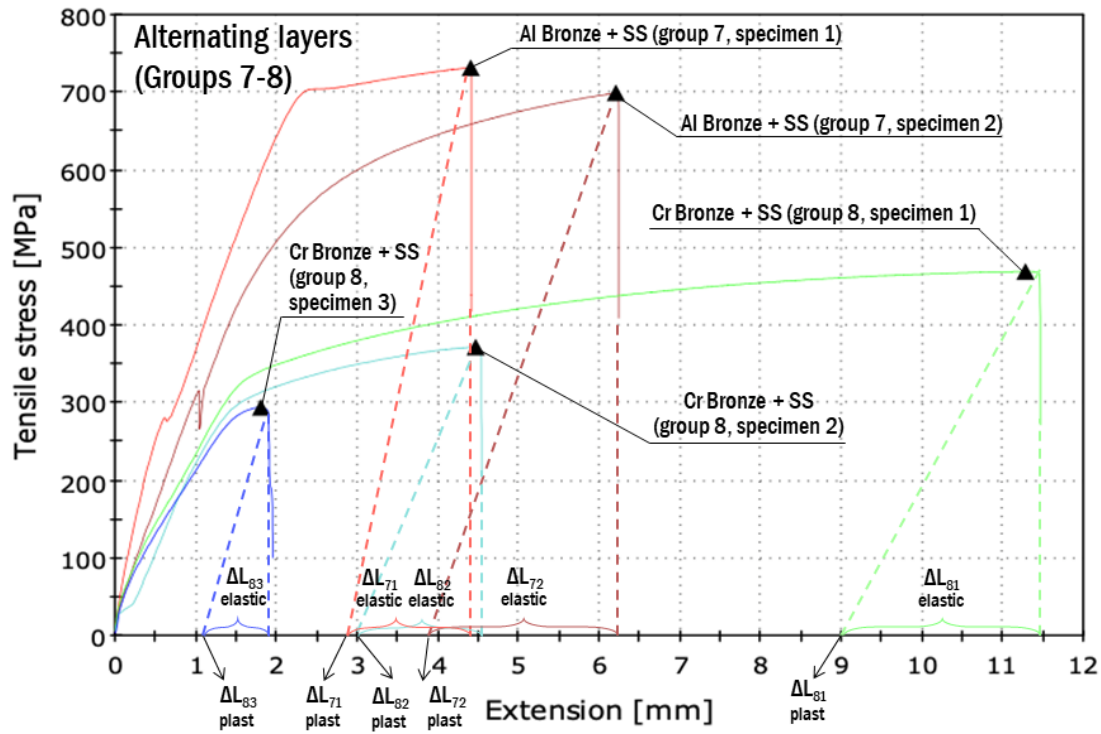

The first sight allows to see that all curves associated with aluminium bronze-based specimens (Groups 1, 4 and 7) are located closer to the left top area of the coordinate plane, than all others. These specimens also have longer linear parts of the stress-extension curves that satisfy the Hooke's law, and the angles between these linear parts and the positive X-axis have the highest values. All mentioned facts evidently point at the better tensile mechanical characteristics of specimens of these groups. Quantitatively, the resulted ultimate tensile stress (UTS) values of all specimens, obtained from the Figures 10-12, are shown in **Figure 13.**

**Figure 13.** UTS values of all specimens of Groups 1-8, [MPa]

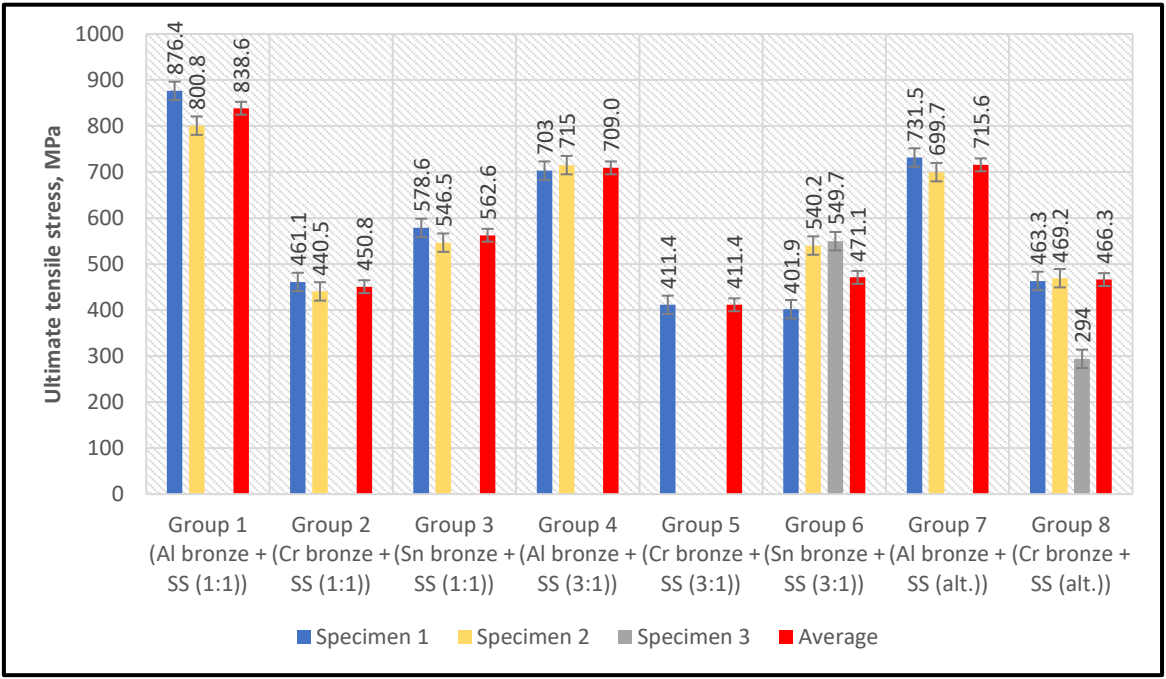

It is seen from Figures 10-12 that the UTS equals the breaking stress in all cases, but it should be mentioned that generally the UTS (stress value associated with the maximal overcome load) and the breaking stress (stress value attributed to mechanical fracture of the specimen) can be different.

It is seen that the specimens of Group 1 demonstrate the highest value of UTS (876.4 MPa). Also, the good results are shown by Group 4 (703–715 MPa) and Group 7 (731.5 MPa). The lowest mechanical strength was associated with chromium bronze-based parts: Group 5 (411.4 MPa), Group 2 (440.5–461.1 MPa), and Group 8 (294.0–469.2 MPa). Therefore, it is seen that binary and gradient materials based on a stainless steel are more prospective with aluminium bronze in terms of mechanical tensile strength; the least quality is demonstrated by binary materials with chromium bronze, and the tin bronze-based alloys have intermediate UTS characteristics.

Analysis of the obtained values of UTS and comparison with recently published related works point at the excellent mechanical characteristics of the fabricated aluminium bronze-based parts. The results presented by I. Sen et al. [58] showed the lower values of the engineering UTS of 0wt.% and 3wt.% Cu-added stainless steel 304H, annealed up to 100 h. The highest measured UTS value was equal to 710.5 MPa. The tensile mechanical testing of Cu-SS FG structures, deposited directly and through the Deloro 22 intersections [29], which were also discussed at the beginning of the Section 1 "Introduction", demonstrated 648.2±25.4 MPa UTS for D22-SS FG structures, 720.1±13.3 MPa for pure D22 alloy, and showed comparatively poor results for multi-metallic structures with copper: 226.2±4.4 MPa for Cu-D22-SS specimens, and 124.4±30.2 MPa for Cu-SS. It is higher than SLM-fabricated 99.2wt% Cu (195-211 MPa; 49±10 MPa) [27,29,91] and cold spray-produced commercially pure copper (125 MPa) [29,92], but rather lower than any specimen that were presented above in Figure 13 (including 75% bronze - 25% steel groups, and groups with alternating layers of bronze and steel too). The maraging steel-copper functional bimetal fabricated via SLM combined with subtractive process [30] also showed lower UTS (not more than 557 MPa, what is slightly higher than that of copper [30], but lower than UTS of the Al bronze-based specimens of Groups 1, 4, and 7 – Figure 13).

Moreover, not only AM-fabricated Cu-Fe and commercially pure copper structures showed lower UTS; even the SS 316L specimens, laser deposited using different building directions (0°, 15°, 30°, 90°), demonstrated the lower ultimate strength: less than 700 MPa – in the study of E. Azinpour et al. [68]; from ~220 to ~510 MPa – in the work of S. Kersten et al. [93], that was less than UTS of the bar stock (~600 MPa) [93]. Because the building direction of 90° to the direction of loading commonly exhibits higher UTS [68-69], it is expected to be possible to improve the tensile mechanical characteristics of all discussed Groups 1-8 by variating of their building directions. P. Guo et al. [69] demonstrated ~17% increase of UTS of the high-power laser deposited SS 316L (from 770 to 900 MPa), and Zhang et al. [94] reported about 19% UTS improvement (536 → 639 MPa) provided by changing of the building direction from 0° to 90°. 90° building direction also demonstrates the more preferable microstructure because of higher homogeneity of the alloy, while 0°-built parts commonly consist of large dendritic grains, generated by epitaxial nucleation, due to specificity of the thermal history (the high temperature gradient between the melt pool and the previously deposited areas) [69].

*3.3.5 Modulus of Resilience, Residual Plastic and Elastic Deformation*

The modulus of resilience (specific energy of proportional deformation) can be evaluated by calculating the triangle area under the engineering stress-strain curves in the region of the proportional deformation. Curves in Figures 10-12 are plotted in tensile stress-extension coordinates; curves in coordinates of tensile stress-relative elongation are demonstrated in Figures S1-S17 as it was said earlier**.** Below in **Figure 14** the resulted values of specific energy of proportional deformation for each specimen of each group are demonstrated.

**Figure 14.** Specific energy of proportional deformation of all specimens of Groups 1-8

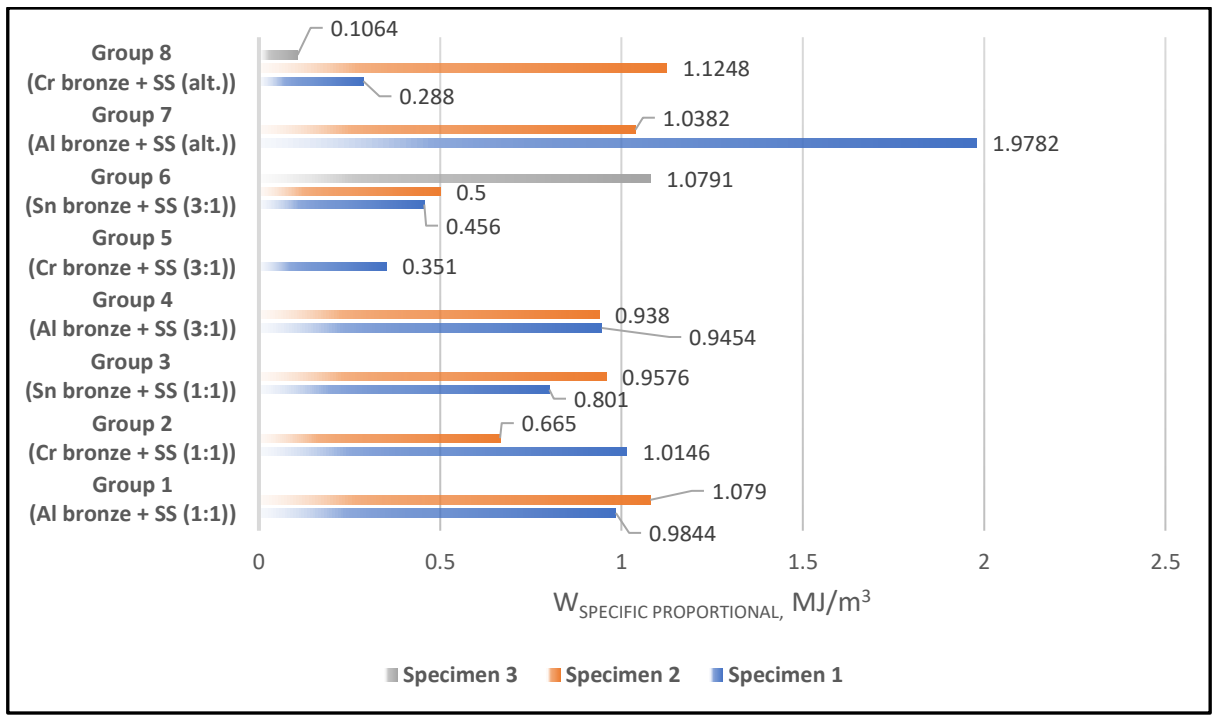

There are two main cases, where such parameter as specific proportional deformation energy is important to study. First of them – traditional pressure shaping processes, such as cold forming, where the deformation energy matters in the light of further mechanical strength and rigidity of the part, and it is known that the process parameters may strongly influence on this value [95]. The second one is associated with dependence between specific energy of proportional tensile deformation and notch toughness (the bigger this energy is, the higher impact load the material can overcome), and it was the purpose to estimate the specific energy of proportional deformation in our case. This problem is not investigated well in case of additive manufacturing, i.a. DED, and is poorly studied in case of the laser deposited Cu-Fe FGMs. From the Figure 14 it is seen that all 3:1 alloys (Groups 4-6) have the lowest average specific proportional deformation energy, alternating layers specimens (Groups 7-8) have high dispersion of results, Groups with chromium bronze (2, 5, 8) are characterized by the lowest values, Groups with aluminum bronze (1, 4, 7) – by the highest. The ultimate value of the specific proportional deformation energy was observed in case of aluminium bronze + stainless steel alternating layers specimen and approximately amounted 1.9782 MJ/m$^3$.

Engineering tensile stress-extension curves demonstrated in Figures 10-12 allow estimation of residual plastic deformation and elastic deformation, which fully disappears when a specimen fails. These values can be observed in Figures 10-12, where $\Delta L_{xy\ plast}$ means residual plastic deformation of the specimen y from the Group x, and $\Delta L_{xy\ elastic}$ means elastic deformation of this specimen. The results of the estimation are summarized in **Figure 15.**

**Figure 15.** Estimated residual plastic and elastic deformation of all specimens of all groups, and plastic-to-elastic deformation ratio

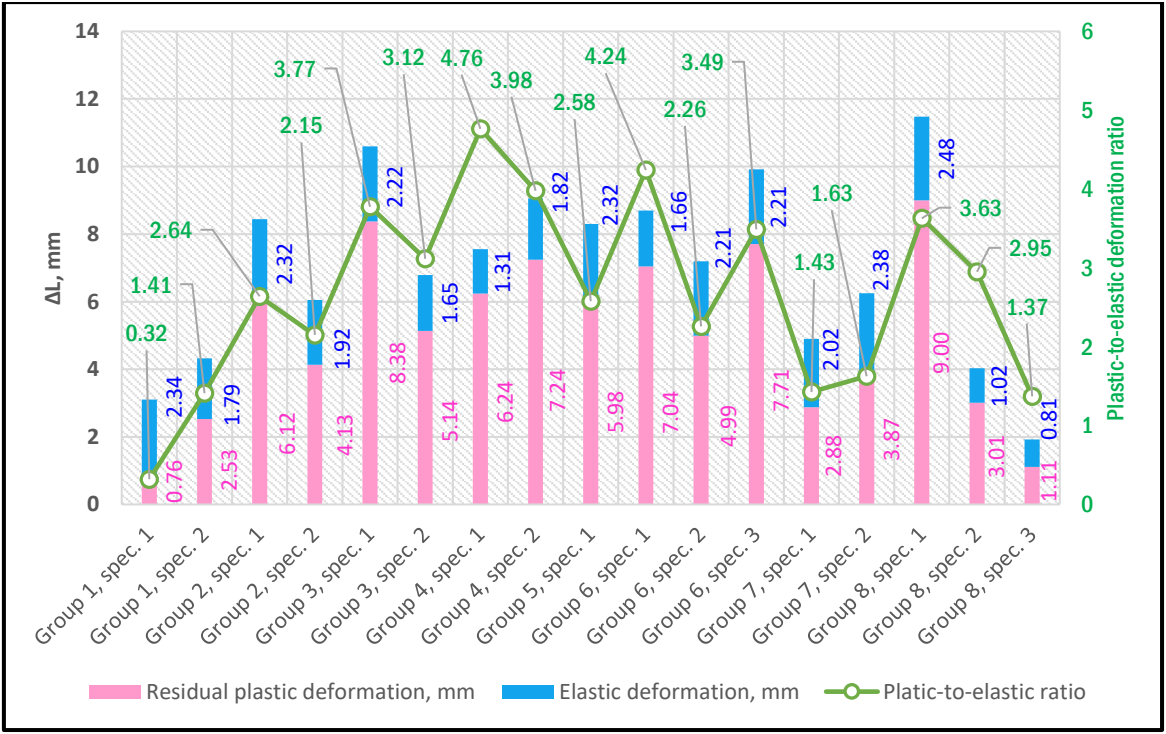

Among others, only Group 1 demonstrated a plastic-to-elastic deformation ratio less than unity. It means that in this case, the failure occurred during low plastic deformation; the applied stretching force caused internal stresses lower than critical values that trigger intercrystalline slip, but higher than stresses, which break interatomic bonds. This phenomenon was observed only one time and only in the group of Al bronze-based specimens.

*3.3.6. Elasticity Modulus*

To estimate the elasticity modulus of the specimens, the precise measurement of angle between linearly proportional part of stress-strain curves and axis of abscissas (see Figures S1-S17) was performed. All resulted elasticity modulus values based on these measurements are presented in **Figure 16**.

**Figure 16.** Elasticity modulus values all specimens of Groups 1-8, [GPa]

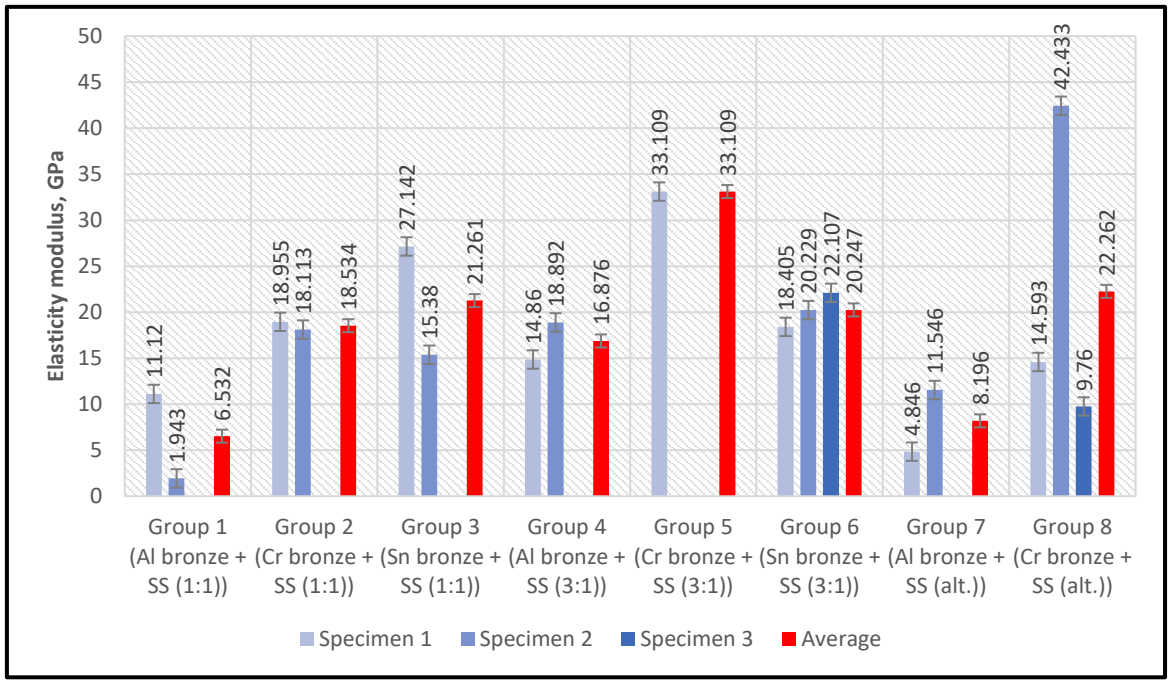

It is seen from the Figure 16 that the highest values of elasticity modulus (up to 42.4 GPa) correspond to specimens of Groups 5 and 8 (both – based on Cr bronze). These materials also had the worst mechanical strength, but the highest britlleness and rigidity (see also §3.3.4). The lowest elasticity modulus is associated with all specimens created from aluminium bronze and SS (Groups 1, 4, and 7). Therefore, assuming the data of §3.3.4, it can be concluded that these materials have both high mechanical strength and high elasticity. Comparing the results with previously measured elasticity modulus of alternating layers structures published in [26], it is seen that the current results are significantly lower (4.8–11.5 GPa versus 35.6–43.2 GPa). This difference could be associated with differences of chemical composition of the bronze powder, its grain particle size and shape, and different amount and size of porosity of as-built parts. Also, it should be noted that measurements discussed in the study [26] were conducted using another method: locally, mostly near the border zone between steel and bronze, which was enriched by intermetallics.

Elasticity modulus of all Cu-Fe structures based on SS 316L (Figure 16), as expected, was found to be lower than the same one of the laser deposited SS 316L (1.943–42.433 GPa versus ~150–200 GPa [93]).

Specimens created using tin bronze demonstrated intermediate values of elasticity modulus in comparison with another groups (from 15.4 to 27.1 GPa).

The related research of the mechanical properties of the DED-fabricated parts of pure SS 316L was conducted by S.-H. Park et al. [96] using almost the same technological installation (InssTek MX-400). The Young's modulus and the Poisson's ratio were indirectly determined by measurement of velocities of the GHz-THz range ultrasonic waves, which occurred in the material during a post-treatment – a surface shallow melting (SSM)-mechanism laser polishing, using the femtosecond laser system. The resulted Young's modulus values were found to be in a range from approximately 209 GPa to 232 GPa (the direct measurement using tensile testing approved accuracy of the indirect method based on the ultrasonic waves). This result is also predictably higher than average results demonstrated above in Figure 16, and even higher than findings of [93]. Therefore, it could be concluded that addition of 50wt.% and more bronze in the SS 316L powder significantly lowers the elasticity modulus of the resulted DED-fabricated binary Cu-Fe system part.

*3.3.7. Yield Stress*

Results of yield stress approximate estimation using the engineering stress-strain curves (Figures S1-S17) are presented in **Figure 17**.

**Figure 17.** Yield stress values of all specimens of Groups 1-8, [MPa]

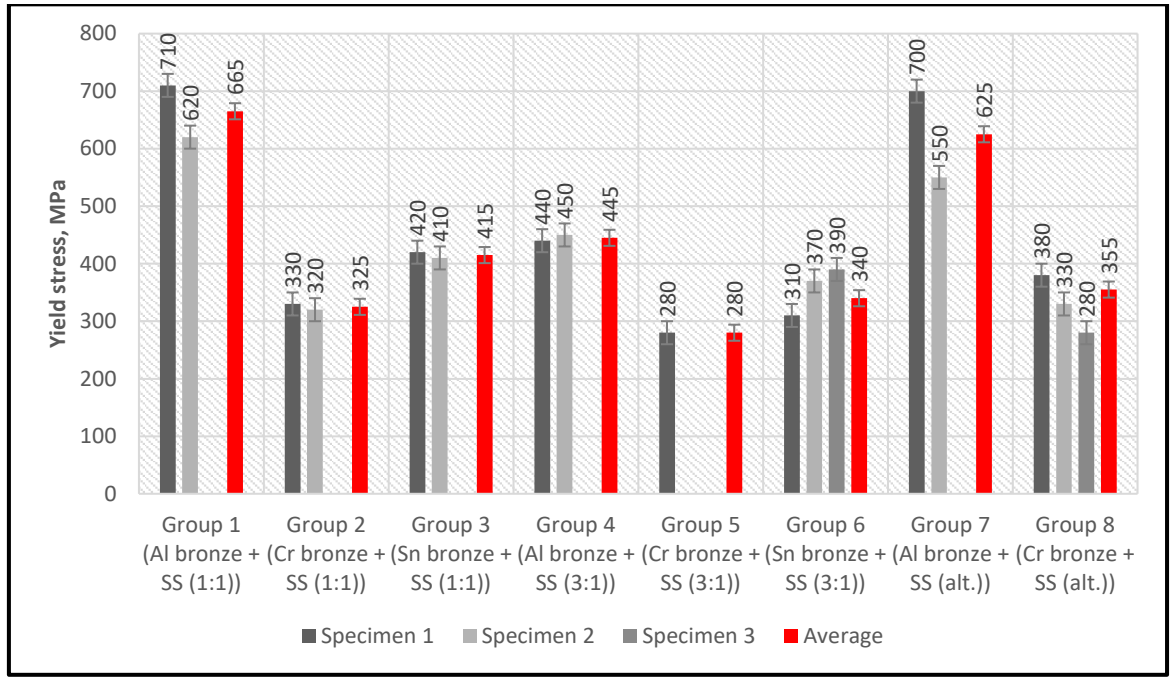

It is observed from Figure 17 that Al bronze-based specimens of Groups 1 and 7 have comparatively high values of the yield stress (550-710 MPa). It is higher than that of the SLM-fabricated SS 316L (440 MPa) [97] and even comparable with additive manufactured TiB2-reinforced SS matrix nanocomposite SMC-5 (~830 MPa at room temperature and ≤520 MPa at temperatures 600°C and higher), and SMC-10 (~980 MPa at room temperature and ≤410 MPa at temperatures 800°C and higher) [98]. Measurements of the 0.2% yield stress of DED-fabricated SS 316L samples, presented by S. Kersten et al. [93], also showed lower average values (from 212.1 MPa to 285.56 MPa in dependence from the build direction). The SS 316L yield stress that was applied in the set of material parameters used for numerical simulation by E. Azinpour et al. [68] was higher than that of [93]: 453.85 MPa – 464.2 MPa in dependence on building direction, but also lower than that of aluminium bronze-based alloys of Groups 1 and 7. This significant property of materials based on aluminium bronze considers their high mechanical strength and provides the possibility to be implemented as the parts of a high-loaded equipment. On the other side, the lowest values of the yield stress are attributed to the Cr bronze-based materials of Groups 2, 5, and 8 (from 250 to 380 MPa). As it was mentioned earlier, these materials are also characterized by the lowest UTS. The Sn bronze-based materials of Groups 3 and 6 exhibit intermediate yield strength in the range from 310 to 420 MPa.

The possible way of increase of the yield strength in AM-fabricated parts of system with positive enthalpy of mixing (such as Cu-Fe) is a grain refinement: A. Zafari and K. Xia [99] reported about than 2.2 times compressive yield strength increase from ~400 MPa to ~900 MPa in SLM-produced immiscible Cu-Fe parts with 250 nm average grain size, associated with liquid separation, monotectic reaction and solid-state phase transformations upon cyclic heating. These parts also showed ultrahigh ultimate compressive strength (UCS) ~1200 MPa.

The notable specificity of groups of FG alternating layers alloys is existence of the distinctive plateau between the elastic and the plastic deformation, that is characterized by increase of absolute elongation with almost constant internal stresses (see Figures S13 and S17). This plateau is associated with 700 MPa internal stress in case of Group 7, and with 280 MPa in case of Group 8, what is equal to the corresponding yield stresses. The visual analysis of the loaded specimens could point at the existence of this plateau by the appearance of the Lueders-Chernov's lines dense network on the flat surfaces of the stressed specimens. These lines are directed at the approximately 45° angle to the axis of load and are attributed to the displacements of the crystal planes with ultimate shear stresses.

*3.3.8. Poisson's Ratio, Bulk Modulus, and Lamé parameters*

A Poisson's ratio for all specimens of all experimental groups was calculated using the results of evaluation of major deformation (Figure 9) and minor deformation through the Vic-3D. Resulted values of a Poisson's ratio are shown in **Figure 18.**

**Figure 18.** Poisson's ratio values

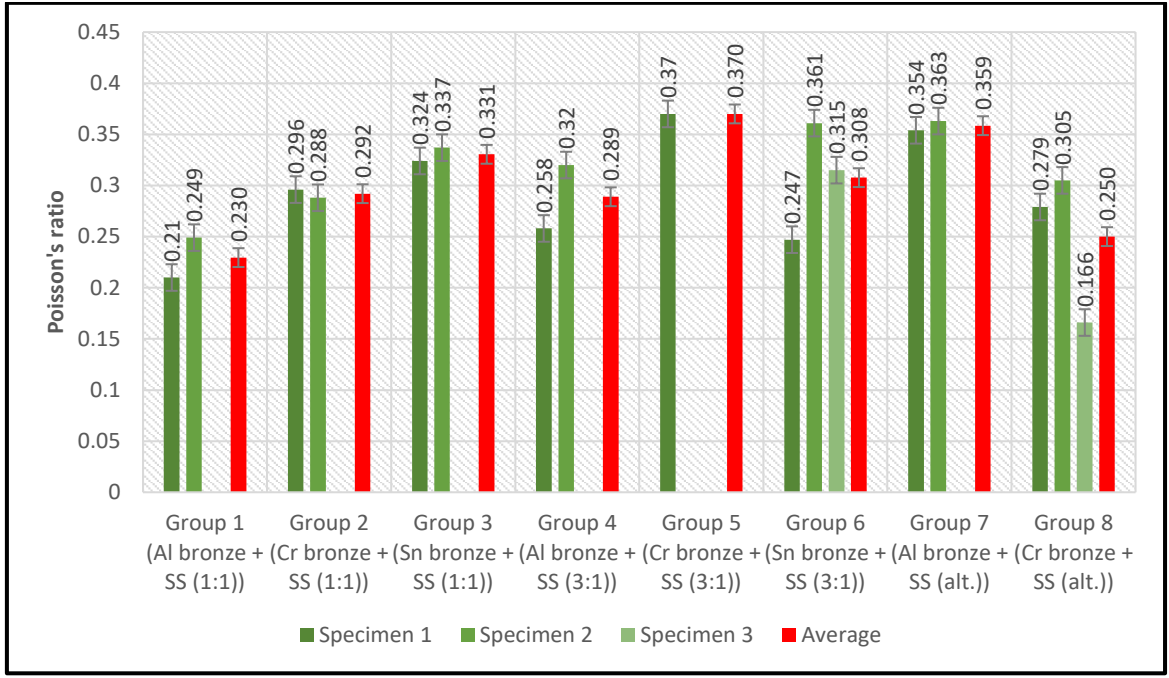

As it could be predicted before experiments, all materials showed a Poisson's ratio in the range of 0.2 – 0.4 average value. A comparison of the resulted values with pure stainless steel fabricated via traditional methods exhibits that its Poisson's ratio (0.28 – 0.31) is close to presented average results in all Groups except №№1, 3, 5, and 7; almost the same Poisson's ratio (0.33) was considered in the numerical simulation of laser deposited SS 316L in the study of E. Azinpour et al. [68]; but the results for the same material, presented in [96], were significantly lower: from ~0.182 to ~0.197, independently from the measurement method (direct measurement via tensile tests or indirect via determination of the ultrasonic velocities).

It is seen from Figure 18, that the highest average Poisson's ratio values (>0.308) are attributed to the Groups 6, 3, 7, and 5 (placed in order of the average Poisson's ratio ascending), and don't have a clear coherence with the bronze-steel ratio and the kind of the applied bronze. The same is observed for the lowest values (≤0.292) that are presented by Groups 2, 4, 8, and 1 (placed in order of the average Poisson's ratio descending).

Evaluation of the average elasticity modulus (Figure 16) and the average Poisson's ratio (Figure 18) allows calculation of all the material constants between the stress tensor and the strain tensor in the homogeneous isotropic elastic medium [100]: the bulk modulus $K$ (9), and the Lamé parameters: a shear modulus $G$ (10) and the second Lamé parameter $\lambda$ (11):

$$K = \frac{E}{3 \cdot (1 - 2 \cdot \mu)} ; \tag{9}$$

$$G = \frac{E}{2 \cdot (1 + \mu)} ; \tag{10}$$

$$\lambda = \frac{E \cdot \mu}{(1 + \mu) \cdot (1 - 2 \cdot \mu)} , \tag{11}$$

where $E$ is an elasticity modulus, and $\mu$ is a Poisson's ratio.

The results of calculation of the bulk modulus and the Lamé parameters are shown in the **Table 7**.

**Table 7.** The bulk modulus and the Lamé parameters values, [GPa]

| Parameter | Group 1 (Al br. + SS 1:1) | Group 2 (Cr br. + SS 1:1) | Group 3 (Sn br. + SS 1:1) | Group 4 (Al br. + SS 3:1) | Group 5 (Cr br. + SS 3:1) | Group 6 (Sn br. + SS 3:1) | Group 7 (Al br. + SS alt.) | Group 8 (Cr br. + SS alt.) |
|---|---|---|---|---|---|---|---|---|
| $K$ | 4.032 | 14.851 | 20.967 | 13.330 | 42.447 | 17.576 | 9.688 | 14.841 |
| $G$ | 2.655 | 7.173 | 7.987 | 6.546 | 12.084 | 7.740 | 3.015 | 8.905 |
| $\lambda$ | 2.262 | 10.069 | 15.643 | 8.966 | 34.392 | 12.416 | 7.678 | 8.905 |

It is seen from the Table 7, that all aluminum bronze-based specimens exhibit the lowest values of the bulk modulus $K$ (from 4.032 GPa to 13.330 GPa) and of the shear modulus $G$ (from 2.655 to 6.546). It points on their rigidity, mechanical strength, and high resistivity to the deformations under the external forces.

Determination of the bulk modulus and all Lamé parameters makes possible to derive the empirical coefficients of the dependence (12) between all components of a stress tensor $\sigma_{ik}$ and components of a strain tensor $u_{ik}$, $u_{ll}$ for isotropic body [100], and determine the equation for the elastic free energy $F$ (13), what is necessary in specific aspects of a theoretical investigation of the stress-strain state of a new material with the unique mechanical properties:

$$\sigma_{ik} = K \cdot u_{ll} \cdot \delta_{ik} + 2 \cdot G \cdot \left( u_{ik} - \frac{1}{3} \cdot \delta_{ik} \cdot u_{ll} \right) ; \tag{12}$$

$$F = G \cdot \left( u_{ik} - \frac{1}{3} \cdot \delta_{ik} \cdot u_{ll} \right)^2 + \frac{K}{2} \cdot u_{ll}^2 . \tag{13}$$

In case of homogeneous strains (strains characterized by constant strain tensor within all volume of the body), the equations (12) and (13) could be expressed in the simpler form:

$$\sigma_{ik} = \frac{E}{1+\mu} \cdot \left( u_{ik} + \frac{\mu}{1-2\cdot\mu} \cdot u_{ll} \cdot \delta_{ik} \right); \quad (14)$$

$$F = \frac{E}{2\cdot(1+\mu)} \cdot \left( u_{ik}^2 + \frac{\mu}{1-2\cdot\mu} \cdot u_{ll}^2 \right). \quad (15)$$

As it was implemented by S.-H. Park et al. [96], and previously shown by Z. Ma et al. [101], P.G. Malischewsky & T.T. Tuan [102], and A. Bayón et al. [103], there is a specific relation (16-17) between the elasticity modulus, the Poisson's ratio, and the ultrasonic (longitudinal (16) and Rayleigh (17)) velocities in homogeneous and isotropic or weakly anisotropic materials such as DED-fabricated parts:

$$V_L = \sqrt{\frac{E\cdot(1-\mu)}{\rho\cdot(1+\mu)\cdot(1-2\cdot\mu)}}; \quad (16)$$

$$V_R = \frac{0.87+1.12\cdot\mu}{1+\mu} \cdot \sqrt{\frac{E}{2\cdot\rho\cdot(1+\mu)}}, \quad (17)$$

The relations (16-17) were used in [96] for non-contact determination of the Young' modulus and the Poisson's ratio; in the current case, these calculations may be conducted vice versa, and the longitudinal $V_L$ and Rayleigh $V_R$ velocities could be determined for all materials (see Table 8) in the assumption that a material density ρ [kg/m$^3$] obeys a rule-of-mixtures equation ρ = ρ$_{SS}$ · ($N_{SS}$,%/100%) + ρ$_{bronze}$ · ($N_{bronze}$,%/100%) [79]. The evaluation of $V_L$ and $V_R$ has practical significance for investigation of parameters of hybrid DED process followed by ultrasonic influence, that could be suggested for further improvement of the quality of DED-fabricated parts: grain refinement, increase of microhardness, wear resistance and microstructure homogeneity, dissolving of oxides, and surface quality improvement [104-107], and for the study of dependence between the resulted characteristics of as-built parts and parameters of the hybrid DED-ultrasonic treatment.

**Table 8.** The ultrasonic waves velocities, [m/s]

| Parameter | Group 1 (Al br. + SS 1:1) | Group 2 (Cr br. + SS 1:1) | Group 3 (Sn br. + SS 1:1) | Group 4 (Al br. + SS 3:1) | Group 5 (Cr br. + SS 3:1) | Group 6 (Sn br. + SS 3:1) | Group 7 (Al br. + SS alt.) | Group 8 (Cr br. + SS alt.) |
|---|---|---|---|---|---|---|---|---|
| $V_L$ | 972.9 | 1746.9 | 1988.0 | 1640.1 | 2672.3 | 1844.4 | 1309.0 | 1827.4 |
| $V_R$ | 528.2 | 877.3 | 931.4 | 827.4 | 1138.1 | 902.4 | 574.7 | 970.6 |

It is seen from the Table 8 that the resulted values of longitudinal (972.9–2672.3 m/s) and Rayleigh (~528–1138 m/s) velocities are significantly lower than that of DED-fabricated SS 316L (5410–5670 m/s and 3000–3170 m/s respectively), discussed in the study [96]. This difference is provided by significantly lower values of elasticity modulus (~7–33 GPa (Figure 16) versus 209–232 GPa [96]).

### 3.4. Fractography

Selected macrofractography images are demonstrated in **Figures 19–26**. It is given 4 images per each group: for group with one specimen (Group 5) – 4 images per specimen; for groups with two specimens (Groups 1-4 and 7) – 2 images per specimen; for Group 6 consisted of three specimens – 2 images for specimen 1, 2 images for specimen 2, and 1 image for specimen 3; for Group 8 also consisted of three specimens – 2 images per specimen for specimens 1-2. Specimen 3 of Group 8 didn't break entirely, so its external surface is demonstrated in separate figure.

White lines in the fractography images show borders of the images taken from different areas of the same specimen fragment or with different focal length of the microscope. If the scale of images is different, it is specified for each image.

**Figure 19.** Selected fractography images of the specimens of Group 1

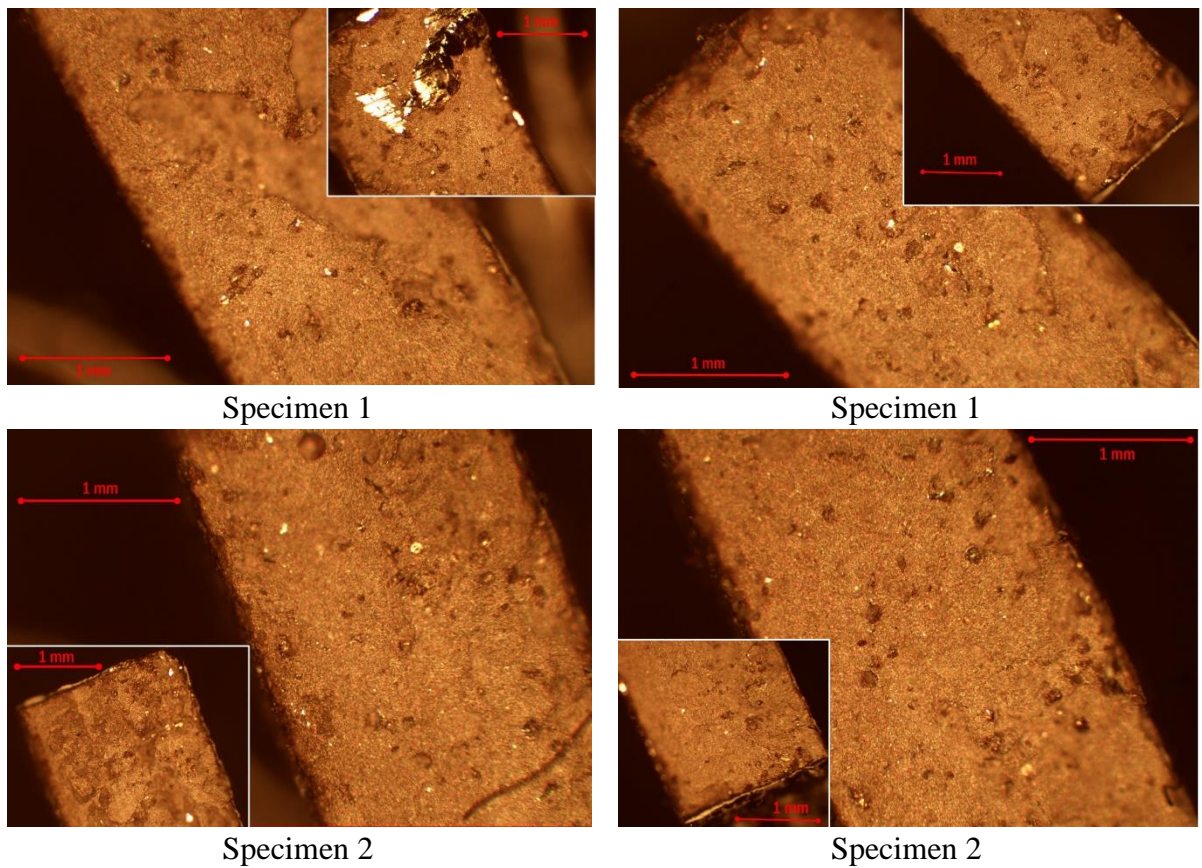

Specimen 1          Specimen 1

Specimen 2          Specimen 2

**Figure 20.** Selected fractography images of the specimens of Group 2

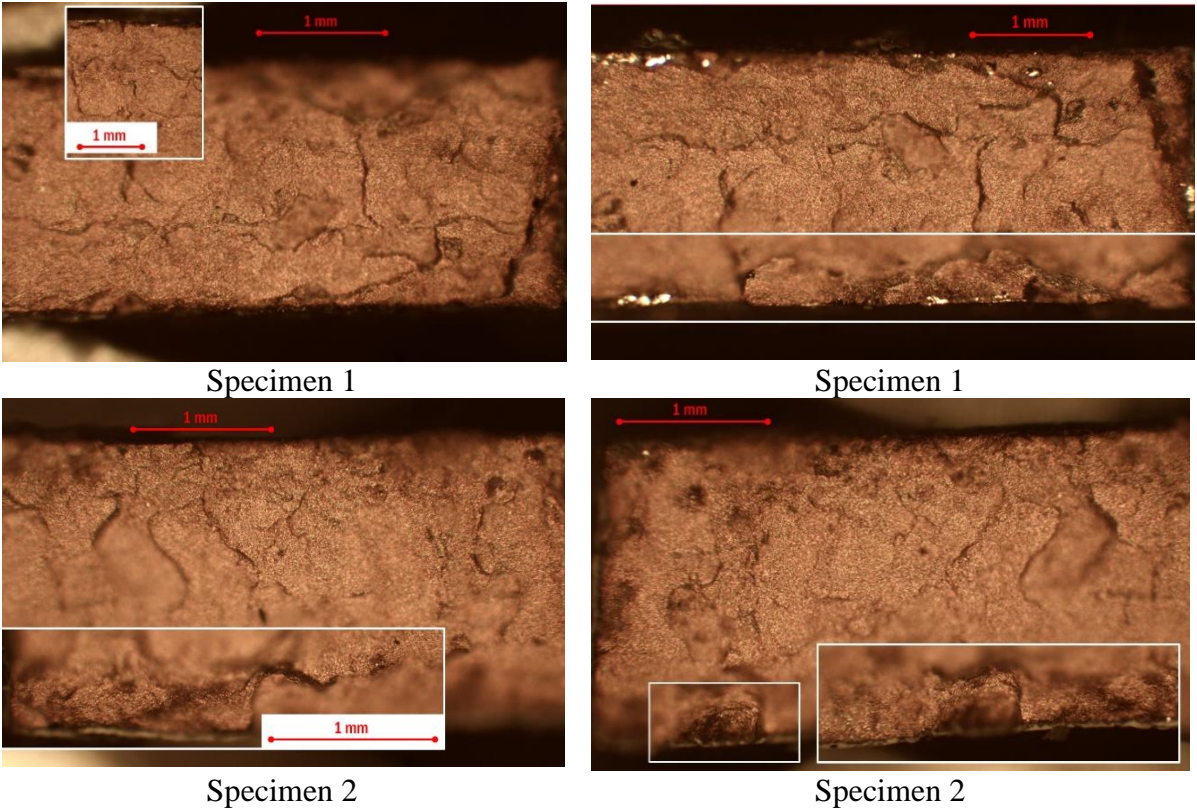

**Figure 21.** Selected fractography images of the specimens of Group 3

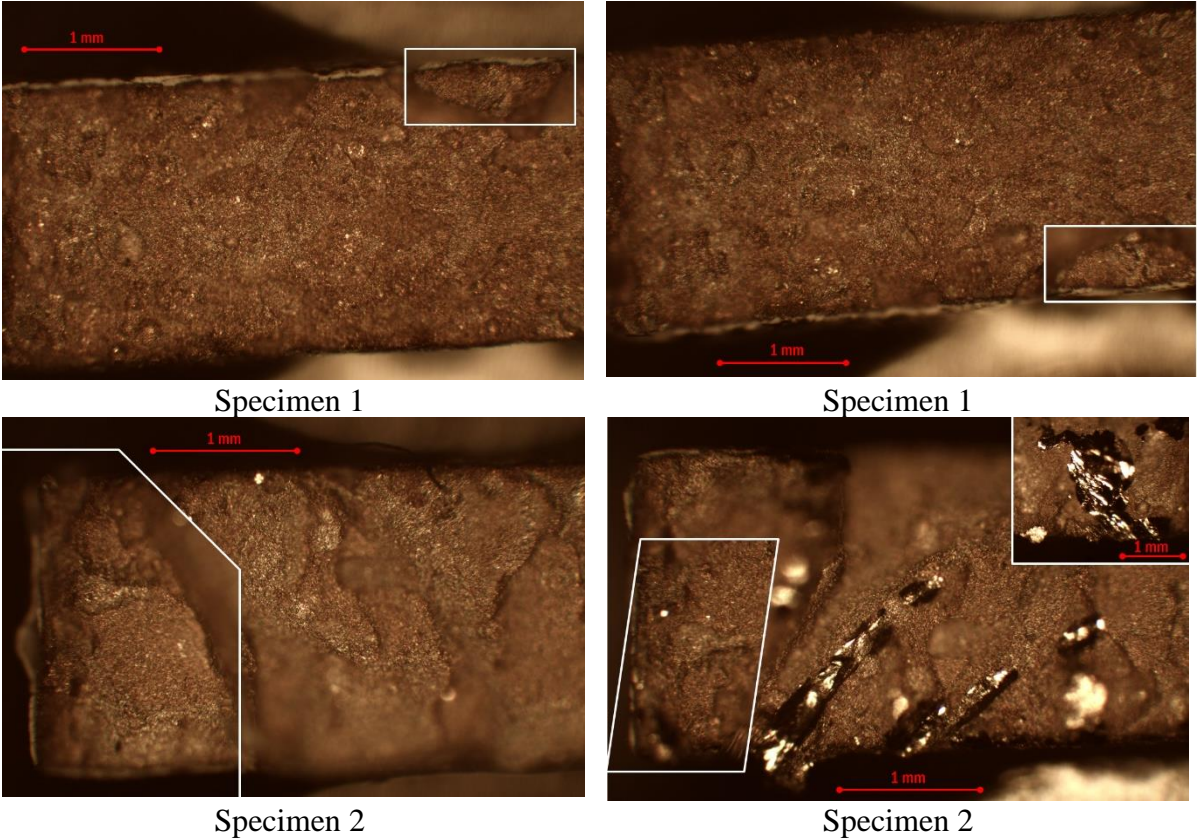

**Figure 22.** Selected fractography images of the specimens of Group 4

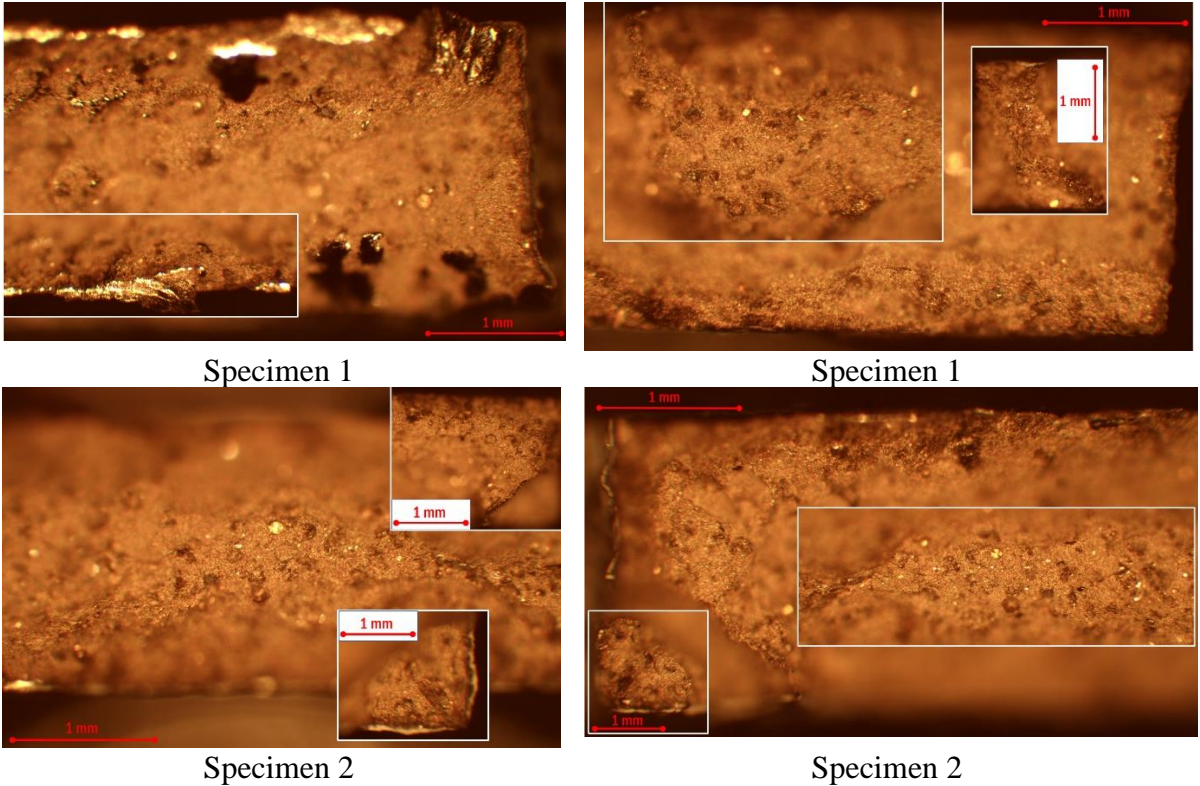

Specimen 1               Specimen 1

Specimen 2               Specimen 2

**Figure 23.** Selected fractography images of the specimens of Group 5

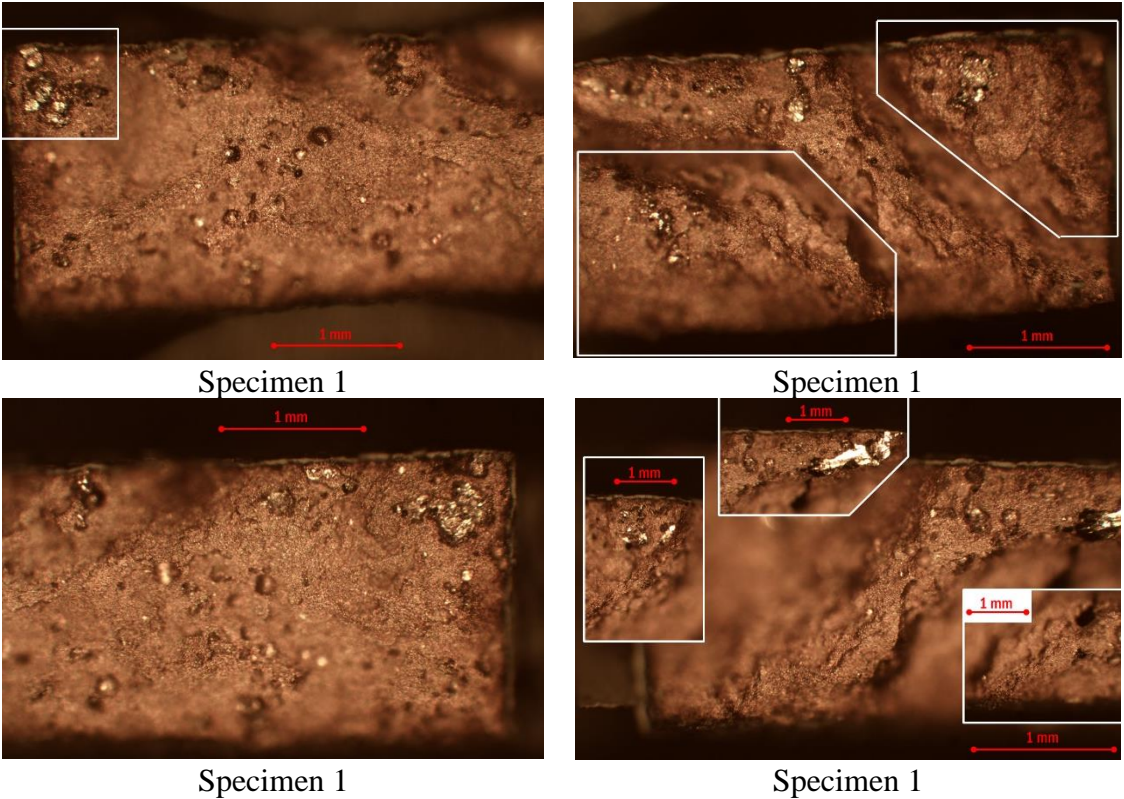

Specimen 1               Specimen 1

Specimen 1               Specimen 1

**Figure 24.** Selected fractography images of the specimens of Group 6

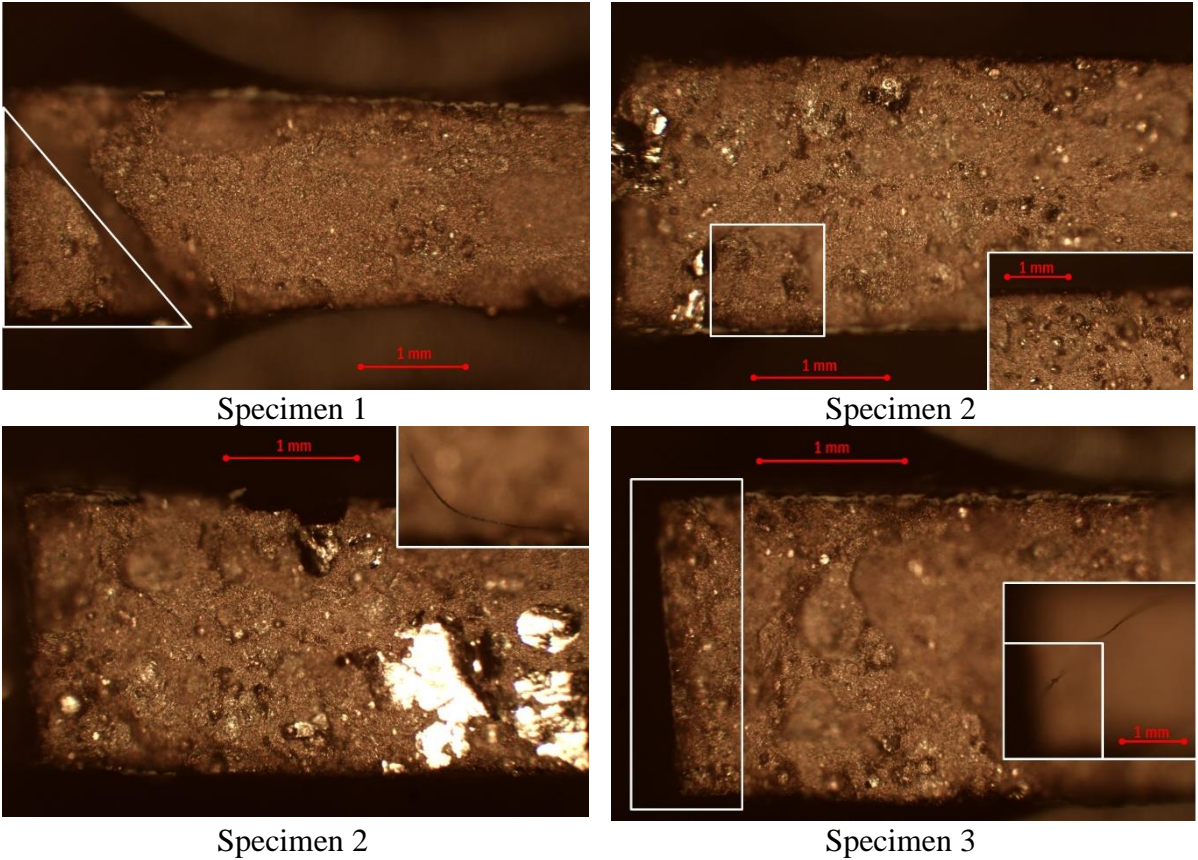

**Figure 25.** Selected fractography images of the specimens of Group 7

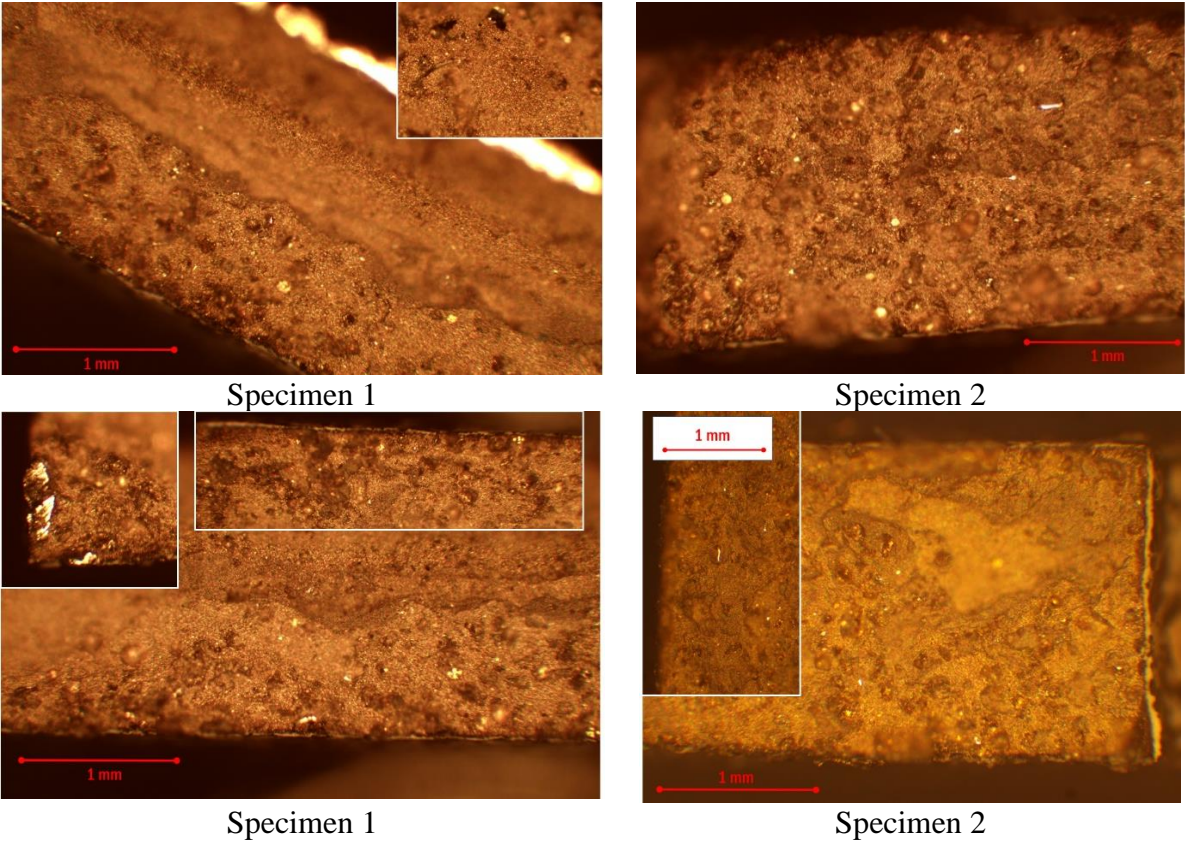

**Figure 26.** Selected fractography images of the specimens of Group 8

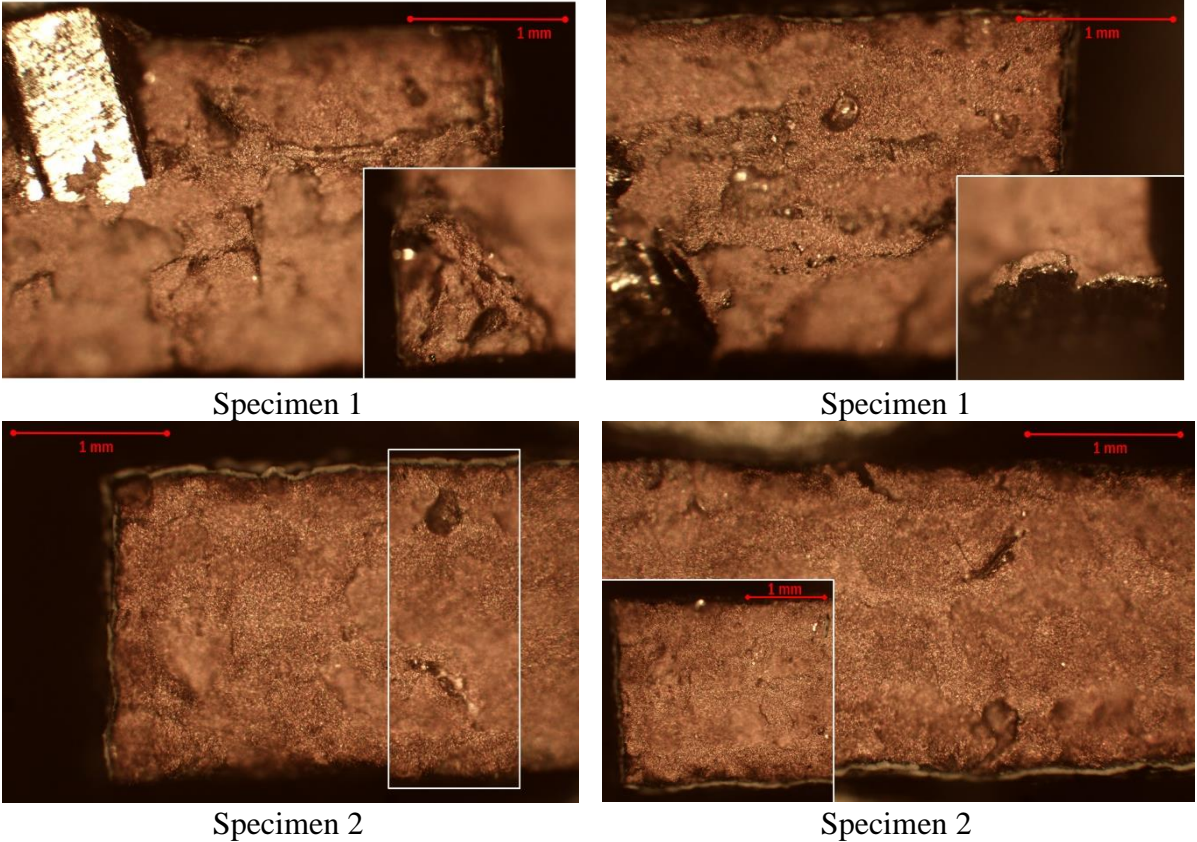

One of the prepared Cr bronze + SS alternating layers specimens didn't fail entirely after tensile test, but only cracked in its middle area. This macroscopic phenomenon of a partial failure has a similar behaviour with a microscopic crack arrest [108] provided by specific elements in the crystalline structure of the specimen such as a δ-ferrite in the binder jet-printed SS 316L, what was demonstrated in the study of T.H. Becker et al [108]. In **Figure 27**, the external pattern of the crack at the surface of the partially failed specimen is demonstrated.

**Figure 27.** Crack at the surface of Group 8 specimen

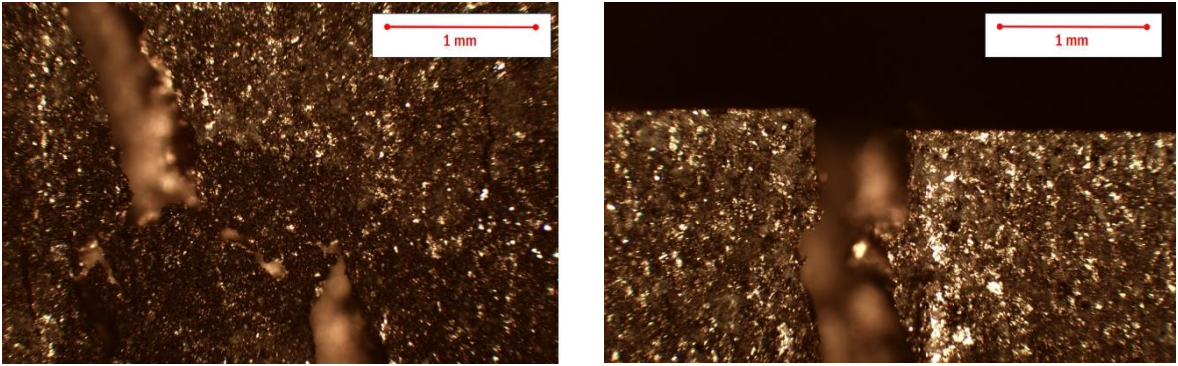

From the macrofractography images (Figures 19-26), it is seen that:

a) Specimens of Group 1 (Figure 19) are characterized by brittle fracture with inclusion of faceted surfaces. They have smooth (in a first approximation) edges around perimeter. This macrostructure pattern of the fracture surfaces proves the suggestion, made in the part 3.3.4, that failure of Group 1 specimen happened during low plastic deformation (the applied stretching force caused internal stresses lower than critical values that trigger intercrystalline slip, but higher than stresses, which break interatomic bonds), because the fracture is characterized predominantly by intracrystalline rupture.

b) Specimens of Group 2 (Figure 20) demonstrate shear fracture with foveolar surface geometry, don't have neither facets nor metallic lustre (except small areas near periphery). Edges are highly jagged. It was mentioned previously in the §3.3.3 that a DIC of Group 2 showed a hyperboloid-like shape of major deformation distribution near the stress concentrator, what is followed by a highly uneven fracture surface near the front side of the fracture surface. This character of failure and such strain distribution isn't attributed to porosity or blisters, but could be caused by lack-of-fusion flaw near the area of ultimate strain [108].

c) Specimens of Group 3 (Figure 21) show brittle fracture, huge faceted surfaces, and smooth edges. It is seen that they are highly affected by porosity.

d) Specimens of Group 4 (Figure 22) have undoubtedly brittle fracture, rough surface, big facets, blisters, and sharp edges. They also have clearly visible defects as huge gas pockets (see left fracture pattern of the specimen 1).

e) Specimen of Group 5 (Figure 23) show brittle fracture, small facets, and highly roughened surface. Specimen have high porosity and demonstrate lithoidal transcrystalline-like rupture.

f) Specimens of Group 6 (Figure 24) are also characterized by brittle fracture, faceted surfaces, foveolar relief, porosity, and jagged edges.

g) Specimens of Group 7 (Figure 25) demonstrate liquation, shear fracture with elements of brittle rupture, a lot of blisters, and foveolar surface. But their periphery is smoother, and fracture type looks more like intracrystalline. The same traits were observed above in the specimen of Group 1 (Figure 19).

h) Specimens of Group 8 (Figure 26) have the biggest facets, unsmooth surface, highly dispersed structure, and particularly show the akin pattern with specimens created from Sn bronze.

It seen that the most common type of breaking of developed specimens is brittle fracture (the same result was demonstrated by P. Guo et al. [69] in case of SS 316L alloy, manufactured by high-power direct laser deposition, but this kind of fracture is not common for the Cu-based AM-fabricated alloys and the alternating layers structures: contrarily, the roll bonded alternating structure [48] described by N. Koga et al., showed that increase of number of layers mainly suppress the brittle fracture presumably due to grain refinement, and that the neighboring Cu layers should suppress the brittle fracture of the Fe layers, that was observed on the pure Fe sheets with 20 μm grain size). However, a shear fracture in current analysis was treated only by Group 2 (chromium bronze + stainless steel (1:1)) and Group 7 (aluminium bronze + stainless steel (alt.)). The same mechanism of a ductile fracture with many voids and dimples was shown by X. Zhang et al. [109] for the Cu-D22-H13 FG alloy, created from the copper and H13 tool steel through the Deloro 22 interlayer via the DED, and in the Cu-SS FGM with ~200 MPa UTS, also fabricated via DED through the interlayers of the same nickel-based alloy [29], that failed in the Cu zone (contrarily, the D22-SS specimens produced without copper exhibited a brittle fracture surface without dimples, but with several microcracks); the H13-Cu FG DMD-fabricated FG alloy discussed by M. Khalid Imran et al. [110] also was characterized by absolute ductile fracture morphology with dimple fractures observed near the crack nucleus. H. Shi et al. [111] described all kinds of rupture: brittle, plastic (shear, ductile), and transcrystalline in case of pre-alloyed Fe-Cu based MMC in dependence from the concrete molten state content (MSC). All of them were also presented in the specimens of Groups 1-8 described below, also dependently from the wt.% chemical composition of the alloy. The transcrystalline fracture was common mostly for Cr bronze-based Groups (such as Group 5), and the intracrystalline rupture was observed in case of specimens based on the Al bronze (Groups 1 and 7).

The most widespread defect, observed during the fractographic analysis, was porosity, that was seen in all specimens, especially – of the 3:1 Groups (№№4-6), and of the Al bronze-based Groups with 1:1 SS-bronze ratio (№1) and alternating layers (№7). It is known that a hot isostatic pressure (HIP) post-treatment can be applied to suppress this defect and simultaneously close internal flaws, cracks, and eliminate the residual stresses [108,112-114] (moreover, the combination of the HIP and heat post-treatment show increase in fracture toughness of the AM-fabricated parts [112]). Another way that could provide the slight decrease of the porosity amount in the DED-fabricated parts is implementation of a static magnetic field during the process of energy deposition: A. Filimonov et al [115] demonstrated such effect in the study of mechanical properties of Inconel 718 structures that were deposited under the superimposed magnetic field of 0.2 T. According to the current state of art, it is seen that such technological feature is not studied well by this day and is worth to be considered specifically.

Nevertheless, besides existence of a high amount and huge size of porosity in the Al bronze-based specimens, there was no lack in their tensile mechanical properties, as it could be seen from the results of the conducted stretching tests.

## 4. Conclusion

The Al bronze-SS binary alloys (50%–50% and 75%–25%) and alternating layers FGM fabricated via the DED in DMT mode exhibited excellent mechanical characteristics:

– UTS: up to 715 MPa (75% bronze, 25% SS); up to 731.5 MPa (alt. layers of bronze and SS); up to 876.4 MPa (50% bronze, 50% SS), what is higher than UTS of laser deposited SS 316L [68,93];

– yield strength: up to 710 MPa (50% bronze, 50% SS); up to 450 MPa (75% bronze, 25% SS); up to 700 MPa (alt. layers of bronze and SS), what is higher than yield strength of the SLM-fabricated SS 316L [97] and comparable with additive manufactured TiB2-reinforced SS matrix nanocomposites SMC-5 and SMC-10 [98].

Chromium bronze alloys demonstrated the lowest tensile mechanical strength parameters (294–463.3 MPa UTS and 280–380 MPa yield strength), tin bronze-based structures had the intermediate values (401.9–578.6 MPa UTS and 310–420 MPa yield strength). The higher homogeneity, regular intermixed structure, absence of visible cracking in as-built state, and low porosity of Cr bronze-based specimens, nevertheless, didn't provide an improvement of their mechanical properties. The possible way of the yield strength increase is a grain refinement [99] that could be associated with liquid separation, monotectic reaction and solid-state phase transformations upon cyclic heating followed by the rapid solidification of 3D-printing process. The UTS of all alloys could be improved by changing of the build direction from 0° to 90° to the load direction, because the 0°-built parts commonly consist of large dendritic grains, generated by epitaxial nucleation, due to the high temperature gradient between the melt pool and the previously deposited areas [68-69,93-94]. The evaluation of the longitudinal $V_L$ and Rayleigh $V_R$ velocities in all alloys, conducted via the elasticity modulus and the Poisson's ratio calculation, will assist in the further implementation and investigation of the ultrasonic-assisted DED process that could be also suggested for improvement of the quality and mechanical characteristics of DED-fabricated Cu-Fe parts, including grain refinement, increase of wear resistance and microstructure homogeneity, decrease of amount of the brittle phases, dissolving of oxides, and the surface quality improvement [96,104-107].

The transcrystalline fracture was common mostly for Cr bronze-based alloys, and the predominantly intracrystalline rupture was observed in case of the specimens based on the Al bronze. This distinction of the Al bronze-based parts, that also provides their mechanical strength, was proved by the low plastic-to-elastic deformation ratio in case of 50%–50% Al bronze-SS alloy. Lowering of this parameter down to values less than unity points at the fact that the failure occurred during low plastic deformation, while the applied stretching force caused internal stresses lower than critical values that trigger an intercrystalline slip, but higher than stresses, which break interatomic bonds.

The most common defect of all specimens, including Al bronze-based groups, was porosity, observed during the fractographic analysis, the prevailing type of failure was a brittle rupture. The possible way of porosity decrease in the parts, printed in the DMT mode, is a HIP post-treatment that can also close the internal flaws, cracks, and eliminate the residual stresses [108,112-114], the combination of the HIP and heat post-treatment that could also increase a fracture toughness [112], and the experimental method of a static magnetic field implementation during the process of laser deposition could be also suggested for slight decrease of porosity [115].

The addition of 50wt.% and more bronze in the SS 316L powder significantly lowers the elasticity modulus of the resulted DED-fabricated binary Cu-Fe system alloy: the elasticity modulus reached the highest values in case of chromium bronze-based materials (up to 42.4 GPa). what is several times lower than Young's modulus of DED-fabricated SS 316L [96], and lower than that of SS 316L components, laser deposited using various build orientations [93]. The Cu-Fe alloys with highest elasticity modulus (Cr bronze-based groups) demonstrated the lowest UTS and yield strength; the lowest average elasticity modulus (approximately 7–17 Gpa) was attributed to materials with highest mechanical strength (Al bronze-based groups). These materials are also expected to have the best notch toughness due to their high specific energy of proportional deformation (modulus of resilience) – up to approximately 1.9782 MJ/m$^3$.

The highest values of LTEC were associated with FGMs created via alternating layers technique (from 1.878±0.15 to 1.959±0.15 K$^{-1}$); therefore, these parts are expected to suffer the highest deformations under the high temperatures, what is an undesired property in many practical applications of Cu-Fe FGMs. Contrarily, the lowest values was attributed to alloys with highest bronze percentage (75%) (the lowest average value was demonstrated in case of 75%–25% Al bronze + SS alloy – 1.212±0.15·10$^{-5}$ K$^{-1}$). These materials are determined to be characterized by the best resistance to the linear thermal deformations under the high temperature influence. Part of Group 5 (Cr bronze + stainless steel), which also had the same bronze percentage (75%), nevertheless, didn't show such a low LTEC. The hypothetic reason of it may be associated with the forming of additional Fe-Cr phases in this alloy and the total absence of Fe-Zn, Fe-Sn, and Fe-Al. The parts with 50%–50% steel-to-bronze ratio exhibited the intermediate values of LTEC (from 1.509±0.15 to 1.572±0.15 K$^{-1}$) between FG alternating layers alloys and 75%–25% Cu-Fe materials. The LTECs of 50%–50% alloys significantly differ from alternating layers parts, what points at impossibility to analytically predict the LTEC of the last ones using the rule-of-mixtures equation. It could be associated with forming of new phases, i.a. intermetallics [26,84] in the interface areas between different metals, which have thermal characteristics (i.a. LTEC) different from pure initial materials – steel and bronze, and the epitaxial growth [47].

An implementation of a three-way interaction regression model considering the parameters of laser treatment, developed for the SLM [71], provided the accurate results for the DED process too. According to this model, the relative thermal expansion $\varepsilon_T$, [μm/m], for SS 316L was evaluated as $\varepsilon_T \cong 219.8$ μm/mm at the room temperature 21°C, and $\varepsilon_T \cong 3913.0$ μm/mm at 220 °C, what is close to the results of [71], provided at the critical laser energy density ($\varepsilon_T \cong 3900$ μm/mm at 220 °C), and higher than the thermal expansion of the parts, fabricated via the SLM at the same laser power and scanning speed, as it was implemented in the DED, but with lower hatch spacing ($h$ = 120 μm, $\varepsilon_T \cong 3100$–3200 μm/m). This fact points at possibility of decrease of LTEC of DED-processed parts through variating of the hatch spacing values.

The results of the conducted research are expected to be applied for the axial [116] and radial [117] DED of the Cu-Fe system FGMs, especially based on the SS – aluminium bronze combination, characterized by the high UTS, yield strength, notch toughness, and low LTEC, performed using the alternating layers technique and the gradient path method.

**Acknowledgements**

The authors express gratitude to Prof. Ivan V. Sergeichev for the detailed review of the manuscript and insightful commentaries, and to students Anna Nikiforova and Dmitry Ivanitsky for their contribution to the conducted research.

**References**

[1] El-Galy, I.M.; Bassiouny, B.I.; Ahmed, M.H. Empirical model for Dry Sliding Wear Behaviour of Centrifugally Cast Functionally Graded Al / SiC$_p$ Composite. *Key Eng. Mater.* 786 **2018**, pp. 276-285.

[2] Ghanavati, R.; Naffakh-Moosavy, H. Additive manufacturing of functionally graded metallic materials: A review of experimental and numerical studies. *J. Mater. Research and Technol.* 13 **2021**, pp. 1628-1664.

[3] Saleh, B.; Jiang, J.; Fathi, R.; Al-hababi, T.; Xu, Q.; Wang, L.; Song, D.; Ma, A. 30 Years of functionally graded materials: An overview of manufacturing methods, Applications and Future Challenges. *Comp. Part* B 201 **2020**, 108376.

[4] Liu, Z.; Meyers, A.M.; Zhang, Z.; Ritchie, R.O. Functional gradients and heterogeneities in biological materials: Design principles, functions, and bioinspired applications *Progress Mater. Sci.* 88 **2017**, pp. 467-498.

[5] Bentov, S.; Zaslansky, P.; Al-Sawalmih, A.; Masic, A.; Fratzl, P.; Sagi, A.; Berman, A.; Aichmayer, B. Enamel-like apatite crown covering amorphous mineral in a crayfish mandible. *Nature Comm.* 3 **2012**, 839.

[6] Wang, B.; Yang, W.; McKittrick, J.; Andre Meyers, M. Keratin: Structure, mechanical properties, occurrence in biological organisms, and efforts at bioinspiration. *Progress Mater. Sci.* 76 **2016**, pp. 229-318.

[7] Kasapi, M.A.; Gosline, J.M. Design complexity and fracture control in the equine hoof wall. *J. Exp. Biol.* 200 **1997**, pp. 1639-1659.

[8] Liu, Z.Q.; Jiao, Z.Y.; Weng, Z.Y.; Zhang, Z.F. Structure and mechanical behaviors of protective armored pangolin scales and effects of hydration and orientation. *J Mech. Behav. Biomed. Mater*. 56 **2016**, pp. 165-174.

[9] Wang, B.; Yang, W.; Sherman, V.R.; Meyers, M.A. Pangolin armor: overlapping, structure, and mechanical properties of the keratinous scales. *Acta Biomater.* 41 **2016**, pp. 60-74.

[10] Chen, P.-Y.; Stokes, A.G.; McKittrick, J. Comparison of the structure and mechanical properties of bovine femur bone and antler of the North American elk (Cervus elaphus canadensis). *Acta Biomater.* 5 **2009**, pp. 693-706.

[11] Chen, P.-Y.; Lin, A.Y.M.; Lin, Y.-S.; Seki, Y.; Stokes, A.G.; Peyras, J.; Olevsky, E.A.; Meyers, M.A.; McKittrick, J. Structure and mechanical properties of selected biological materials. *J. Mech. Behav. Biomed. Mater.* 1 **2008**, pp. 208-226.

[12] Elbaum, R.; Zaltzman, L.; Burgert, I.; Fratzl, P. The role of wheat awns in the seed dispersal unit. *Science* 316 **2007**, pp. 884-886.

[13] Liu, Z.; Meyers, M.A.; Zhang, Z.; Ritchie, R.O. Functional gradients and heterogeneities in biological materials: Design principles, functions, and bioinspired applications. *Progress Mater. Sci.* 88 **2017**, pp. 467-498.

[14] Yan, L.; Chen, Y.; Liou, F. Additive manufacturing of functionally graded metallic materials using laser metal deposition. *Additive Manufacturing* 31 **2020** 100901.

[15] Ghanavati, R.; Naffakh-Moosavy, H. Additive manufacturing of functionally graded metallic materials: A review of experimental and numerical studies. *J. Mater. Research and Tech.* 13 **2021**, pp. 1628-1664.

[16] Singh, D.D.; Arjula, S.; Reddy, A.R. Functionally Graded Materials Manufactured by Direct Energy Deposition: A review. *Mater. Today: Proc.* Article in press. Accessed online August 28, 2021 from

https://reader.elsevier.com/reader/sd/pii/S2214785321034611?token=881E605ACF30D9A224318409F3722F0A0E2DE17115030741E00AD500066FBC6C8FC274ACB42F82433E6897AE6F174A0D&originRegion=eu-west-1&originCreation=20210828152626

[17] Heer, B.; Bandyopadhyay, A. Compositionally graded magnetic-nonmagnetic bimetallic structure using laser engineered net shaping. *Mater. Lett.* 216 **2018**, pp. 16-19.

[18] Bandyopadhyay, A.; Heer, B. Additive manufacturing of multi-material structures. *Mater. Sci. Eng.* R 129 **2018**, pp. 1-16.

[19] Zhang, Y.; Bandyopadhyay, A. Direct fabrication of compositionally graded Ti-Al2O3 multi-material structures using Laser Engineered Net Shaping. *Additive Manuf.* 21 **2018**, pp. 104-111.

[20] Makarenko, K.; Dubinin, O.; Shishkovsky, I. Analytical Evaluation of the Dendritic Structure Parameters and Crystallization Rate of Laser-Deposited Cu-Fe Functionally Graded Materials. *Materials* **2020**, *13*, 5665. https://doi.org/10.3390/ma13245665

[21] Jamroziak, K.; Roik, T.; Gavrish, O.; Vitsuik, I.; Lesiuk, G.; Correia, J.A.F.O.; De Jesus, A. Improved manufacturing performance of a new antifriction composite parts based on copper. *Eng. Failure Analysis* **2018**, *91*, 225-233

[22] Karnati, S.; Sparks, T.E.; Liou, F.; Newkirk, J.W.; Taminger, K.M.B.; Seufzer, W.J. Laser metal deposition of functionally gradient materials from elemental copper and nickel powders. In Proceedings of the 26th Solid Freeform Fabrication Symposium, Austin, TX, USA, 10–12 August 2015; Bourell, D.L., Ed.; The University of Texas: Austin, TX, USA, 2015; pp. 789–802.

[23] Moharana, B.R.; Sahu, S.K.; Maiti, A.; Sahoo, S.K.; Moharana, T.K. An experimental study on joining of AISI 304 SS to Cu by Nd-YAG laser welding process. *Mater. Today Proc.* **2020**.

[24] Makarenko, K.; Shishkovsky, I. Direct Energy Deposition of Cu-Fe System Functionally Graded Structures. *IOP Conf. Ser.: Mater. Sci. Eng.* **2020**, *969*, 012104.

[25] Tan, C.; Chew, Y.; Bi, G.; Wang, D.; Ma, W.; Yang, Y.; Zhou, K. Additive manufacturing of steel–copper functionally graded material with ultrahigh bonding strength. *J. Mater. Sci. Technol.* 72 **2021**, pp. 217-222.

[26] Makarenko, K.; Dubinin, O.; Shornikov, P.; Shishkovsky, I. Specific aspects of the transitional layer forming in the aluminium bronze—stainless steel functionally graded structures after laser metal deposition. *Proc. CIRP* **2020**, *94*, 346–351.

[27] Liu, Z.H.; Zhang, D.Q.; Sing, S.L.; Chua, C.K.; Loh, L.E. Interfacial characterization of SLM parts in multi-material processing: Metallurgical diffusion between 316L stainless steel and C18400 copper alloy. *Mater. Char.* 94 **2014**, pp. 116-125.

[28] Bai, Y.; Zhang, J.; Zhao, C.; Li, C.; Wang, H. Dual interfacial characterization and property in multi-material selective laser melting of 316L stainless steel and C52400 copper alloy. *Mater. Characterization* 167 **2020**, 110489.

[29] Zhang, X.; Pan, T.; Flood, A.; Chen, Y.; Zhang, Y.; Liou, F. Investigation of copper/stainless steel multi-metallic materials fabricated by laser metal deposition. *Mater. Sci. Eng.* A 811 **2021** 141071.

[30] Tan, C.; Zhou, K.; Ma, W.; Min, L. Interfacial characteristic and mechanical performance of maraging steel-copper functional bimetal produced by selective laser melting based hybrid manufacture. *Mater. Design* 155 **2018**, pp. 77-85.

[31] Kulish, A.; Salvatore, F.; Rech, J.; Courbon, J.; Munoz, J.; Joffre, T. Effect of direct energy deposition parameters on morphology, residual stresses, density, and microstructure of 1.2709 maraging steel. *Int. J. Adv. Manuf. Technol.* **2021**.

[32] Onuike, B.; Heer, B.; Bandyopadhyay, A. Additive manufacturing of Inconel 718—Copper alloy bimetallic structure using laser engineered net shaping (LENS™). *Additive Manuf.* 21 **2018**, pp. 133-140.

[33] Zhang, C.H.; Zhang, H.; Wu, C.L.; Zhang, S.; Sun, Z.L.; Dong, S.Y. Multi-layer functional graded stainless steel fabricated by laser melting deposition. *Vacuum* 141 **2017**, pp. 181-187.

[34] Gualtieri, T.; Bandyopadhyay, A. Niobium carbide composite coatings on SS304 using laser engineered net shaping (LENS™). *Mater. Lett.* 189 **2017**, pp. 89-92.

[35] Bandyopadhyay, A.; Dittrick, S.; Gualtieri, S.; Wu, J.; Bose, S. Calcium phosphate–titanium composites for articulating surfaces of load-bearing implants. *J. Mech. Behav. Biomed. Mater.* 57 **2016**, pp. 280-288.

[36] Vamsi Krishna, B.; Xue, W.; Bose, S.; Bandyopadhyay, A. Functionally graded Co–Cr–Mo coating on Ti–6Al–4V alloy structures. *Acta Biomater.* 4 **2008**, pp. 697-706.

[37] Sahasrabudhe, H.; Harrison, R.; Carpenter, C.; Bandyopadhyay, A. Stainless steel to titanium bimetallic structure using LENS$^{TM}$. *Additive Manuf.* 5 **2015**, pp. 1-8.

[38] Ghosh, M.; Chatterjee, S.; Mishra, B. The effect of intermetallics on the strength properties of diffusion bonds formed between Ti–5.5Al–2.4V and 304 stainless steel. *Mater. Sci. Eng.* A363 **2003**, pp. 268-274.

[39] Qin, B.; Sheng, G.M.; Huang, J.W.; Zhou, B.; Qiu, S.Y.; Li, C. Phase transformation diffusion bonding of titanium alloy with stainless steel. *Mater. Character.* 56 **2006**, pp. 32-38.

[40] Kundu, S.; Chatterjee, S.; Olson, D.; Mishra, B. Effects of Intermetallic Phases on the Bond Strength of Diffusion-Bonded Joints between Titanium and 304 Stainless Steel Using Nickel Interlayer. *Metallurg. Mater. Transactions* A 38 **2007**, pp. 2053-2060.

[41] Elrefaey, A.; Tillmann, W. Solid state diffusion bonding of titanium to steel using a copper base alloy as interlayer. *J. Mater. Proc. Technol.* 209 **2009**, pp. 2746-2752.

[42] Li, W.; Karnati, S.; Kriewall, C.; Liou, F.; Newkirk, J.; Taminger, K.M.B.; Seufzer, W.J. Fabrication and characterization of a functionally graded material from Ti-6Al-4V to SS316 by laser metal deposition. *Additive Manuf.* 14 **2017**, pp. 95-104.

[43] Bi, Y.; Xu, Y.; Zhang, Y.; Xue, R.; Bao, Y. Single-pass laser welding of TC4 Ti alloy to 304 SS with V interlayer and V/Cu bilayer. *Mater. Lett.* 285 **2021**, 129072.

[44] Tan, C.; Chew, Y.; Duan, R.; Weng, F.; Sui, S.; Lan Ng, F.; Du, Z.; Bi, G. Additive manufacturing of multi-scale heterostructured high-strength steels. *Mater. Res. Lett.* 9 **2021**, pp. 291-299.

[45] Melzer, D.; Džugan, J.; Koukolíková, M.; Rzepa, S.; Vavřík, J. Structural integrity and mechanical properties of the functionally graded material based on 316L/IN718 processed by DED technology. *Mater. Sci. Eng.* A 811 **2021**, 141038.

[46] Feenstra, D.R.; Banerjee, R.; Fraser, H.L.; Huang, A.; Molotnikov, A.; Birbilis, N. Critical review of the state of the art in multi-material fabrication via directed energy deposition. *Curr. Opin. Solid State Mater. Sci.* 25 **2021**, 100924.

[47] Chen, M.; Khan, H.A.; Wan, Z.; Lippert, J.; Sun, H.; Shang, S.-L.; Liu, Z.-K.; Li, J. Microstructural characteristics and crack formation in additively manufactured bimetal material of 316L stainless steel and Inconel 625. *Addit. Manuf.* 32 **2020**, 101037.

[48] Koga, N.; Tomono, S.; Umezawa, O. Low-temperature tensile properties of Cu-Fe laminated sheets with various number of layers. *Mater. Sci. Eng.* A 811 **2021**, 141066.

[49] Koga, N.; Zhang, W.; Umezawa, O.; Tschan, V.; Sas, J.; Weiss, K.P. Temperature dependence on tensile properties of Cu-40mass%Fe dual phase alloy. *IOP Conf. Ser.: Mater. Sci. Eng.* 279 **2017**, 012004.

[50] Park, J.; Park, K.; Kim, J.; Jeong, Y.; Kawasaki, A.; Kwon, H. Fabrication of a Functionally Graded Copper-Zinc Sulfide Phosphor *Sci. Rep.* 6 **2016**, pp. 1-6.

[51] Lin, X.; Yue, T.M.; Yang, H.O.; Huang, W.D. Laser rapid forming of SS316L/Rene88DT graded material. *Mater. Sci. Eng.* A 391 **2005**, pp. 325-336.

[52] Carroll, B.E.; Otis, R.A.; Borgonia, J.P.; Suh, J.; Dillon, R.P.; Shapiro, A.A.; Hoffmann, D.C.; Liu, Z.-K.; Beese, A.M. Functionally graded material of 304L stainless steel and inconel 625 fabricated by directed energy deposition: Characterization and thermodynamic modeling. *Acta Mater.* 108 **2016**, pp. 46-54.

[53] Li, L.; Wang, J.; Lin, P.; Liu, H. Microstructure and mechanical properties of functionally graded TiCp/Ti6Al4V composite fabricated by laser melting deposition. *Ceramics Int.* 43 **2017**, pp. 16638-16651.

[54] Gualtieri, T.; Bandyopadhyay, A. Additive manufacturing of compositionally gradient metal-ceramic structures: Stainless steel to vanadium carbide. *Mater. and Design* 139 **2018**, pp. 419-428.

[55] Bobbio, L.D.; Bocklund, D.; Otis, R.; Borgonia, J.P.; Dillon, R.P.; Shapiro, A.A.; McEnerney, B.; Liu, Z.-K.; Beese, A.M. Characterization of a functionally graded material of Ti-6Al-4V to 304L stainless steel with an intermediate V section. *J. Alloys and Compounds* 742 **2018**, pp. 1031-1036.

[56] Reichardt, A.; Dillon, R.P.; Borgonia, J.P.; Shapiro, A.A.; McEnerney, B.W.; Momose, T.; Hosemann, P. Development and characterization of Ti-6Al-4V to 304L stainless steel gradient components fabricated with laser deposition additive manufacturing. *Mater. Design* 104 **2016**, pp. 404-413.

[57] Futamura, Y.; Tsuchiyama, T.; Takaki, S. Proc 1st Int. Conf. Adv. Struct. Steels ICASS **2002**, 201.

[58] Sen, I.; Amankwah, E.; Kumar, N.S.; Fleury, E.; Oh-ishi, K.; Hono, K.; Ramamurty, U. Microstructure and mechanical properties of annealed SUS 304H austenitic stainless steel with copper. *Mater. Sci. Eng.* A 528 **2011**, 4491-4499.

[59] Maziasz, J.; Shingledecker, J.P.; Evans, N.D.; Pollard, M.P. Developing new castaustenitic stainless steels with high-temperature creep resistance. *J. Pressure Vessel Technol.* 131/051404 (1) **2006** 1.

[60] Viswanathan, R.; Purget, R.; Rawls, P. Coal-fired power materials *Adv. Mater. Processes* **2008** 47.

[61] Viswanathan, R.; Coleman, K. Boiler materials for ultra-super critical (USC) coal power plants. USC quarterly report, **2004** 5.

[62] Kan, T.; Sawaragi, Y.; Yamadera, Y.; Okada, H. *Proc. of 6th Liege Conf on Materials for Advanced Power Engineering*, part-1 **1998**, 441.

[63] Tohyama, A.; Minami, Y. Development of the high temperature materials for ultrasuper critical boilers. *NKK technical report* 84 **2001**, 30.

[64] Sourmail, T. *Mater. Sci. Technol.* 17 *1* **2001** 1.

[65] Laha, K.; Kyono, J.; Shinya, N. *Scripta Mater.* 56 **2007** 915.

[66] Caminada, S.; Cumino, G.; Cipolla, L.; Venditti, D.; Gianfrancesco, A.D.; Minami, Y.; Ono, T. *Int. J. Pressure Vessels Piping* 87 **2010** 336.

[67] Fleury, E.; Hong, S.M.; Shim, J.H.; Lee, S.C. *Proc. of the 3rd Symposium on Heat Resistant Steels and Alloys for High Efficiency USC Power Plants 2009*, Tsukuba, Japan, June 2–4, **2009**.

[68] Azinpour, E.; Darabi, R.; De Sa, J.C.; Santos, A.; Hodek, J.; Dzugan, J. Fracture analysis in directed energy deposition (DED) manufactured 316Lstainless steel using a phase-field approach. *Finite Elem. Analysis Design* 177 **2020**, 103417.

[69] Guo, P.; Zou, B.; Huang, C.; Gao, H. Study on microstructure, mechanical properties and machinability of efficiently additive manufactured AISI 316L stainless steel by high-power direct laser deposition. *J. Mater. Process. Technol.* 240 **2017**, pp. 12-22.

[70] ASTM International: ASTM E8/E8M-16a, Standard Test Methods for Tension Testing of Metallic Materials. URL: https://www.astm.org/DATABASE.CART/HISTORICAL/E8E8M-16A.htm (Access date: August 28, 2021).

[71] Yakout, M.; Elbestawi, M.A.; Veldhuis, S.C. A study of thermal expansion coefficients and microstructure during selective laser melting of Invar 36 and stainless steel 316L. *Additive Manuf.* 24 **2018**, pp. 405-418.

[72] Liu, M.; Kumar, A.; Bukkapatnam, S.; Kuttolamadom, M. A Review of the Anomalies in Directed Energy Deposition (DED) Processes & Potential Solutions - Part Quality & Defects. *Proc. Manuf.* 53 **2021**, pp. 507-518.

[73] Liu, Z.; Kim, H.; Liu, W.; Cong, W.; Jiang, Q.; Zhang, H. Influence of energy density on macro/micro structures and mechanical properties of as-deposited Inconel 718 parts fabricated by laser engineered net shaping. *J. Manuf. Processes* 42 **2019**, pp. 96-105.

[74] DebRoy, T.; Wei, H.L.; Zuback, J.S.; Mukherjee, T.; Elmer, J.W.; Milewski, J.O.; Beese, A.M.; Wilson-Heid, A.; De, A.; Zhang, W. Additive manufacturing of metallic components – Process, structure and properties. *Progress Mater. Sci.* 92 **2018**, pp. 112-224.

[75] Li, R.; Liu, J.; Shi, Y.; Wang, L.; Jiang, W. Balling behavior of stainless steel and nickel powder during selective laser melting process. *Int. J. Adv. Manuf. Technol.* 59 **2012**, pp. 1025-1035.

[76] Arisoy, Y.M.; Criales, L.E.; Özel, T.; Lane, B.; Moylan, S.; Domnez, A. Influence of scan strategy and process parameters on microstructure and its optimization in additively manufactured nickel alloy 625 via laser powder bed fusion. *Int. J. Adv. Manuf. Technol.* 90 **2017**, pp. 1393-1417.

[77] Butler, C.; Babu, S.; Lundy, R.; O'Reilly Meehan, R.; Punch, J.; Jeffers, N. Effects of processing parameters and heat treatment on thermal conductivity of additively manufactured AlSi10Mg by selective laser melting. *Mater. Characterization* 173 **2021** 110945.

[78] Pavlovic, A.S.; Suresh Babu, V.; Seehra, M.S. High-temperature thermal expansion of binary alloys of Ni with Cr, Mo and Re: a comparison with molecular dynamics simulations. *J. Phys. Condens. Matter* 8 **1996**, pp. 3139-3149.

[79] Ferreira, A.J.M.; Batra, R.C.; Roque, C.M.C.; Qian, L.F.; Martins, P.A.L.S. Static Analysis of Functionally Graded Plates Using Third-Order Shear Deformation Theory and a Meshless Method. *Comp. Struct.* 69 **2005**, pp. 449-457.

[80] Large-Area Picosecond Photo-Detectors Project Database. Chapter 17. Material Expansion Coefficients. Linear Thermal Expansion Coefficients of Metals and Alloys // The University of Chicago, Argonne, Fermilab and Berkeley. Accessed online August 22, 2021 from https://psec.uchicago.edu/thermal_coefficients/cte_metals_05517-90143.pdf

[81] AZo Materials. High-Leaded Tin Bronze UNS C93200. Accessed online August 22, 2021 from https://www.azom.com/article.aspx?ArticleID=6552

[82] Dubenko, I.S.; Gaidukova, I.Yu.; Granovsky, S.A.; Gratz, E.; Gurjazkas, D.; Markosyan, A.S.; Müller, H. Strongly Enhanced Thermal Expansion in Binary Y-Mn Intermetallic Compounds. *Solid State Comm.* 103 *8* **1997**, pp. 495-499.

[83] Chung, Y.-L.; Chang, H.-X. Mechanical Behavior of Rectangular Plates with Functionally Graded Coefficient of Thermal Expansion Subjected to Thermal Loading. *J. Therm. Stress* 31 **2008**, pp. 368-388.

[84] Wang, W.; Takata, N.; Suzuki, A.; Kobashi, M.; Kato, M. Formation of multiple intermetallic phases in a hypereutectic Al-Fe binary alloy additively manufactured by laser powder bed fusion. *Intermetallics* 125 **2020** 106892.

[85] Toupin, R.A. (1966) On St. Venant's principle. In: Görtler H. (eds) Applied Mechanics. Springer, Berlin, Heidelberg. https://doi.org/10.1007/978-3-662-29364-5_15

[86] INSTRON 5900 Series Universal Testing Systems up to 50 kN. 5900 Series Premier Testing Systems Brochure. Accessed online August 22, 2021 from https://www.instron.com/-/media/literature-library/products/2012/12/5900-series-general-brochure.pdf?la=en

[87] Andreatta, F.; Lanzutti, A.; Vaglio, E.; Totis, G.; Sortino, M.; Fedrizzi, L. Corrosion behaviour of 316L stainless steel manufactured by selective laser melting. *Mater. Corrosion* 70 **2019**, pp. 1633-1645.

[88] Susan, D.F.; Puskar, J.D.; Brooks, J.A.; Robino, C.V. Quantitative characterization of porosity in stainless steel LENS powders and deposits. *Mater. Char.* 57 **2006**, pp. 36-43.

[89] Wang, L.; Pratt, P.; Felicelli, S.D.; Kadiri, H.E.; Berry, J.T.; Wang, P.T.; Horstemeyer, M.F. Pore Formation in Laser-Assisted Powder Deposition Process. *J. Manuf. Sci. Eng.* 131(5) **2009**, 051008.

[90] Wolff, S.J.; Lin, S.; Faierson, E.J.; Liu, W.K.; Wagner, G.J.; Cao, J. A framework to link localized cooling and properties of directed energy deposition (DED)-processed Ti-6Al-4V. *Acta Mater.* 132 **2017**, pp. 106-117.

[91] Popovich, A.; Sufiiarov, V.; Polozov, I.; Borisov, E.; Masaylo, D.; Orlov, A. Microstructure and mechanical properties of additive manufactured copper alloy. *Mater. Lett.* 179 **2016**, pp. 38–41.

[92] Yang, K.; Li, W.; Guo, X.; Yang, X.; Xu, Y. Characterizations and anisotropy of cold spraying additive-manufactured copper bulk. *J. Mater. Sci. Technol.* 34 **2018**, pp. 1570–1579.

[93] Kersten, S.; Praniewicz, M.; Kurfess, T.; Saldana, C. Build Orientation Effects on Mechanical Properties of 316SS Components Produced by Directed Energy Deposition. *Proc. Manuf.* 48 **2020**, pp. 730-736.

[94] Zhang, K.; Wang, S.; Liu, W.; Shang, X. Characterization of stainless steel parts by laser metal deposition shaping. *Mater. Des.* 55 **2014**, pp. 104-119.

[95] El Mehtedi, M.; Buonadonna, P.; D'Annibale, A.; Di Ilio, A. Effects of process parameters on the deformation energy in a sheet-bulk metal forming process for an automotive component. *Proc. CIRP* 99 **2021**, 248-253.

[96] Park, S.-H.; Liu, P.; Yi, K.; Choi, G.; Jhang, K.-Y.; Sohn, H. Mechanical properties estimation of additively manufactured metal components using femtosecond laser ultrasonics and laser polishing. *Int. J. Mach. Tools Manuf.* 166 **2021**, 103745.

[97] Yin, Y.J.; Sun, J.Q.; Guo, J.; Kan, X.F.; Yang, D.C. Mechanism of high yield strength and yield ratio of 316 L stainless steel by additive manufacturing. *Mater. Sci. Eng.* A 744 **2019**, 773-777.

[98] Al Mangour, B.; Kim, Y.-K.; Grzesiak, D.; Lee, K.-A. Novel $TiB_2$-reinforced 316L stainless steel nanocomposites with excellent room- and high-temperature yield strength developed by additive manufacturing. *Comp. Part* B 156 **2019**, 51-63.

[99] Zafari, A.; Xia, K. Nano/ultrafine grained immiscible Fe-Cu alloy with ultrahigh strength produced by selective laser melting. *Mater. Res. Lett.* 9 **2021**, *6*, pp. 247-254.

[100] Landau, L.D.; Lifshitz, E.M. (1986). Theory of Elasticity. Vol. 7 (3rd ed.). Butterworth-Heinemann. ISBN 978-0-7506-2633-0.

[101] Ma, Z.; Zhang, W.; Du, P.; Zhu, X.; Krishnaswamy, S.; Lin, L.; Lei, M. Nondestructive measurement of elastic modulus for thermally sprayed WC-Ni coatings based on acoustic wave mode conversion by small angle incidence. *NDT Int.* 94 **2018**, pp. 38-46.

[102] Malischewsky, P.G.; Tuan, T.T. A special relation between Young's modulus, Rayleigh-wave velocity, and Poisson's ratio. *J. Acoust. Soc. Am.* 126 (6) **2009**, pp. 2851-2853.

[103] Bayón, A.; Gascón, F.; Nieves, F.G. Estimation of dynamic elastic constants from the amplitude and velocity of Rayleigh waves. *J. Acoust. Soc. Am.* 117 (6) **2005**, pp. 3469-3477.

[104] Yan, S.; Wu, D.; Huang, Y.; Liu, N.; Zhang, Y.; Niu, F.; Ma, G. C fiber toughening $Al_2O_3$-$ZrO_2$ eutectic via ultrasound-assisted directed laser deposition. *Mater. Lett.* 235 **2019**, pp. 228-231.

[105] Zhang, D.; Li, Y.; Wang, H.; Cong, W. Ultrasonic vibration-assisted laser directed energy deposition in-situ synthesis of NiTi alloys: Effects on microstructure and mechanical properties. *J. Manuf. Proc.* 60 **2020**, pp. 328-339.

[106] Wang, X.H.; Liu, S.S.; Zhao, G.L.; Zhang, M.; Ying, W.L. In-situ formation ceramic particles reinforced Fe-based composite coatings produced by ultrasonic assisted laser melting deposition process. *Opt. Las. Technology* 136 **2021** 106746.

[107] Todaro, C.J.; Easton, M.A.; Qiu, D; Zhang, D.; Bermingham, M.J.; Lui, E.W.; Brandt, M.; StJohn, D.H.; Qian, M. Grain structure control during metal 3D printing by high-intensity ultrasound. *Nat Commun*. 11 **2020**, 142.

[108] Becker, T.H.; Kumar, P.; Ramamurty, U. Fracture and fatigue in additively manufactured metals. *Acta Mater.* https://doi.org/10.1016/j.actamat.2021.117240

[109] Zhang, X.; Sun, C.; Pan, T.; Flood, A.; Zhang, Y.; Li, L.; Liou, F. Additive manufacturing of copper–H13 tool steel bi-metallic structures via Ni-based multi-interlayer. *Additive Manuf.* 36 **2020**, 101474.

[110] Khalid Imran, M.; Masood, S.H.; Brandt, M.; Bhattacharya, S.; Mazumder, J. Direct metal deposition (DMD) of H13 tool steel on copper alloy substrate: Evaluation of mechanical properties. *Mater. Sci. Eng.* A 528 **2011**, pp. 3342-3349.

[111] Shi, H.; Duan, L.; Tan, S.; Fang, X. Influence of pre-alloying on Fe-Cu based metal matrix composite. *J. Alloys of and Compounds* 868 **2021**, 159134.

[112] Khosravani, M.R.; Berto, F.; Ayatollahi, M.R.; Reinicke, T. Fracture behavior of additively manufactured components: A review. *Theor. Appl. Fract. Mech.* 109 **2020**, 102763.

[113] Carter, L.N.; Attalah, M.M.; Reed, R.C. Laser Powder Bed Fabrication of nickel-base superalloys: influence of parameters; characterization, quantification and mitigation of cracking. *Superalloys* **2012**, pp. 577-586.

[114] Sutton, A.T.; Kriewall, C.S.; Leu, M.C.; Newkirk, J.W. Powder characterization techniques and effects of powder characteristics on part properties in powder-bed fusion processes. *Virtual Phys. Prototyp.* 12 **2017**, pp. 3-29.

[115] Filimonov, A.M.; Rogozin, O.A.; Dubinin, O.N.; Kuzminova, Y.O.; Shibalova, A.A.; Okulov, I.V.; Akhatov, I.S.; Evlashin, S.A. Modification of Mechanical Properties in Directed Energy Deposition by a Static Magnetic Field: Experimental and Theoretical Analysis. *Materials* 14 **2021**, 5190.

[116] Osipovich, K.S.; Gurianov, D.A.; Chumaevsky, A.V. Influence of 3D-Printing Parameters on Bimetallic Products Manufacturing Process of Cu-Fe System. *IOP Conf. Ser.: Mater. Sci. Eng.* 1079 **2021**, 042089.

[117] Hofmann, D.C.; Roberts, S.; Otis, R.; Kolodziejska, J.; Dillon, R.P.; Suh, J.; Shapiro, A.A.; Liu, Z.; Borgonia, J. Developing gradient metal alloys through radial deposition additive manufacturing. *Sci. Rep.* 4 **2014** 5357.

# Supporting Information

**Physical Properties of Cu-Fe System Functionally Graded and Multimaterial Energy Deposited Structures after the DED**

*Konstantin Makarenko\*, Oleg Dubinin, Stepan Konev, and Igor Shishkovsky*

**Figure S1.** Engineering stress-strain curve of Al bronze + stainless steel (1:1) (Group 1, specimen 1)

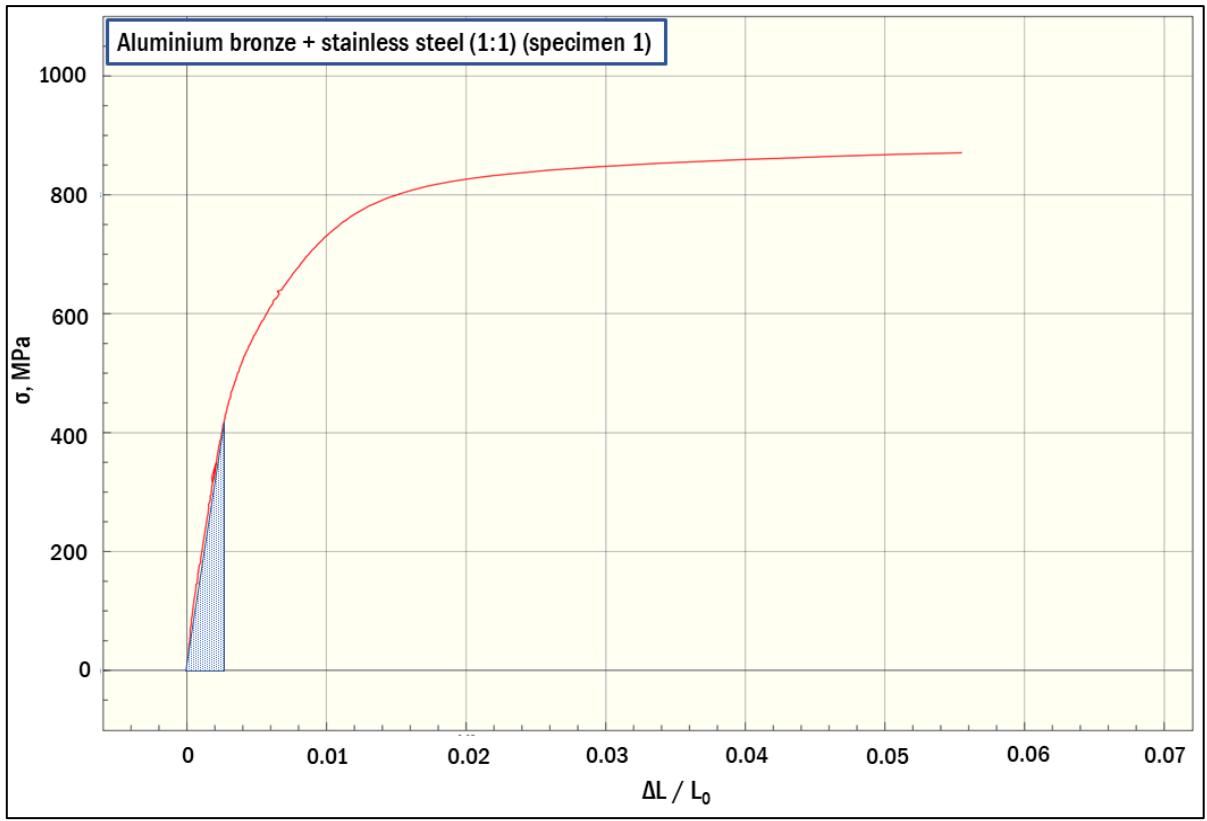

**Figure S2.** Engineering stress-strain curve of Al bronze + stainless steel (1:1) (Group 1, specimen 2)

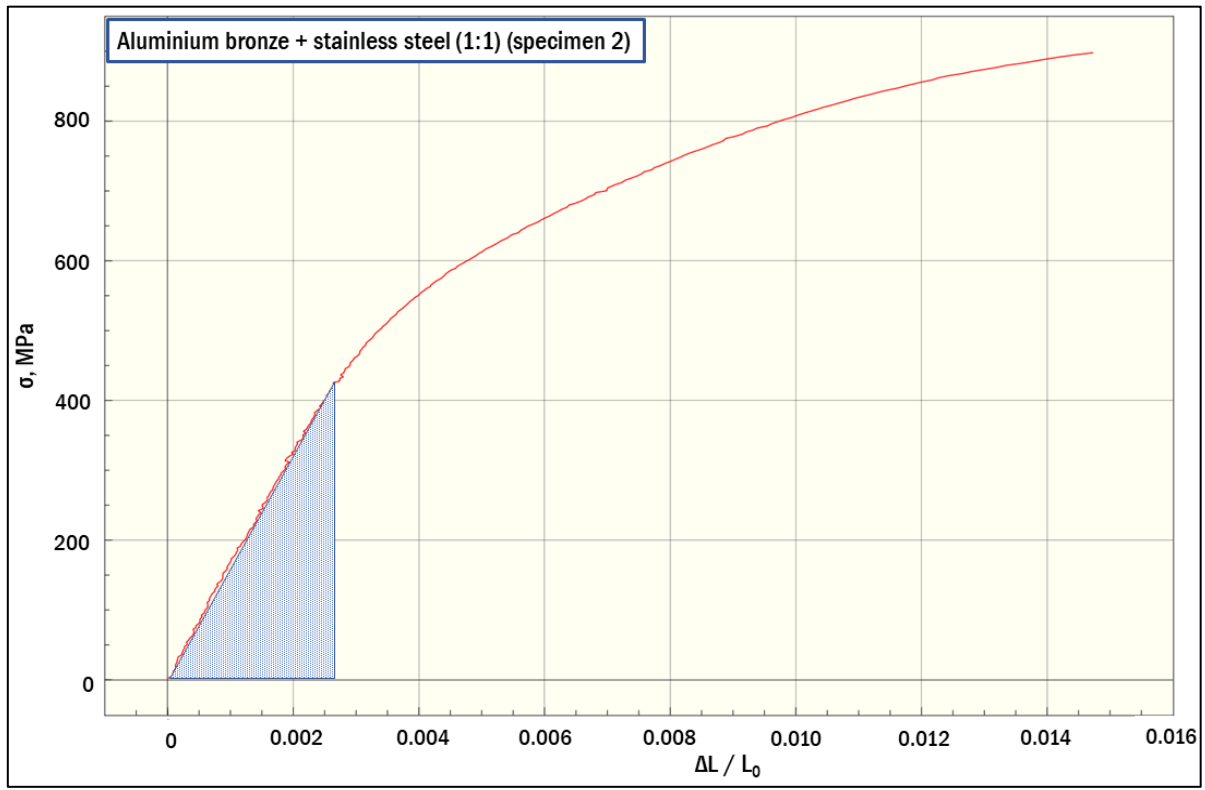

**Figure S3.** Engineering stress-strain curve of Cr bronze + stainless steel (1:1) (Group 2, specimen 1)

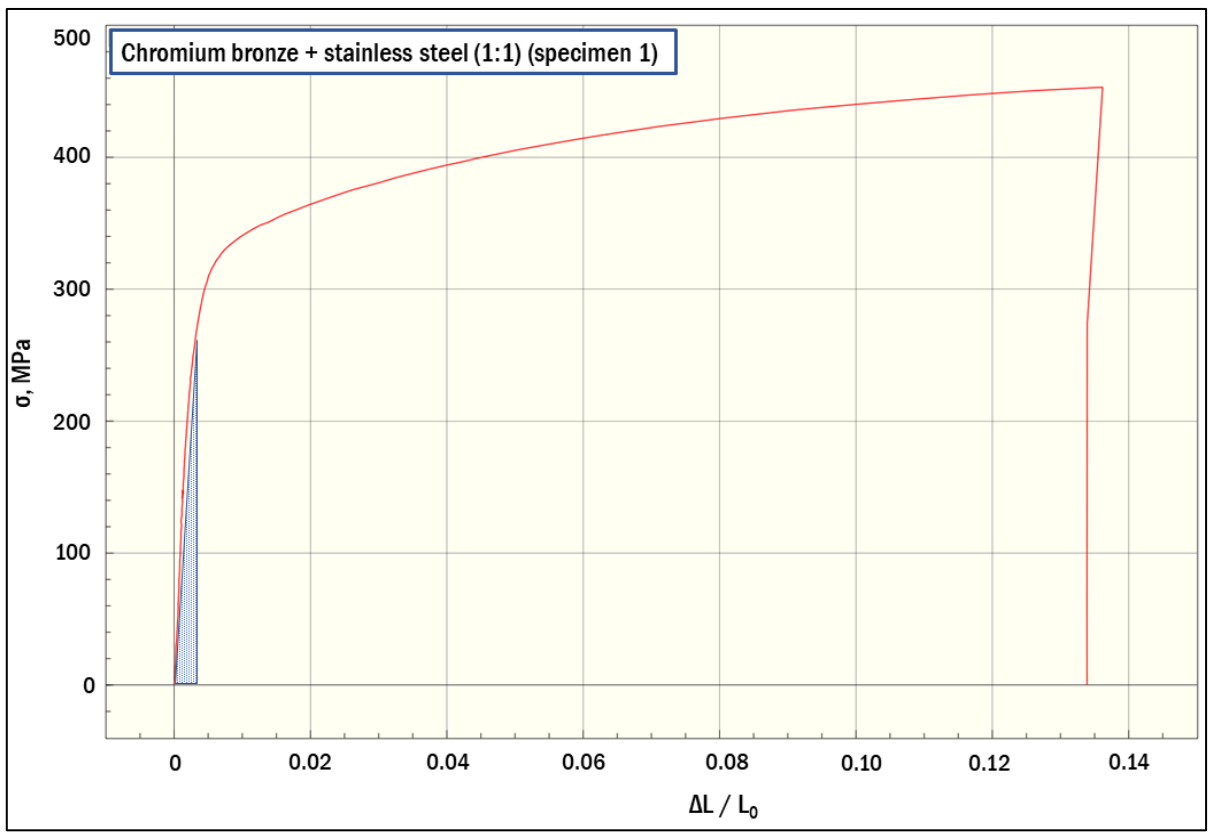

**Figure S4.** Engineering stress-strain curve of Cr bronze + stainless steel (1:1) (Group 2, specimen 2)

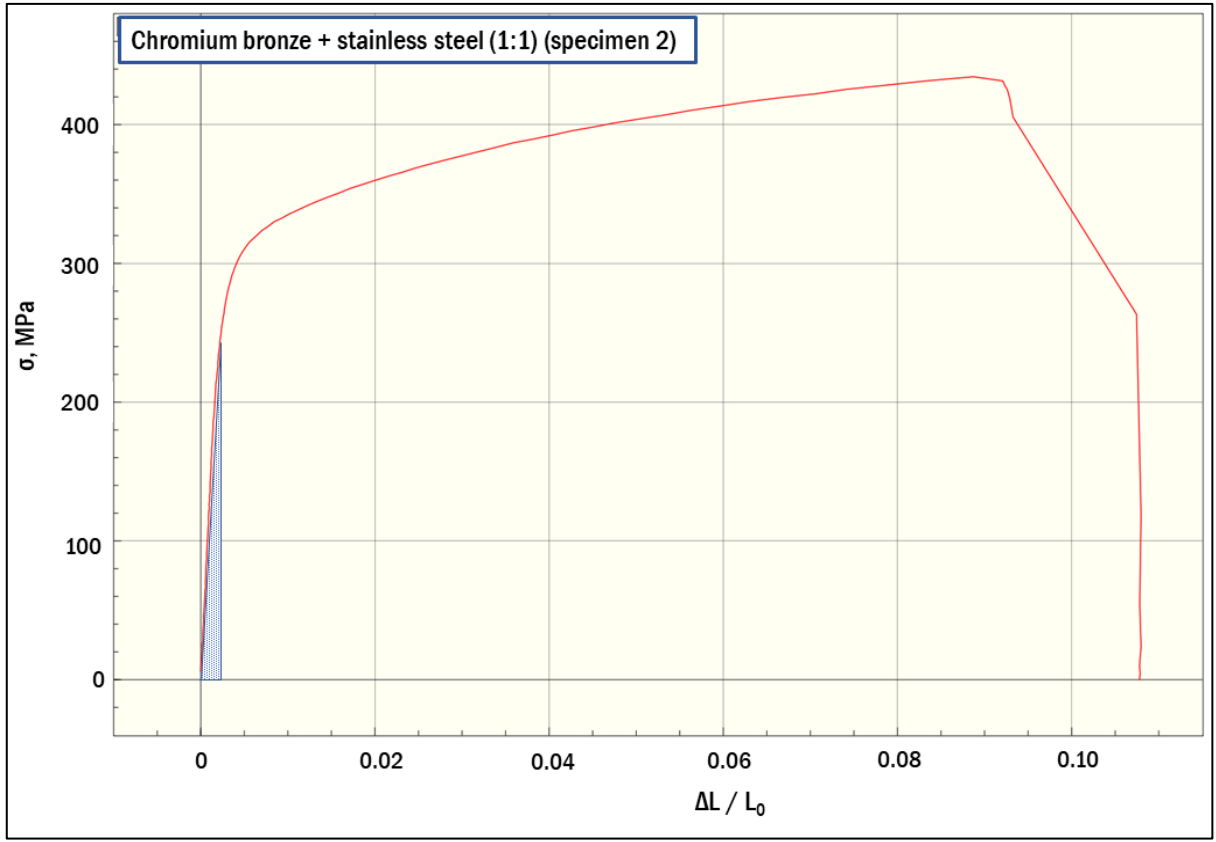

**Figure S5.** Engineering stress-strain curve of Sn bronze + stainless steel (1:1) (Group 3, specimen 1)

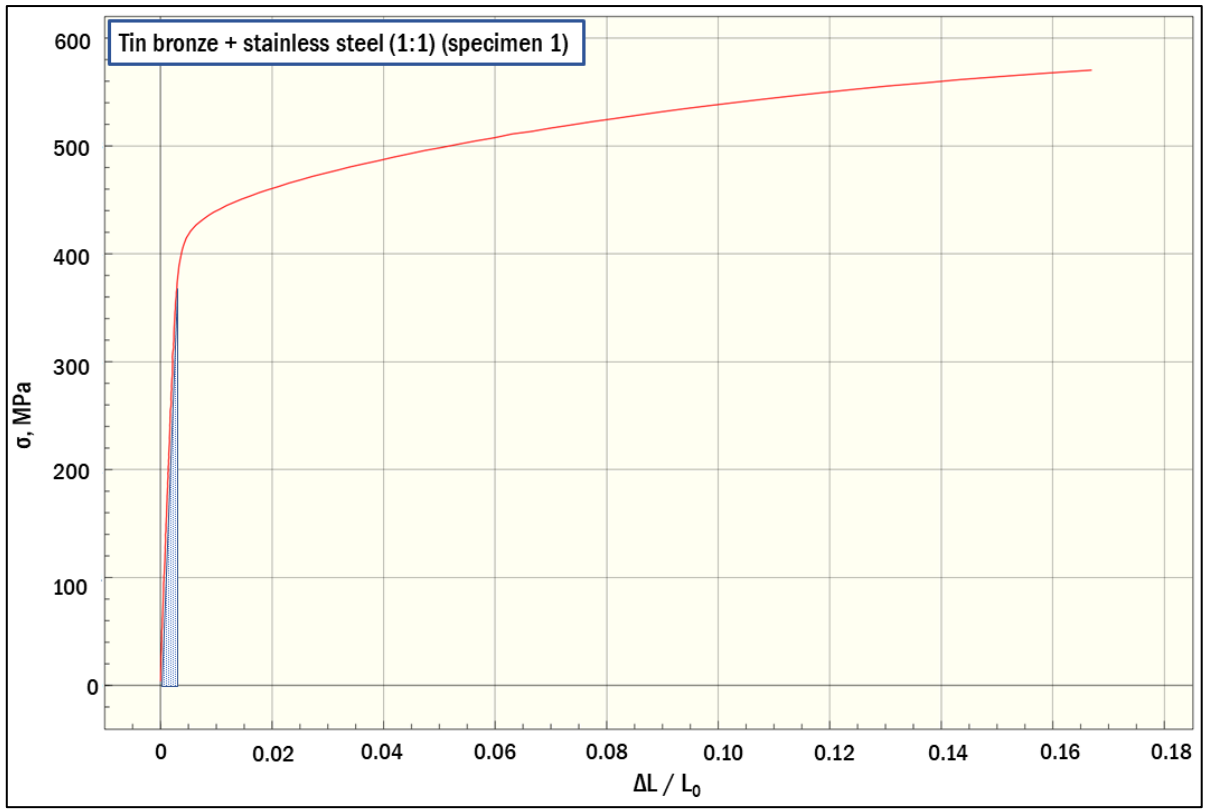

**Figure S6.** Engineering stress-strain curve of Sn bronze + stainless steel (1:1) (Group 3, specimen 2)

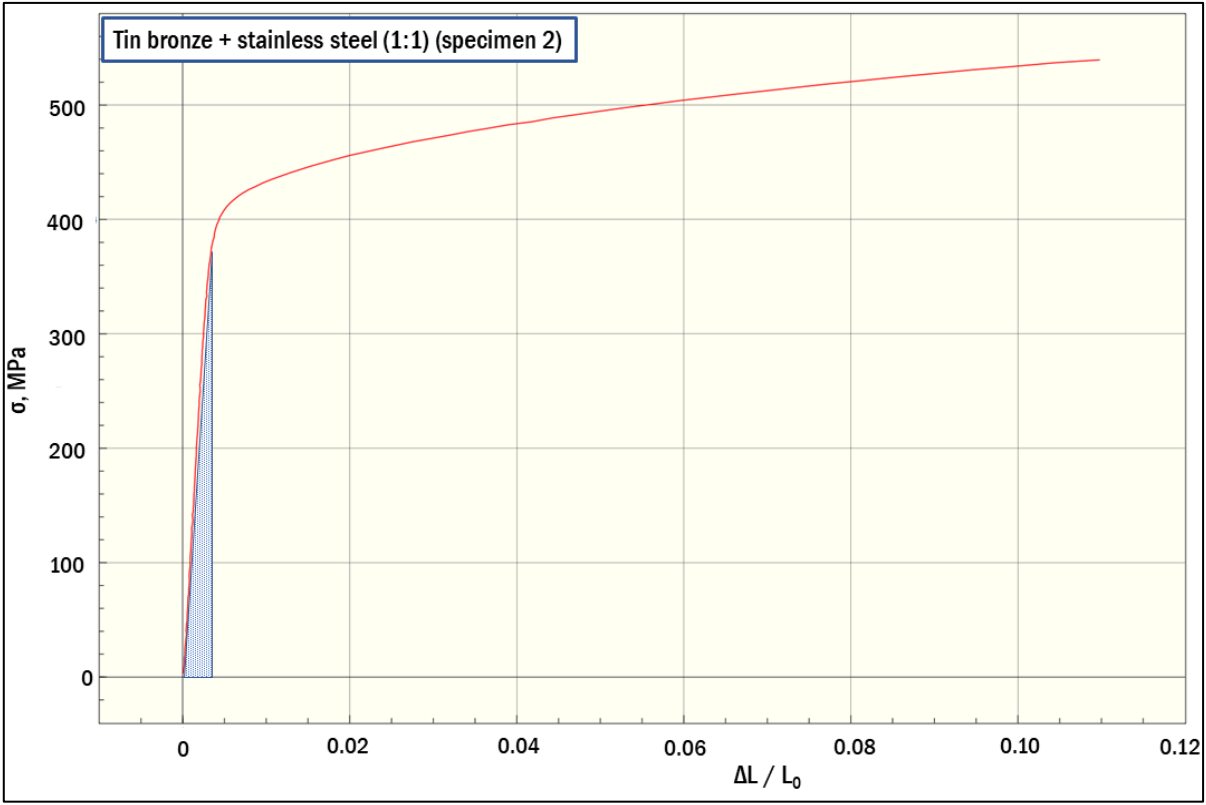

**Figure S7.** Engineering stress-strain curve of Al bronze + stainless steel (3:1) (Group 4, specimen 1)

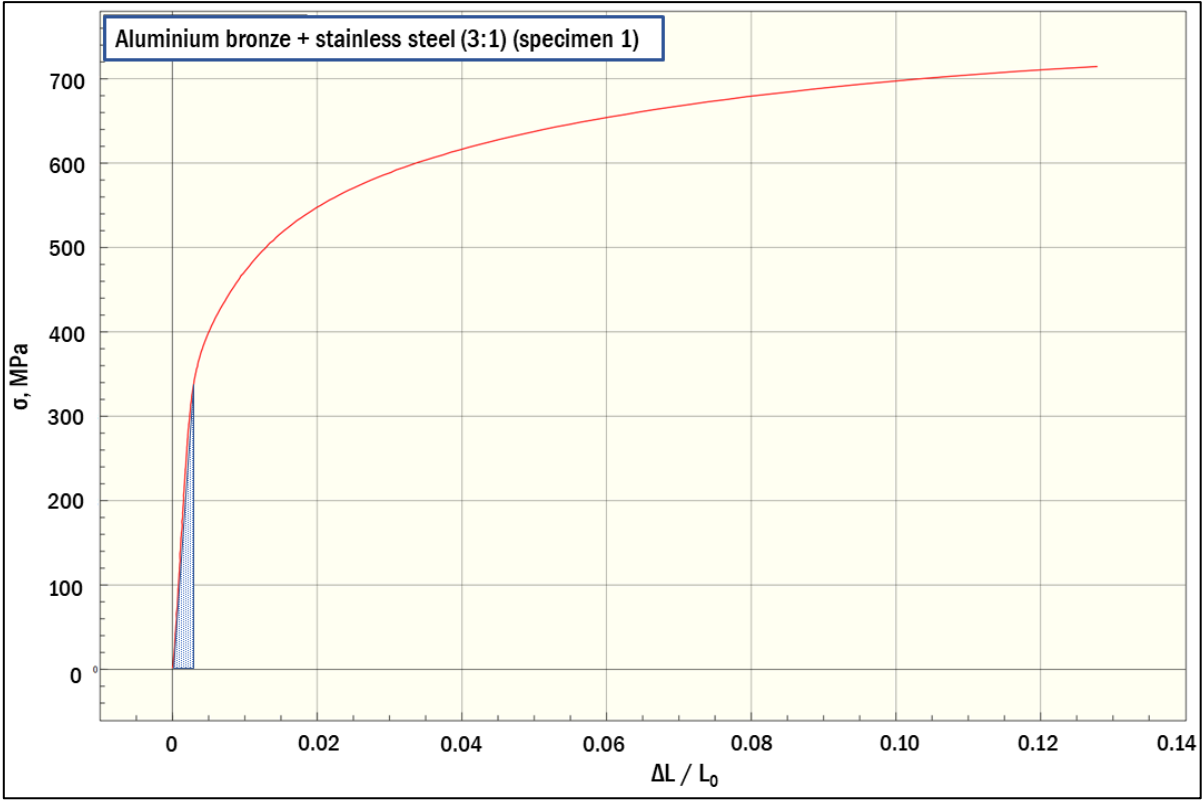

**Figure S8.** Engineering stress-strain curve of Al bronze + stainless steel (3:1) (Group 4, specimen 2)

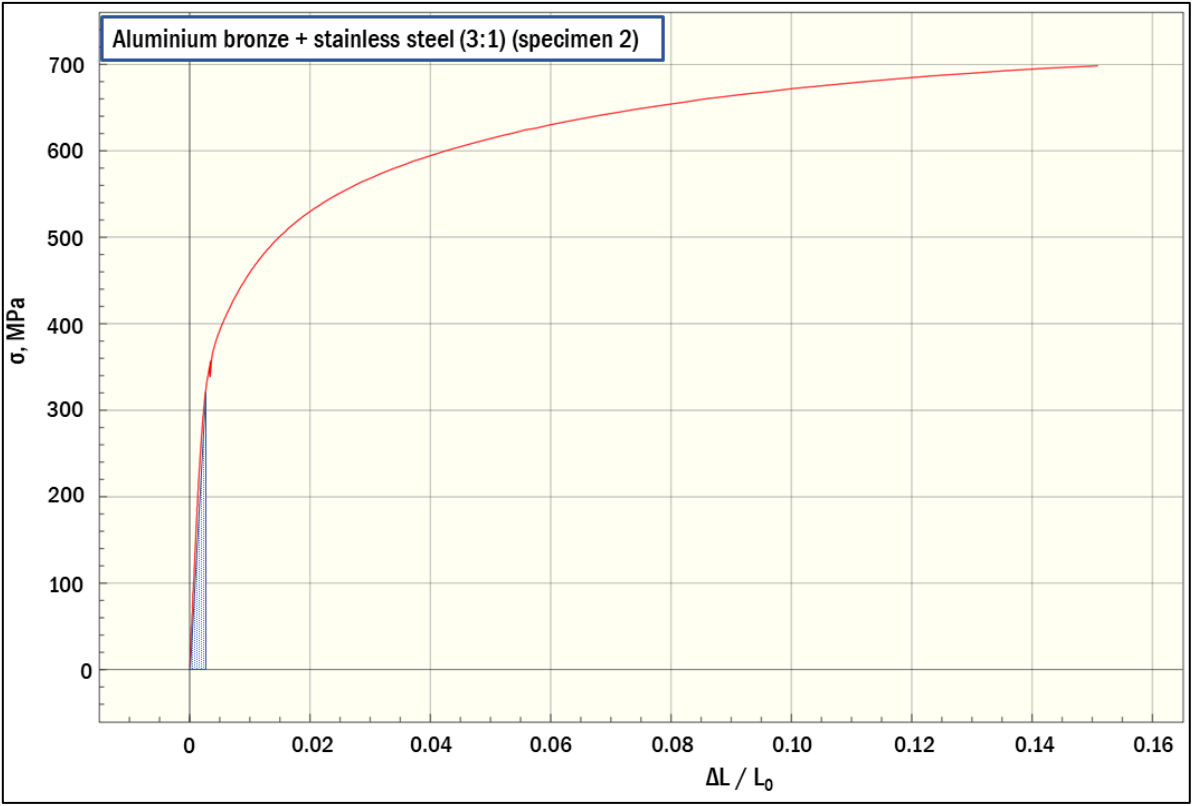

**Figure S9.** Engineering stress-strain curve of Cr bronze + stainless steel (3:1) (Group 5, specimen 1)

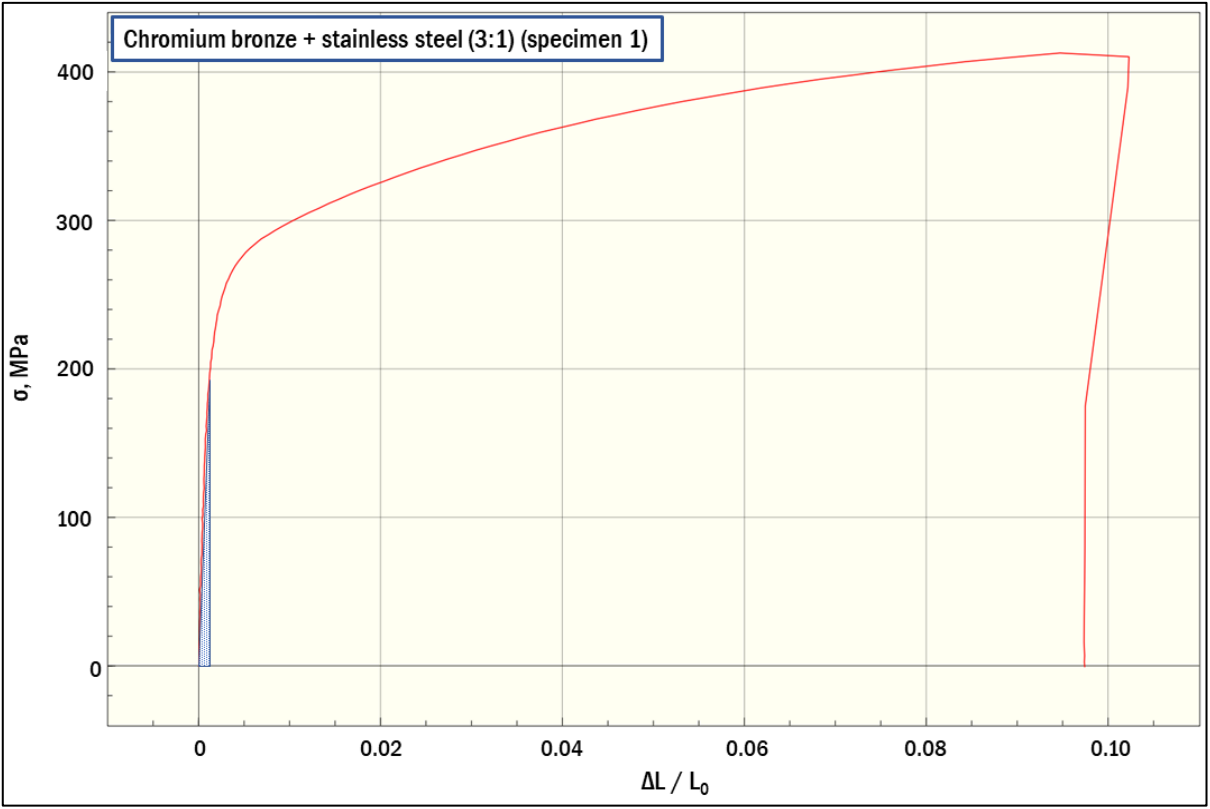

**Figure S10.** Engineering stress-strain curve of Sn bronze + stainless steel (3:1) (Group 6, Specimen 1)

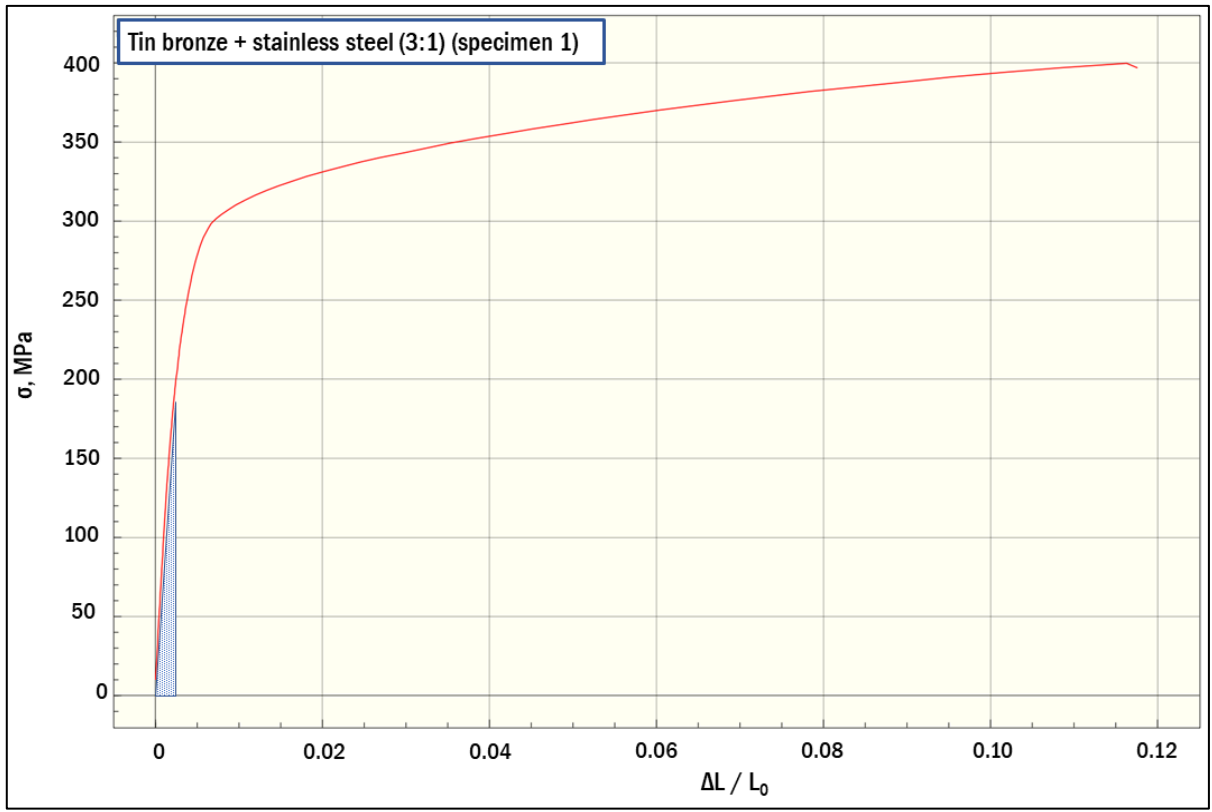

**Figure S11.** Engineering stress-strain curve of Sn bronze + stainless steel (3:1) (Group 6, specimen 2)

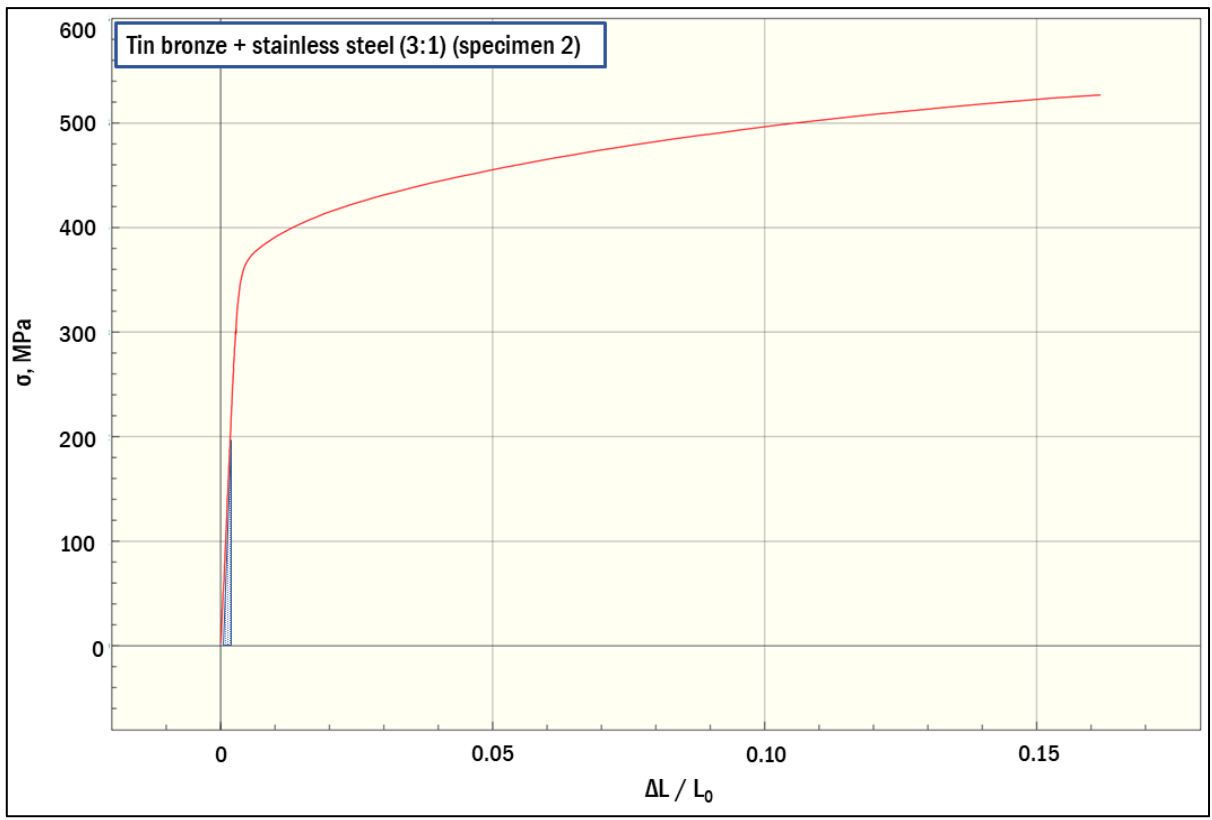

**Figure S12.** Engineering stress-strain curve of Sn bronze + stainless steel (3:1) (Group 6, specimen 3)

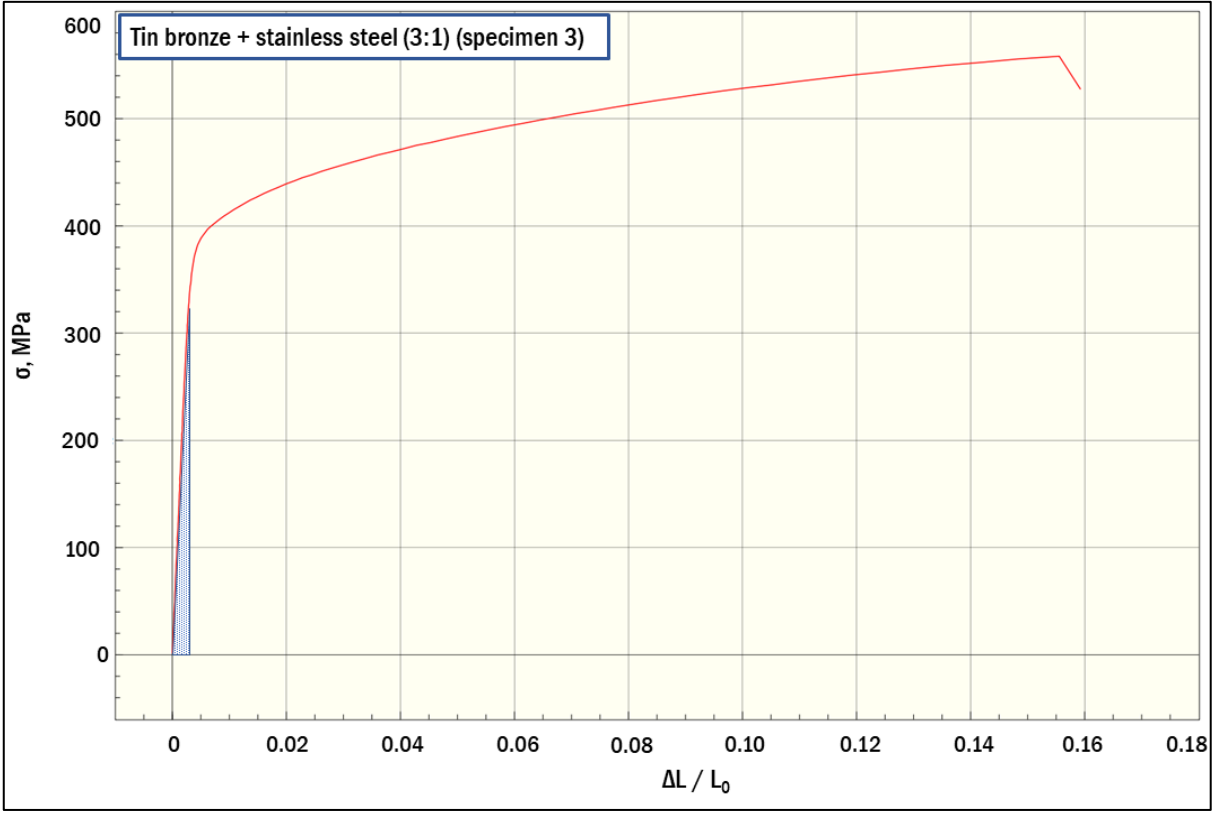

**Figure S13.** Engineering stress-strain curve of Al bronze + stainless steel (alt.) (Group 7, specimen 1)

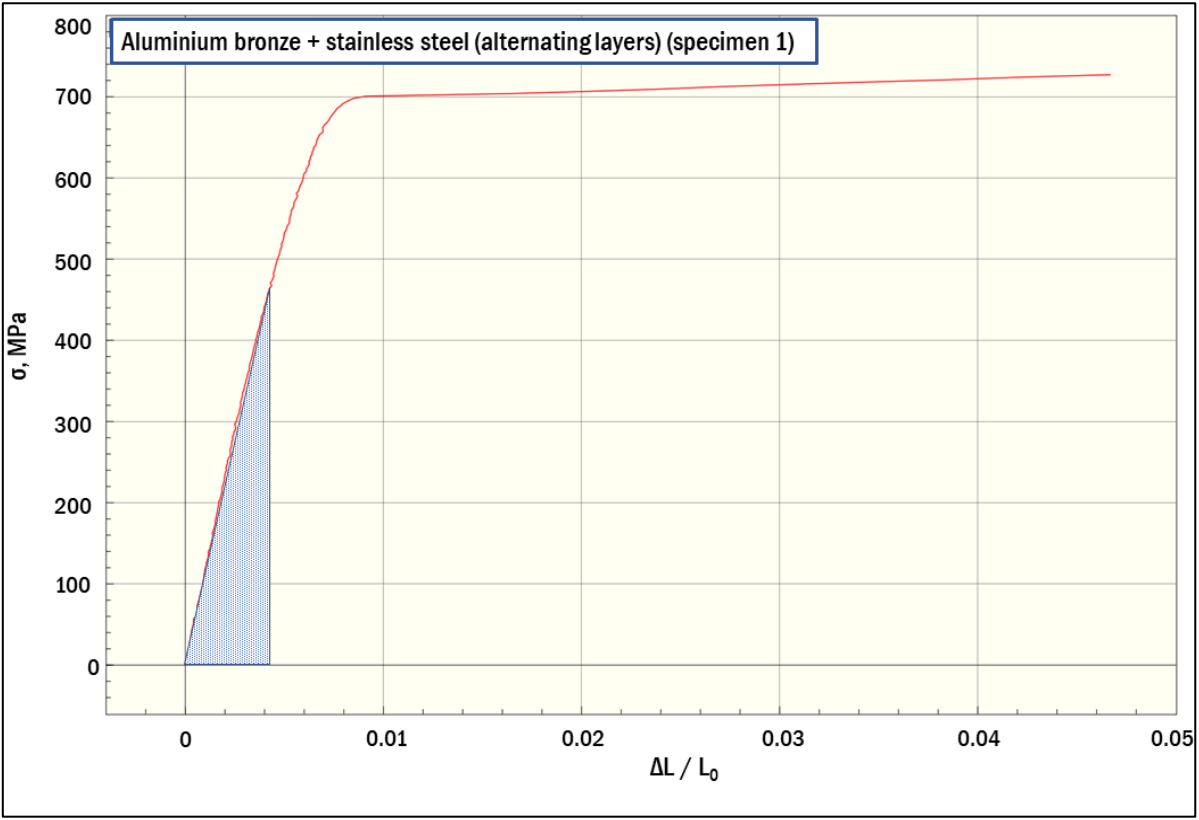

**Figure S14.** Engineering stress-strain curve of Al bronze + stainless steel (alt.) (Group 7, specimen 2)

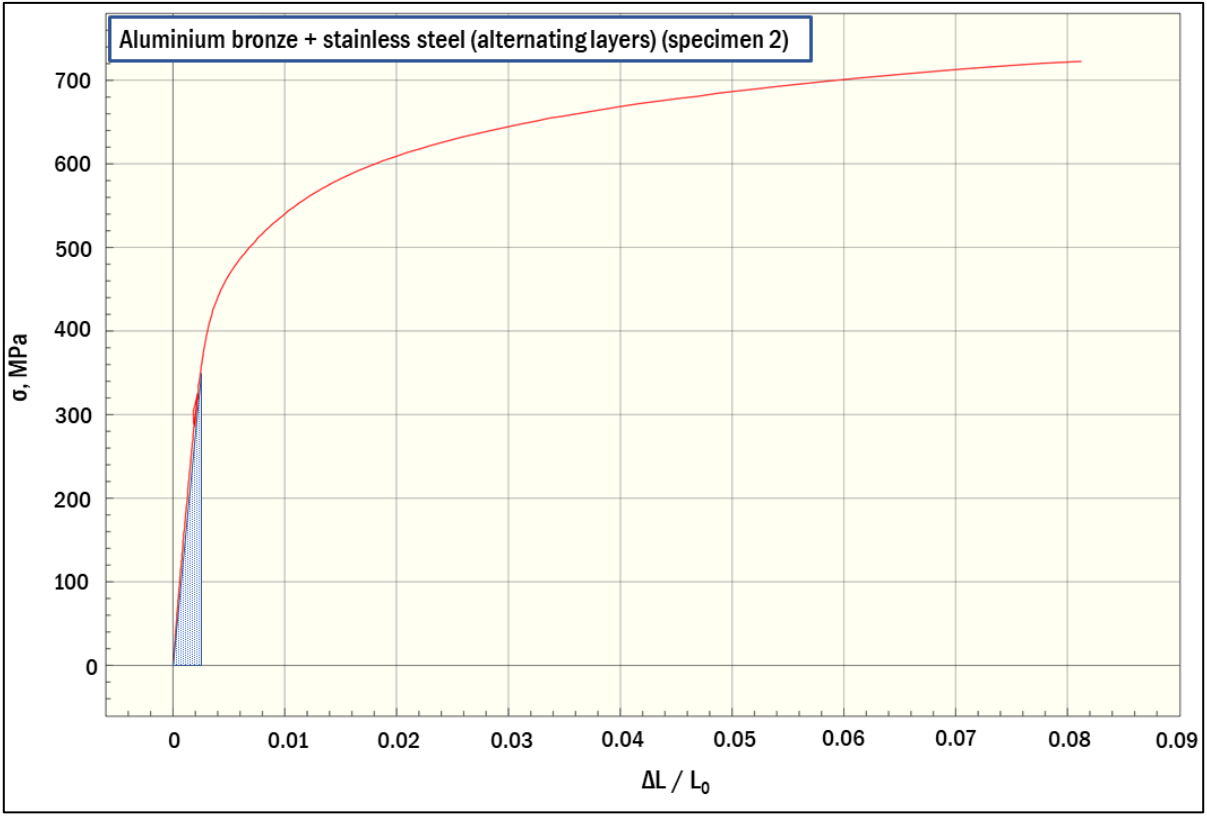

**Figure S15.** Engineering stress-strain curve of Cr bronze + stainless steel (alt.) (Group 8, specimen 1)

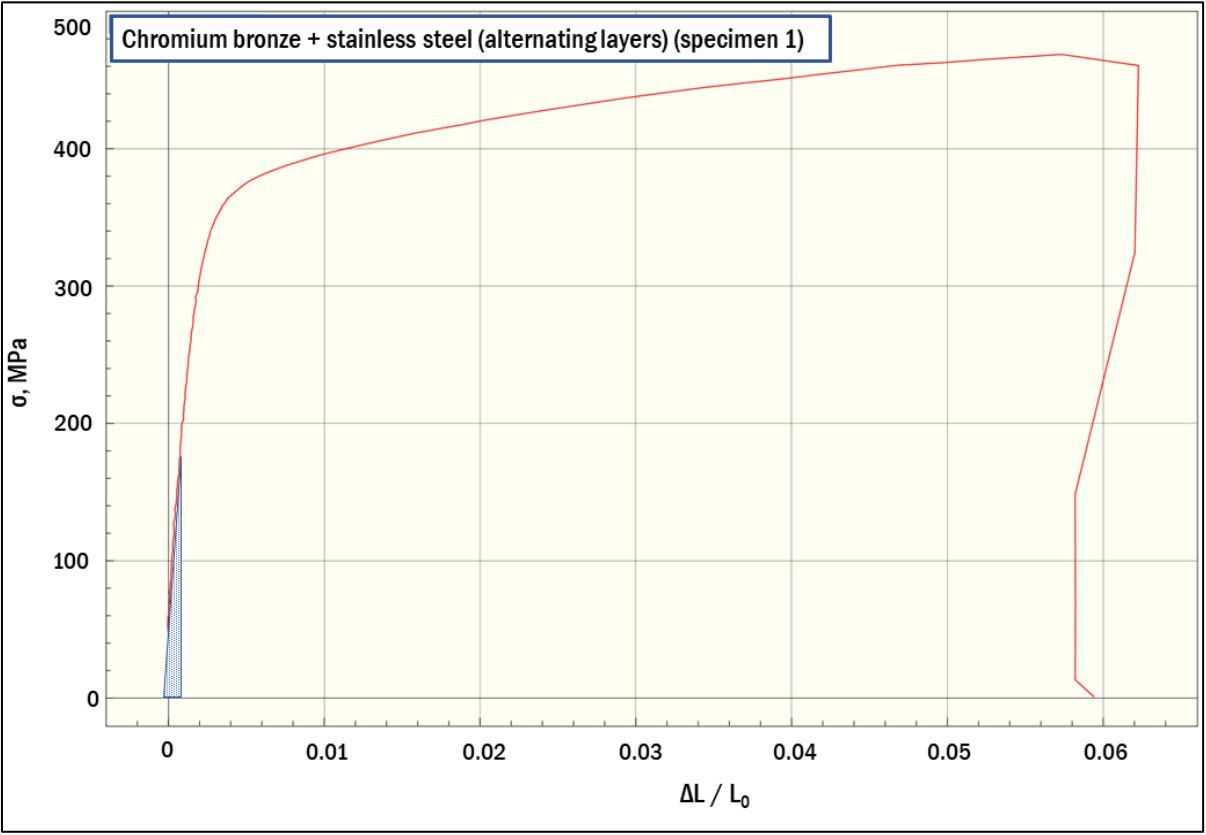

**Figure S16.** Engineering stress-strain curve of Cr bronze + stainless steel (alt.) (Group 8, specimen 2)

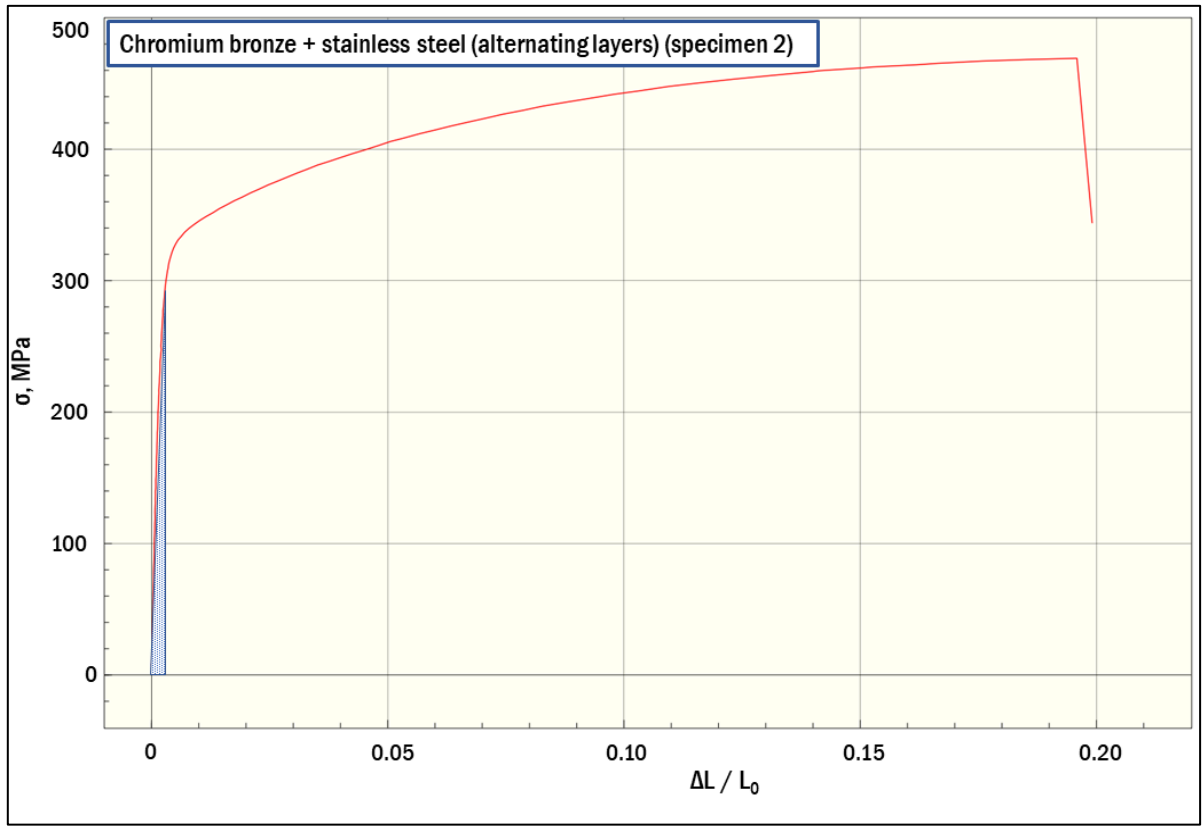

**Figure S17.** Engineering stress-strain curve of Cr bronze + stainless steel (alt.) (Group 8, specimen 3)

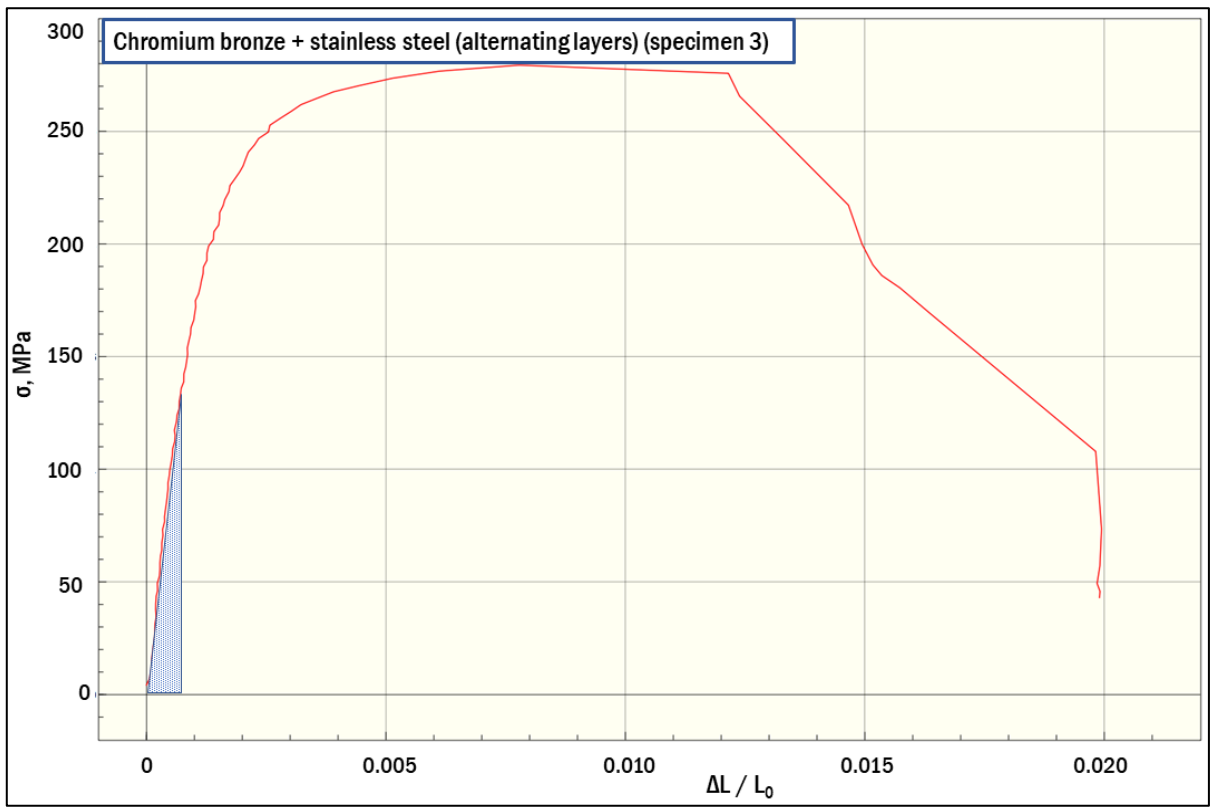